\documentclass[a4paper,11pt]{article}
\usepackage{jheppub} 
\usepackage{graphicx} 
\usepackage{array}
\usepackage{multirow}
\usepackage{multicol}
\usepackage{colortbl}
\usepackage{tcolorbox}
\usepackage{capt-of}
\usepackage{bm}		    
\usepackage[normalem]{ulem}
\usepackage{slashed}
\usepackage{bigints}
\usepackage[LGR,T1]{fontenc}
\usepackage[latin9]{inputenc}
\usepackage{textcomp}
\usepackage{mathrsfs}
\usepackage{amsmath}
\usepackage{amssymb}
\usepackage{multicol}
\usepackage{braket}
\usepackage{array}
\usepackage{makecell}

\definecolor{bgrey}{rgb}{0.4, 0.6, 0.8}

\newcommand{\dvec}{\bm \nabla}

\def\be{\begin{equation}}
\def\ee{\end{equation}}
\def\figs/B{B}
\def\bea{\begin{eqnarray}}
\def\eea{\end{eqnarray}}
\def\bg{\begin{eqnarray}}
\def\nd{\end{eqnarray}}

\newcommand{\xvec}{{\bm x}}
\newcommand{\kvec}{{\bm k}}
\newcommand{\qvec}{{\bm q}}
\newcommand{\early}{{\mbox{\tiny early}}}
\newcommand{\late}{{\mbox{\tiny late}}}
\newcommand{\ii}{\mathrm{i}}
\newcommand{\dd}{\mathrm{d}}

\def\at{{\tilde{a}}}
\def\kt{{\tilde{k}}}
\def\mt{{\tilde{m}}}
\def\Ht{{\tilde{H}}}
\def\Rt{{\tilde{R}}}
\def\etat{{\tilde{\eta}}}
\def\omegat{{\tilde{\omega}}}
\def\nt{{\tilde{n}}}
\def\ant{{\tilde{a}^3\tilde{n}_k}}
\def\chit{{\tilde{\chi}}}
\def\chitk{{\tilde{\chi}_k}}

\newcommand{\MPl}{{M_\mathrm{Pl}}}
\newcommand{\GeV}{{\mathrm{GeV}}}

\newcommand{\rd}{{\mbox{\tiny RD}}}
\newcommand{\md}{{\mbox{\tiny MD}}}
\newcommand{\rh}{{\mbox{\tiny RH}}}
\usepackage{titlesec}
\titleformat{\paragraph}[hang]{\normalfont\normalsize\bfseries}{\theparagraph}{1em}{}
\titlespacing*{\paragraph}{0pt}{3.25ex plus 1ex minus .2ex}{1.5ex plus .2ex}

\title{\boldmath \center Gravitational Particle Production of Scalars: \\ Analytic and Numerical Approaches \\ Including Early Reheating}

\author[a]{Leah Jenks,}
\author[a,b]{Edward W. Kolb,}
\author[a]{Keyer Thyme}

\affiliation[a]{Kavli Institute for Cosmological Physics, \\
The University of Chicago,  5640 South Ellis Avenue, Chicago, IL 60637, U.S.A.}
\affiliation[b]{Enrico Fermi Institute,\\
The University of Chicago,  5640 South Ellis Avenue, Chicago, IL 60637, U.S.A.}

\emailAdd{ljenks@uchicago.edu}
\emailAdd{Rocky.Kolb@uchicago.edu}
\emailAdd{thyme@uchicago.edu}

\abstract{Cosmological gravitational particle production (GPP) is a generic mechanism by which particles are produced during the inflationary epoch. In this work we consider the GPP of massive scalars in an effort to fully understand the spectrum of the produced particles. We consider scalars which are conformally or minimally coupled to gravity, as well as models of both early and late reheating. We numerically calculate the particle production in each scenario and compare the results to analytic approximations from boundary matching, Stokes phenomenon, steepest descent, and inflaton scattering methods. For each method, we describe the regime of validity and show that there is good agreement between the analytic and numerical results.}

\begin{document}
\maketitle
\flushbottom

\section{Introduction}
\label{sec:intro}

 It is well known that particle production can occur in the early universe as a result of the rapid expansion of spacetime during inflation. This effect is known as cosmological gravitational particle production (GPP) \cite{Parker:1969au, Parker:1971pt, Ford:2021syk,Chung:1998zb,Chung:1998ua},and arises from the dynamics of a quantum field on a curved spacetime background \cite{DeWitt:1975ys, Birrell:1982ix, Parker:2009uva}. GPP is a generic process with very few necessary ingredients, only requiring a massive spectator field on top of an inflating spacetime background. GPP is particularly relevant in the context of dark matter, which is known to make up a quarter of the energy density of the universe, though its exact particle nature is unknown \cite{Planck:2018jri}. As it requires no coupling to standard model particles, GPP is an attractive candidate to produce some or all of the dark matter, yielding `completely dark' matter. GPP has been studied for a wide variety of inflation models and for particles ranging from scalars to spin-s, e.g., \cite{Chung:1998zb, Graham:2015rva, Ema:2015dka, Ema:2016hlw, Ema:2018ucl, Ema:2019yrd, Kolb:2020fwh,Ahmed:2020fhc, Alexander:2020gmv, Kolb:2021xfn, Kolb:2021nob,Kolb:2022eyn,Kolb:2023dzp, Capanelli:2023uwv,Kaneta:2023uwi, Maleknejad:2022gyf, Capanelli:2024pzd,Capanelli:2024nkf, Racco:2024aac, Choi:2024bdn, Verner:2024agh} (see \cite{Kolb:2023ydq} and references therein for a more detailed review). For each of these models there is a region of parameter space in which particles produced gravitationally can be the dark matter, but GPP can also be a relevant production mechanism for other cosmological relics.
 
 A variety of methods have been employed to study GPP. In some cases, it is possible to approximate particle production using various analytical methods, but for a full analysis, numerical methods must be employed. In order to understand the spectra of produced particles, it is thus important to fully understand where each analytic method is valid and how these approximations compare to the full numerical calculation. Previous work in the literature have discussed the aspects of the various analytic methods individually, but without detailed discussion of when they are valid. The relevant analytic methods are as follows: first, one can trace out the evolution of each mode as it transitions from the inflationary quasi-de Sitter regime into matter and/or radiation domination by solving the mode equation in each region and matching at the boundaries. This method we refer to as `boundary matching.' One can also employ the Stokes phenomenon  \cite{Hashiba:2021npn} and steepest descent \cite{PhysRevD.67.083514} methods to approximate GPP. Finally, GPP can be analytically approximated by considering inflaton scattering processes, as discussed in e.g. \cite{Tang_2017, Basso_2022}.  In this work, we perform a detailed analysis of the GPP for scalars, considering both analytic and numerical methods for the GPP of both conformally and minimally-coupled scalars. For the background inflationary dynamics we consider quadratic inflation and both early and late reheating scenarios, the former of which do not yet appear in the literature. We analytically approximate the GPP in each of these cases, and outline the regime of validity of each analytic method. We additionally numerically calculate the GPP and compare with the analytic results, finding good agreement between the two. We aim to present a comprehensive picture of the analytic and numerical methods for calculating GPP, with a particular focus on the full spectrum of analytic approximations.
 
 The structure of the paper is as follows: in Sec.~\ref{sec:GPP} we provide a brief overview of the GPP procedure and the setup for a massive scalar field on an inflationary background. In Sec.~\ref{sec:numerical}, we present comprehensive numerical results for the GPP of scalars, considering both conformal and minimal couplings as well as both early and late reheating. We then turn to a comparison with analytic results, summarizing the analytic methods and results in Sec.~\ref{sec:methods-results}. Finally we conclude with a discussion in Sec.~\ref{sec:conclusions}. Technical details of the analytic calculations can be found in the appendices. 

 Throughout the paper, the following conventions are used: we use a mostly minus metric signature and employ natural units such that $c = \hbar = 1$. Greek indices, $\mu,\nu,\ldots$ indicate a sum over all four spacetime indices, while Latin indices indicate a spatial sum only. We furthermore take $\MPl=(8\pi G)^{-1/2}$.

\section{Inflation and Gravitational Particle Production}
\label{sec:GPP}

In this section we provide a brief overview of the mechanics of GPP and the assumptions of the background evolution of the Universe during inflation. Detailed descriptions of GPP be found in several textbooks, such as \cite{Birrell:1982ix, Mukhanov:2005sc,Mukhanov:2007zz,Parker:2009uva,Baumann:2022mni}, as well as more recent reviews, such as \cite{Kolb:2023ydq,Ford:2021syk,Armendariz-Picon:2023gyl}, so here we focus just on the key features.

GPP is a semiclassical process, meaning that the background gravitational field is classical, whereas the spectator field is quantized. For the background evolution, we assume a cosmological spacetime that is homogeneous, isotropic, and expanding. This background is described by the spatially flat Friedmann-Robertson-Walker (FRW) metric with line element given by $ds^2=a^2(\eta)\left(d\eta^2-d\xvec^2\right)$, where $\eta$ denotes conformal time, $a$ denotes scale factor, and the two quantities are related by $a\,d\eta=dt$. For the early evolution of the Universe, we assume starting in a (quasi-)de Sitter (dS) phase of accelerated expansion (inflation), followed by a period of evolution as a matter-dominated (MD) universe, then ``reheating'' to a radiation-dominated (RD) Universe. 

This evolution is conveniently engineered by assuming that there is a single scalar field - the inflaton field $\varphi$, whose potential energy drives an inflationary phase with $\ddot{a}>0$ (dot is derivative with respect to coordinate time). Inflation ends when the kinetic energy of the inflaton field dominates the potential energy and $\ddot{a}$ first becomes negative. The inflaton field then undergoes coherent oscillations about the minimum of its potential, and its energy density redshifts as matter ($\rho_\varphi\propto a^{-3}$) when averaged over several oscillations. During this MD era, the energy density evolves as $\exp{(-\Gamma_\varphi t)}$, where the decay width, $\Gamma_\varphi$, is the rate at which the inflaton decays into relativistic standard model particles. Eventually, the radiation dominates the inflaton energy density, at which point the Universe ``reheats'' and enters the RD epoch. 

 We consider two reheating scenarios: early reheating and late reheating. Late reheating occurs when $(a_\rh/a_e)>(m/H_e)^{-2/3}$, i.e., when $H_\rh < m$. In numerical calculations, late reheating is implemented by setting $\Gamma_\varphi=0$. This corresponds to a reheat time of $a_{\rh} = \infty$; we assume that all of the particle production occurs sufficiently before the reheating epoch such that the particle production arises from the quasi-de Sitter and matter-dominated eras and any details of reheating are irrelevant. For early reheating, there exists a limiting case of ``immediate reheating,'' where reheating occurs instantaneously at the end of inflation. 

For reheating we follow the analysis in Ref.\ \cite{PhysRevD.64.023508}. Although the conversion of the inflaton energy density into radiation is effective throughout the MD and RD eras, we use ``reheating'' to refer to the moment where $\rho_\varphi=\rho_\mathrm{R}$. The radiation energy density at reheating can be expressed as $\rho_\rh=(\pi^2g_\star/30) T_\rh^4$, where $g_\star$ is the effective number of relativistic degrees of freedom at $T_\rh$. From the Friedmann equation, we obtain $3\MPl^2H_\rh^2=\rho_\mathrm{R}(T_\rh)$. By setting $\rho_\varphi=\rho_R$ and $H=\Gamma_\varphi$, the reheat temperature can be expressed as
\be
T_\rh = \left(\Gamma_\varphi\MPl\right)^{1/2}\left(\frac{90}{\pi^2 g_\star}\right)^{1/4} \ .
\ee
The production of neutrinos during nucleosynthesis places a lower bound of $T_{\rh}^{\rm min} \approx 5$ MeV \cite{de_Salas_2015}.  The maximum $T_{\rh}$ depends the energy scale of inflation, as  
\be
T_\rh < \left(H_e\MPl\right)^{1/2}\left(\frac{90}{\pi^2 g_\star}\right)^{1/4}
\ee 
 
For inflation itself, there are a wide range of models describing the inflationary dynamics (see, e.g., \cite{martin2023encyclopaedia} for a summary and review). For simplicity, we here we adopt a simple model of single-field quadratic inflation. The action for the inflaton field in the quadratic model is 
\be
S_\varphi = \int \dd^4x \sqrt{-g}\left[\tfrac{1}{2}g^{\mu\nu} \nabla_\mu\varphi\nabla_\nu\varphi 
- \tfrac{1}{2}m_\varphi^2\varphi^2  \right] \ . \label{eq:inflatonaction}
\ee
The energy density $\rho_\varphi$ and pressure $p_\varphi$ are given by 
\be
\rho_\varphi    = \frac{1}{2}\dot{\varphi}^2 + \frac{1}{2}m_\varphi^2\varphi^2 \ ; 
\quad p_\varphi = \frac{1}{2}\dot{\varphi}^2 - \tfrac{1}{2}m_\varphi^2\varphi^2 \ ,
\ee
and the equation of motion for the inflaton field is 
\be
\ddot{\varphi} + 3H\dot{\varphi} + m_\varphi^2\varphi + \Gamma_\varphi\dot{\varphi} = 0
\label{eq:inflatoneom}
\ee
The Hubble expansion rate is $H=\dot{a}/a = a^\prime/a^2$ (prime denotes derivative with respect to conformal time), and the Ricci curvature is $R = -6\ddot{a}/a - 6\dot{a}^2/a^2 = -6a^{\prime\prime}/a^3$. As discussed above, the decay width of the inflaton, $\Gamma_\varphi$, sets the dynamics of reheating.

We use the subscript ``$e$'' to denote the end of inflation, and the subscript ``RH'' to denote reheating. For example, $a_\rh$ denotes the value of the scale factor at reheating, and $a_{e}$ denotes the value of the scale factor at the end inflation. The dS phase lasts between $\eta\rightarrow -\infty$ and $\eta_e$, the MD phases lasts between $\eta_e$ and $\eta_\rh$, and the RD phase commences at $\eta_\rh$.

For convenience, we will often rescale common variables relative to $a_e$, $H_e$, and $\eta_e$, which respectively denote the values at the end of inflation of the scale factor, Hubble constant, and conformal time. We define the following dimensionless quantities: $\at=a/a_{e}$, $\mt=m/H_{e}$, $\kt=k/a_{e}H_{e}$, $\etat=a_{e}H_{e}(\eta-\eta_{e})-1$, $\omegat_k=\omega_k/a_{e}H_{e}$, $\Ht=H/H_{e}$, $\nt_k=n_k/H_e^3,\nt=n/H_e^3$, and $\tilde{R}= R/6H_e^2$. 

For our analytic calculations, we will often approximate the evolution of the background spacetime.  In Table \ref{tab:etaRelation} we summarize how the scale factor $\tilde{a}$, the scalar curvature $\tilde{R}$, and the Hubble parameter $\tilde{H}$ depend on the conformal time $\tilde{\eta}$ in de Sitter evolution, MD evolution, and RD evolution. Our conventions are such that at the end of inflation, $\tilde{\eta}_e=-1$, and $\tilde{a}_e=1$.

\renewcommand*\arraystretch{2.0}
\begin{table}
\begin{center}
\begin{tabular}{|c|c|c|c|}
\hline 
 & de Sitter (dS) & Matter domination (MD) & Radiation domination (RD)\tabularnewline [1ex]
\hline 
$\tilde{a}$ & $-\tilde{\eta}^{-1}$ & $\left(\dfrac{3}{2}+\dfrac{\tilde{\eta}}{2}\right)^{2}$ & $\at_\rh + \at_\rh^{1/2}(\etat-\etat_\rh) \approx \at_\rh^{1/2}\etat \quad (\etat\gg\etat_\rh)$ \\ [1ex]
\hline 
$\tilde{R}$ & $-2$ & $-\dfrac{1}{2}\tilde{a}^{-3}=-\dfrac{1}{2}\left(\dfrac{3}{2}+\dfrac{\tilde{\eta}}{2}\right)^{-6}$ & $0$\tabularnewline [1ex]
\hline 
$\tilde{H}$ & $1$ & $\tilde{a}^{-3/2}=\left(\dfrac{3}{2}+\dfrac{\tilde{\eta}}{2}\right)^{-3}$ & $\tilde{a}_\rh^{1/2}\tilde{a}^{-2}\approx \tilde{a}_\rh^{-1/2}\tilde{\eta}^{-2}\quad (\etat\gg\etat_\rh)$\tabularnewline [1ex]
\hline 
\end{tabular}
\captionof{table}{Dependence of scale factor $\at$, Hubble constant $\Ht$, and scalar curvature $\tilde{R}$ on conformal time $\etat$ during the epochs of quasi de Sitter (dS), matter domination (MD), and radiation domination (RD).}
\label{tab:etaRelation}
\end{center}
\end{table}

In addition to the inflaton field determining the evolution of the background spacetime, we consider an additional ``spectator'' scalar field $\sigma$ which does not effect the evolution of the background.  The action is 
\be
S_\sigma = \int \dd^4x \sqrt{-g}\left[\tfrac{1}{2}g^{\mu\nu} \nabla_\mu\sigma\nabla_\nu\sigma
- \tfrac{1}{2}m_\sigma^2\sigma^2 + \tfrac{1}{2}\xi R\sigma^2 \right] \ . \label{eq:spectatornaction}
\ee

We will consider two types of coupling to gravity: minimal, where the coupling constant $\xi=0$, and conformal, where the coupling constant $\xi=1/6$. 

In an FRW spacetime the kinetic term for $\sigma$ is noncanonical, so one performs a change of variables $\sigma(\eta,\xvec)=a^{-1}(\eta)\phi(\eta,\xvec)$ and the action becomes
\begin{align}
    S^\mathrm{FRW}[\phi(\eta,\xvec)] & = \int \! \dd \eta \int \! \dd^3\xvec \, \Bigl[ \tfrac{1}{2} (\phi^\prime)^2 - \tfrac{1}{2} |\dvec \phi|^2 - \tfrac{1}{2} a^2 m_\mathrm{eff}^2 \phi^2  \Bigr] \, ,\\
    m_\mathrm{eff}^2(\eta) & = m^2 + (\tfrac{1}{6} - \xi) R(\eta) \ ,
\end{align}
where $\dvec = (\partial_x, \partial_y, \partial_z)$ denotes derivatives with respect to comoving spatial coordinates. Note that the effective mass squared $a^2(\eta) m_\mathrm{eff}^2(\eta)$ is time-dependent and the sign can be positive or negative.  In conformal time, the equation of motion is 
\be
\label{eq:wave_eqn}
    \phi^{\prime\prime} - \dvec^2 \phi + a^2 m_\mathrm{eff}^2 \phi = 0 \ , 
\ee
which is of the form of a second-order linear wave equation with a time-dependent effective mass. 

Next, we promote $\phi$ to a quantum operator with canonical commutation relations, and write the solution to the linear wave equation in terms of mode functions $\chi_\kvec(\eta)$:
\be
\label{eq:ansatz}
\hat{\phi}(\eta,\xvec) = \int \! \! \frac{\dd^3\kvec}{(2\pi)^3} \Bigl[ \hat{a}_\kvec \, \chi_\kvec(\eta) \, e^{\ii \kvec \cdot \xvec} 
+ \hat{a}_\kvec^\dagger \, \chi_\kvec^*(\eta) \, e^{-\ii \kvec \cdot \xvec}  \Bigr] \ .
\ee
The ladder operators $\hat{a}_\kvec$ and $\hat{a}_\kvec^\dagger$ satisfy the usual commutation relations $[\hat{a}_\kvec,\hat{a}^\dagger_\qvec]=(2\pi)^3\delta(\kvec-\qvec)$, $[\hat{a}_\kvec,\hat{a}_\qvec]=[\hat{a}_\kvec^\dagger,\hat{a}_\qvec^\dagger]=0$ .  The mode equation becomes\footnote{Now we take advantage of the isotropy of the background spacetime and note that we need only consider $k\equiv|\kvec|$.}
\be
\chi_k^{\prime\prime}(\eta) + \omega_k^2(\eta)\chi_k(\eta) = 0 \ .
\label{eq:modeeq}
\ee
This is a harmonic oscillator equation for each wavenumber $k$ with a time dependent frequency given by
\be
\omega_k^2(\eta)  = k^2 + a^2(\eta) m^2 + \bigl( \tfrac{1}{6} - \xi \bigr) a^2(\eta) R(\eta) \ .
\label{eq:dispersion}
\ee

Gravitational particle production arises if the initial (early) and final (late)\footnote{Note that in this context, `early' and `late' refer to the initial and final states and are distinct from early and late reheating.} creation and annihilation operators are not the same. We define the initial vacuum state with early-time creation and annihilation operators $\hat{a}_k^{\early\dagger}$ and $\hat{a}_k^{\early}$ by
\be 
\hat{a}_k^\early  \ket{0^\early} = 0\ket{0^\early } \ . 
\ee 
It is related to the late-time operators by 
\begin{align}
\hat{a}_k^\early = \alpha_k^* \hat{a}_k^\late - \beta_k^* \hat{a}_{-k}^{\late\dagger},
\end{align}
where $\alpha_k$ and $\beta_k$ are  Bogoliubov coefficients. The value of $\beta_k$ can be found from the mode equations such that 
\be 
|\beta_{k}|^2 = \lim_{\eta \rightarrow \infty} \left[\frac{\omega_k}{2}|\chi_k|^2 + \frac{1}{2\omega_k}|\partial_\eta\chi_k|^2 - \frac{1}{2}\right] \ , \label{eq:betak2}
\ee 
where $\omega_k$ is the frequency of the mode and the factor of 1/2 arises from the normalization of the field. Note that the expression Eq.~\eqref{eq:betak2} is only valid for $\omega_k^2>0$. From $|\beta_k|^2$ we can construct the spectrum of particles produced \cite{Kolb:2020fwh}
\begin{equation}
n_k =\frac{k^3}{2\pi^2}|\beta_k|^2 \ , 
\end{equation}
where $k$ is the comoving wavenumber. Then, the comoving number density, $a^3 n$ is given by 
\be 
a^3n = \int \frac{dk}{k} n_k \ .
\label{eq:dknk}
\ee 
 From the comoving number density, one can determine the relic density.

We will take the Bunch-Davies initial condition to solve for particle production. In the early time limit $a\rightarrow 0$ and $a^2R\rightarrow$0, which implies that $\omega_k^2\rightarrow k^2$. In this limit, the mode equation has solution
\begin{equation}
    \lim _{k\eta \rightarrow - \infty} \chi_k(\eta)=\frac{1}{\sqrt{2 k}} e^{-i k \eta} \ .
\end{equation}

\section{Numerical Results}
\label{sec:numerical}

In this section we present our numerical GPP results. We first describe the distinctions between the early and late reheating scenarios, then present results for conformally-coupled scalars in Sec.~\ref{sec:numerical-conformal} and minimally-coupled scalars in Sec.~\ref{sec:numerical-minimal}.

\begin{figure}[htb!]
\begin{center}
\includegraphics[width=0.8\linewidth]{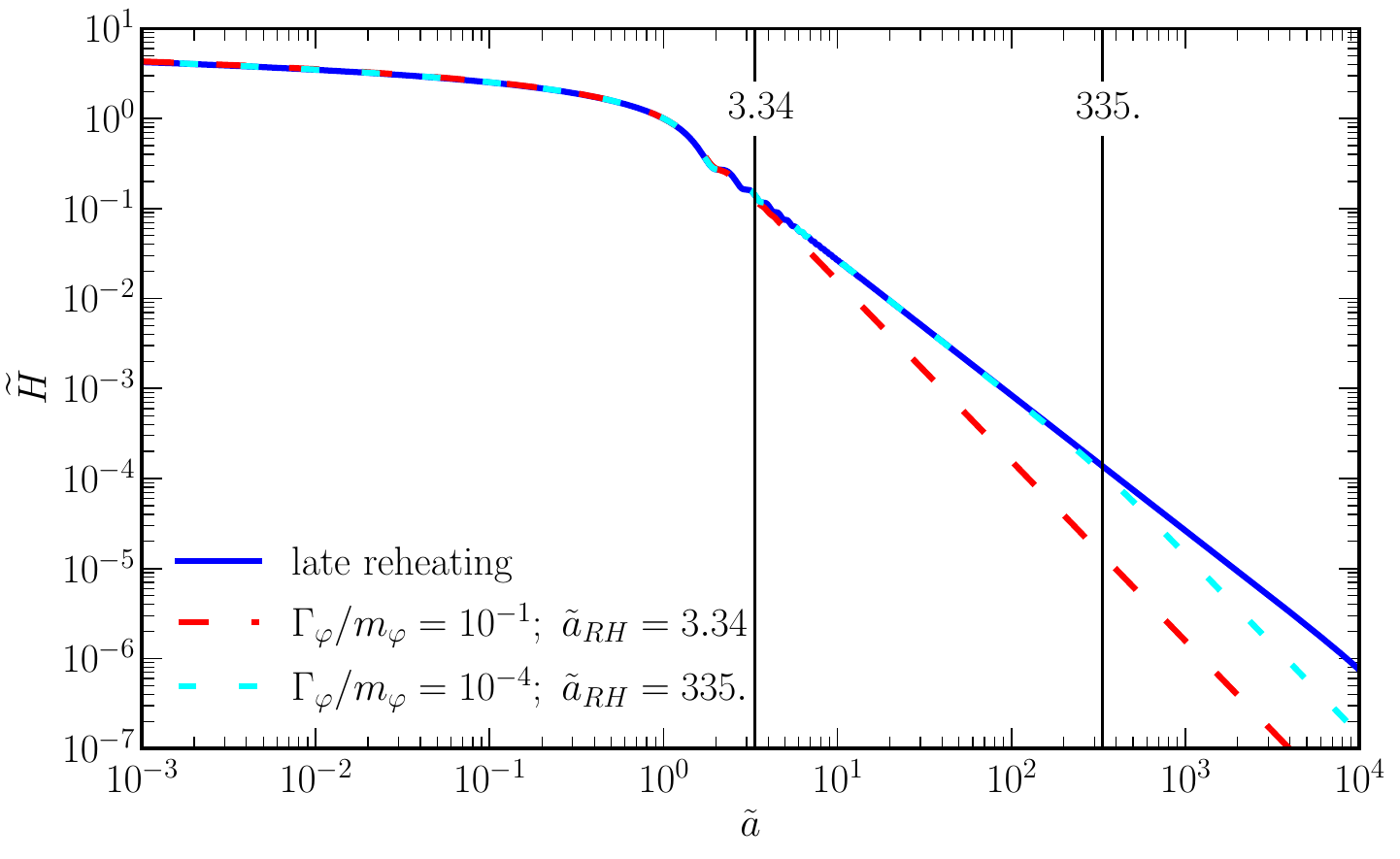} 
\caption{Comparison of $H/H_e$ for late (blue solid line) and early (red and cyan dashed lines) reheating.  }
\label{fig:earlylate}
\end{center}
\end{figure}

In Fig.~\ref{fig:earlylate}, we compare the evolution of $\Ht$ for three different values of $\Gamma_\varphi$, which correspond to late reheating ($\Gamma_\varphi =0$) and two nonzero values for $\Gamma_\varphi$, corresponding to early reheating.
We emphasize that though we have defined `late reheating' to correspond to $\Gamma_\varphi = 0$, in reality it corresponds to a range of parameters such that $\Ht_\rh < m$. One can appreciate this from Fig.~\ref{fig:earlylate}, in which the reheating scenarios diverge from each other at $\at = \at_\rh$.

The difference between early and late reheating is largely driven by the difference in the late time evolution of $H$. A later $\tilde{a}_\rh$ therefore corresponds to an early reheating background that is closer to that of late reheating. For the remainder of this work, we take $\tilde{a}_\rh = 3.34$, corresponding to a decay width of $\Gamma_\varphi/m_\varphi=0.1$, for our early reheating numerical analysis. For the analytic analysis, $a_\rh$ will be a free parameter.

To determine the evolution of $\tilde{a}^3\tilde{n}_k$, we solve the differential equation for $\chi_k$ and $\chi_k^\prime$ in $\eta$, starting from deep in the relativistic region ($\tilde{k}>\tilde{a}\tilde{m}$) with Bunch-Davies initial conditions and ending where the oscillations in the Bogoliubov coefficient (and thus the value of $\tilde{a}^{3}\tilde{n}_{k}$) have ``frozen out'' to a constant value, which is used as the final value for $\tilde{a}^{3}\tilde{n}_{k}$.

\begin{figure}[htb!]
\begin{center}
\includegraphics[height=2.0in]{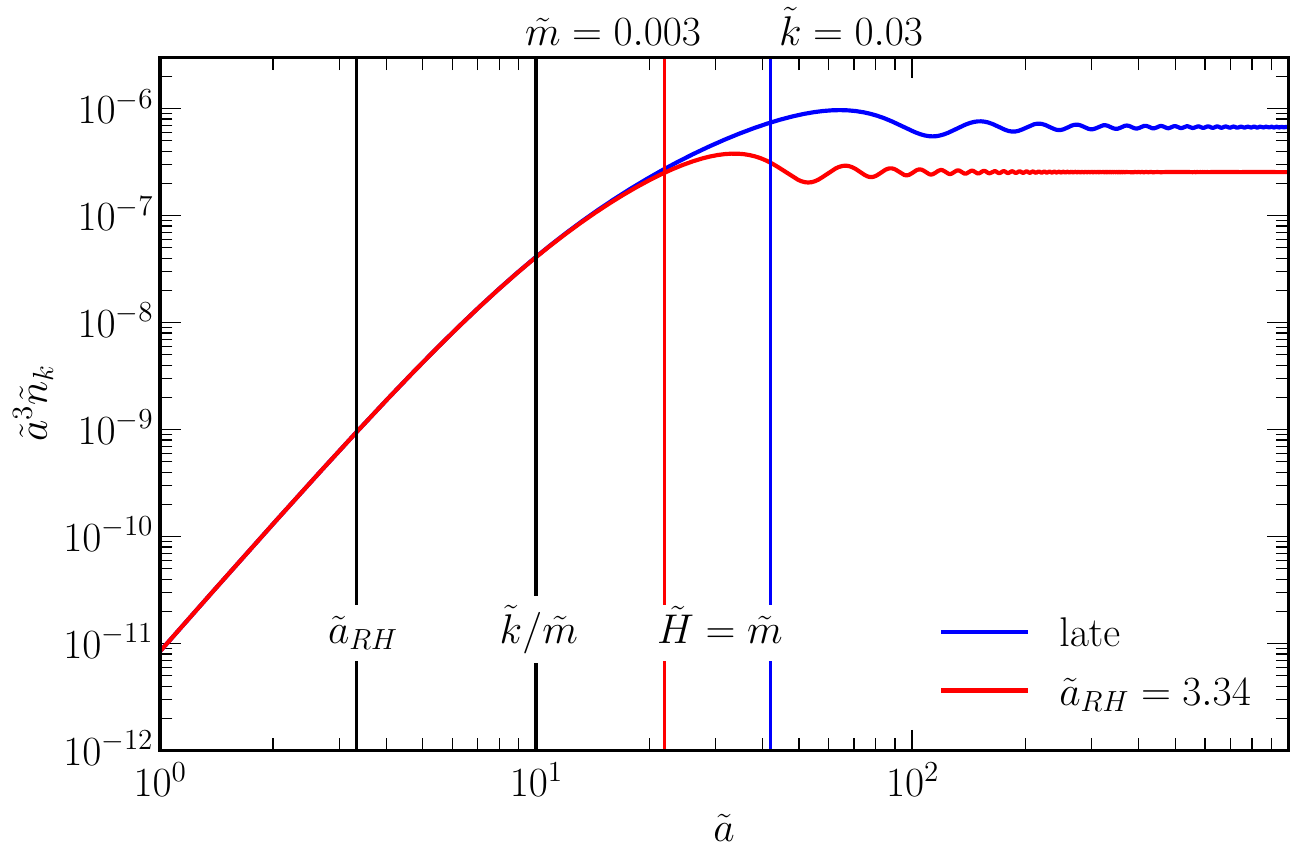}
\includegraphics[height=2.0in]{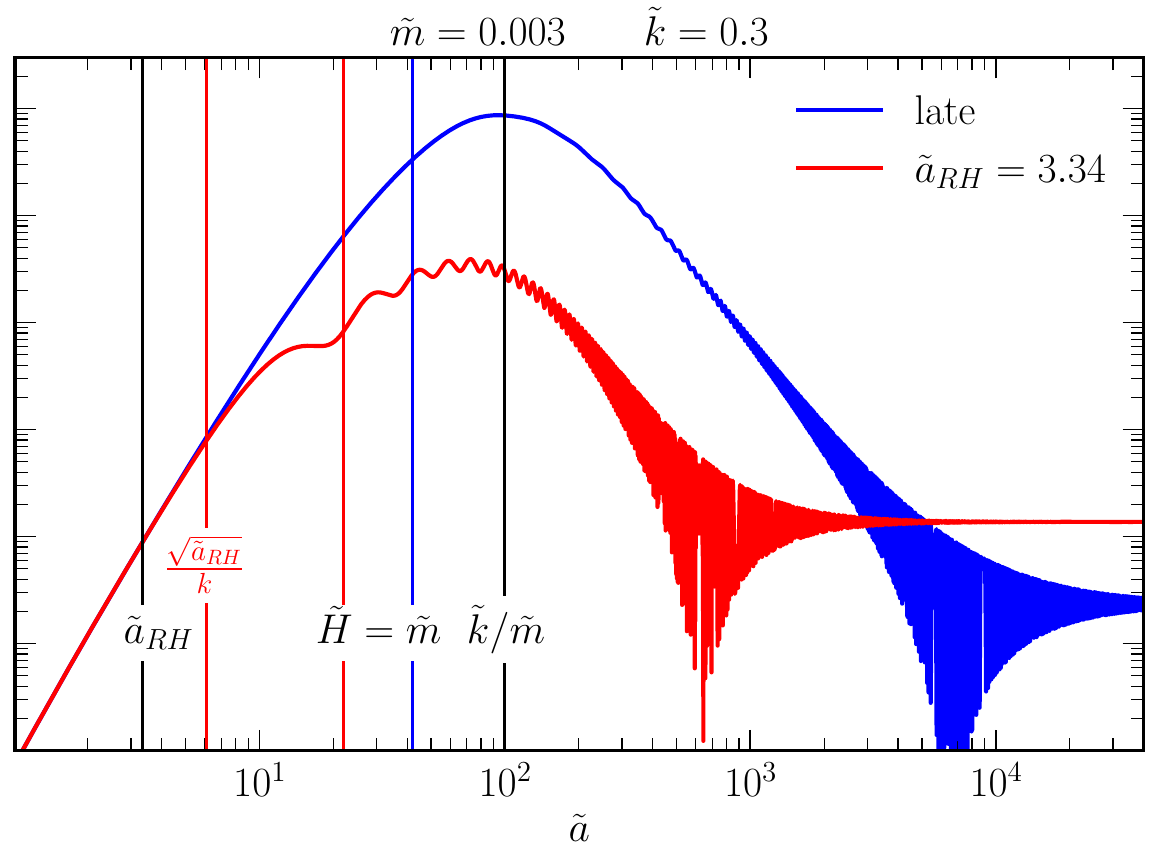}
\caption{Comparison of the evolution of $\tilde{a}^3\tilde{n}_k$ in $\tilde{a}$ for late and early reheating, for conformally coupled scalars with $\tilde{m}=0.003$ and the indicated values of $\tilde{k}$.  } \label{fig:n_kearlylate}
\end{center}
\end{figure}

Example numerical results of the evolution of $\at^3\nt_k$ in $\at$ are shown in Fig.~\ref{fig:n_kearlylate} for $\mt=0.003$ and two choices of $\kt=0.03$ and $\kt=0.3$ for both late and early reheating.  In both cases conformal coupling ($\xi=1/6$) was assumed. The left figure is $\ant$ versus $\tilde{a}$ for $\tilde{k}=0.03$ (so $\tilde{k}<\tilde{m}^{1/3}$), and the right figure is $\ant$ versus $\at$  for $\kt=0.3$ (so $\tilde{k}>\tilde{m}^{1/3}$). For $\tilde{m}\geq 1$, the spectra would be the same for early and late reheating. 

For minimally-coupled scalars, the calculation is slightly more subtle, because $\omega_k$ (Eq.~\eqref{eq:dispersion}), which enters Eq.~\ref{eq:betak2}, can be imaginary if the $R$ term dominates.  Because Eq.~\eqref{eq:betak2} is only valid for $\omega_k^2 > 0$, we emphasize that $\tilde{a}^{3}\tilde{n}_{k}$ only has physical meaning at late times (see \cite{Kolb:2023ydq} for more discussion). However, it still exhibits the same ``freeze out'' behavior, and the same numerical procedure described above is used to determine $\tilde{a}^{3}\tilde{n}_{k}$ values for minimally-coupled scalars. 

\subsection{Conformal coupling}
\label{sec:numerical-conformal}
\begin{figure}[htb!]
\begin{center}
\includegraphics[width=0.8\linewidth]
{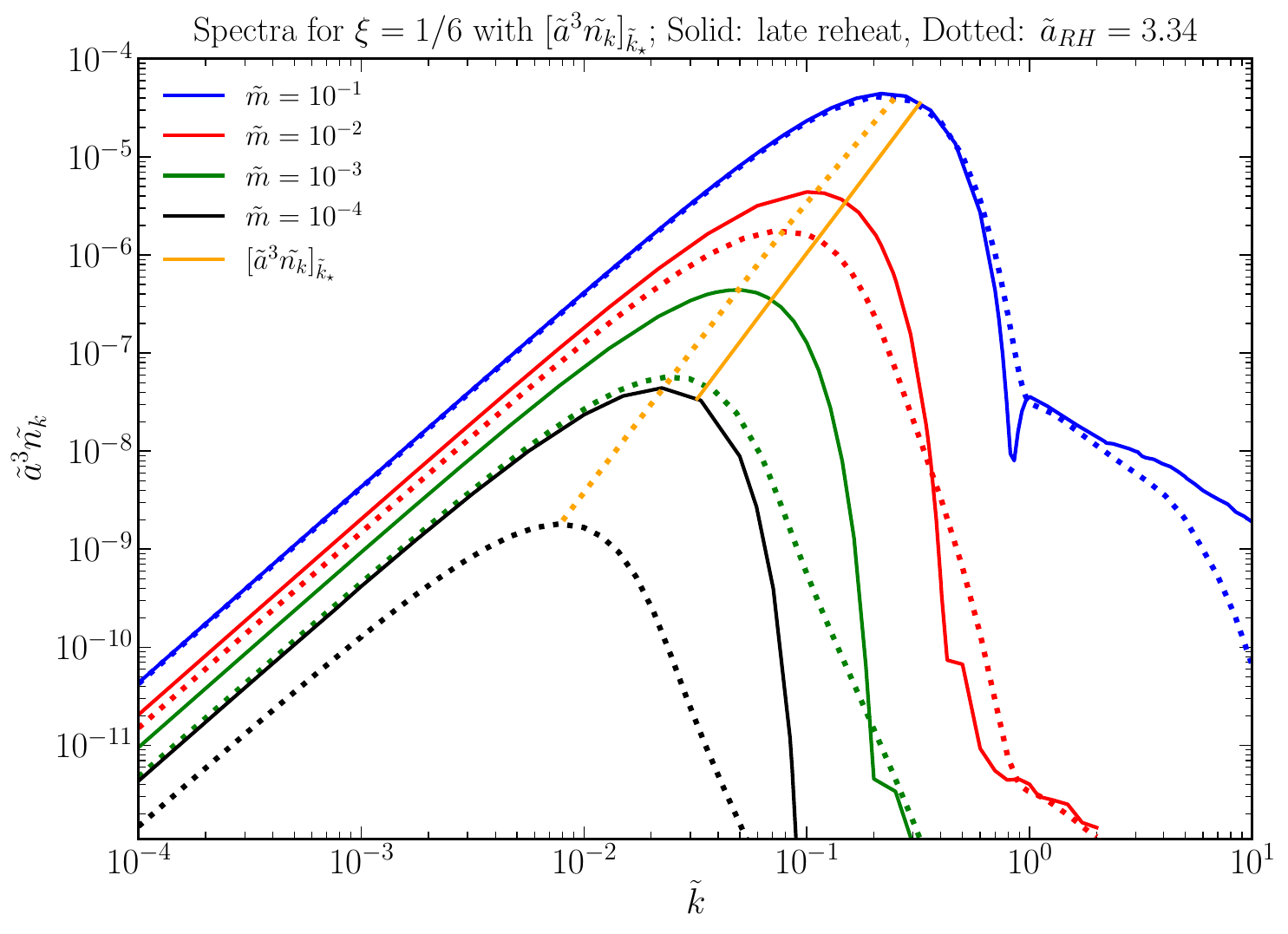}
    \caption{Numerical $\tilde{a}^3\tilde{n}_k$ versus $\tilde{k}$ spectra of conformally-coupled scalars. Solid and dotted lines represent late and early reheating, respectively.  Note that at low $\kt$, late reheating spectra are dominant over those of early reheating, but become subdominant in the red-tilted region. The orange curves trace the value of the spectra evaluated at the ``dominant modes'' $\kt_\star$, which approximate the peak values of the spectra. }
\label{fig:doubleConformal}
\end{center}
\end{figure}

Let us now discuss the features of the spectrum in detail. We first consider conformally-coupled scalars. Shown in Fig.~\ref{fig:doubleConformal} is an example of the $\tilde{a}^{3}\tilde{n}_{k}$ versus $\tilde{k}$ spectra for conformally-coupled scalars with masses ranging from $\tilde{m}=10^{-4}$ to $\tilde{m}=10^{-1}$. Late and early reheating are denoted by solid and dotted lines, respectively.  As expected, the early and late results approach each other as $\tilde{m}$ approaches $1$. We also show $\tilde{a}^3\tilde{n}_k$ evaluated at the ``dominant modes'' $\kt_\star$ in orange, which approximate the peak values of the spectra. We will discuss this further in Sec.~\ref{sssec:dominantmodes}.

After obtaining  $\tilde{a}^{3}\tilde{n}_{k}$, we can find the comoving number density, $\tilde{a}^{3}\tilde{n}$. For completeness, one should perform the full integral over $\tilde{n}_k$ as in Eq.~\eqref{eq:dknk}. However, for conformally-coupled scalars, one can simply approximate the full integral as the peak value of $\tilde{a}^3 \tilde{n}_k$, which we denote as $ \left[\ant\right]_\mathrm{peak}$.  

\begin{figure}[htb!]
\begin{center}   
\includegraphics[width=0.8\linewidth]
{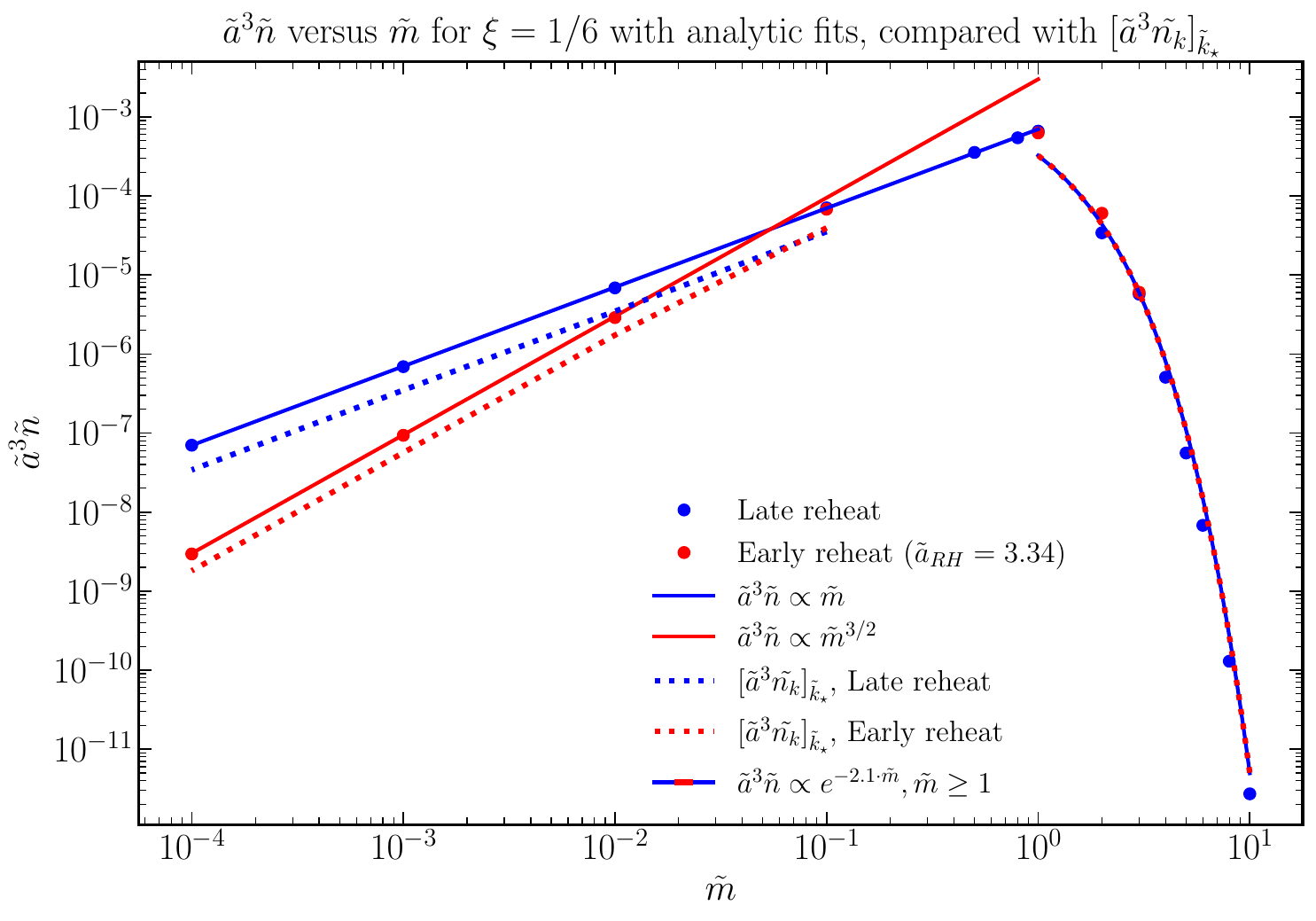}
\caption{Comoving number density ($\tilde{a}^3\tilde{n}$) for a range of masses, with analytic fits that approximate the data points. Late reheating (red) and early reheating (blue) values agree for $\tilde{m}\geq 0.1$. The dotted lines represent $\ant$ evaluated at $\kt_\star$, which is discussed in Sec. \ref{sssec:dominantmodes}.}  
\label{fig:n_vs_m_Conformal}
\end{center}
\end{figure} 

Figure \ref{fig:n_vs_m_Conformal} shows the comoving number density $\tilde{a}^{3}\tilde{n}$ for a range of masses from $\tilde{m}=10^{-4}$ to $\tilde{m}=10$. We show the numerical results for late and early reheating values as red dots and blue dots, respectively, as well as the best fit lines to the numerical results.  For small masses ($\tilde{m}\leq 1$), the comoving number density for late (early) reheating scales as $\tilde{m}$ ($\tilde{m}^{3/2}$), until $\tilde{m} \approx 0.1$, where the results for late and early reheating converge. For $\tilde{m}\geq 1$, the comoving number density for both late and early reheating scale as $e^{-2.1\tilde{m}}$. For each mass, we also plot the value of $\ant$ evaluated at the dominant mode, $k_\star$. We see that the numerically integrated value of $\at^3\nt$ is well approximated (up to a factor of two) by assuming $\tilde{a}^{3}\tilde{n} \approx \left[\ant\right]_{\kt_\star}$.

\subsubsection{Dominant modes}
\label{sssec:dominantmodes}

Conformally-coupled scalars have spectra that are strongly peaked, so that $\tilde{a}^{3}\tilde{n}$ can be approximated by the value of $\tilde{a}^{3}\tilde{n}_k$ at $\tilde{k}_{\mathrm{peak}}$. These are the modes that carry most of the comoving number density, and can in turn be approximated by the ``dominant mode'' of the spectrum, which we denote by $\tilde{k}_\star$. Dominant modes exist for conformally-coupled light mass scalars ($\tilde{m}<1$ and $\xi=1/6$). These are modes that re-enter the horizon after inflation at $\at=\at_\star$ when $3\tilde{H}(\at_\star)=\tilde{m}$, and can be found as $\kt_\star=\at_\star \mt/3$. 

During matter domination, we approximate $\tilde{H}(\at_\star)=\tilde{a}_\star^{-3/2}$, yielding $\at_\star=(\mt/3)^{-2/3}$ and $\kt_\star=(\mt/3)^{1/3}$. During radiation domination, we approximate $\tilde{H}_\star=\tilde{a}_\rh^{1/2}\tilde{a}_\star^{-2}$, yielding $\at_\star=(3\at_\rh^{1/2}/\mt)^{1/2}$ and $\kt_\star = \at_\rh^{1/4}(\mt/3)^{1/2}$.

\begin{figure}[htb!]
\begin{center}  
\includegraphics[width=0.8\linewidth]{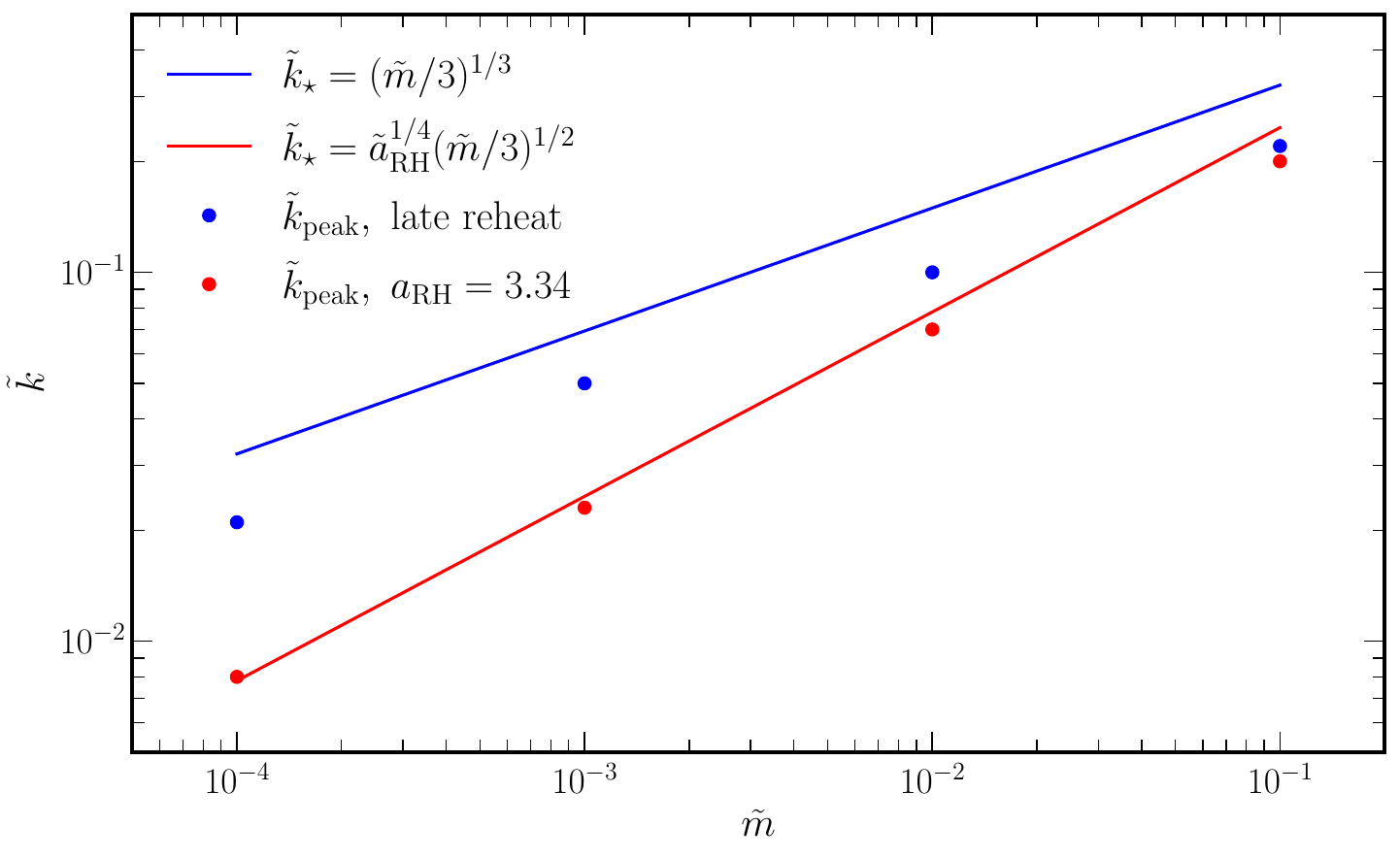}
\caption{Comparison the the results for the dominant mode ($\kt_\star$--solid curves) and numerical results for the peak value (points) for late (blue) and early (red) reheating scenarios. } 
\label{fig:dominantmodes}
\end{center}
\end{figure}

Figure \ref{fig:dominantmodes} compares the numerically determined peak value in $\kt$ of $\tilde{a}^{3}\tilde{n}_{k}$ and the value of the dominant mode $\tilde{k}_\star$ for four values of $\tilde{m}$ from $\tilde{m}=10^{-4}$ to $\tilde{m}=10^{-1}$. Indicated in blue are modes that re-enter horizon during matter domination, and represented by red are the results for early reheating spectra.  Up to order unity, there is agreement between the peaks found via the numerical spectra and the dominant modes derived above. Finally, we note that we will usually drop the factors of $3$ in the definitions of $\kt_\star$, and refer to $\kt_\star=\mt^{1/3}$ and $\kt_\star = \at_\rh^{1/4}\mt^{1/2}$ as the dominant modes for late and early reheating respectively.

\subsection{Minimal coupling}
\label{sec:numerical-minimal}
\begin{figure}[htb!]
\begin{center}  
\includegraphics[width=0.8\linewidth]{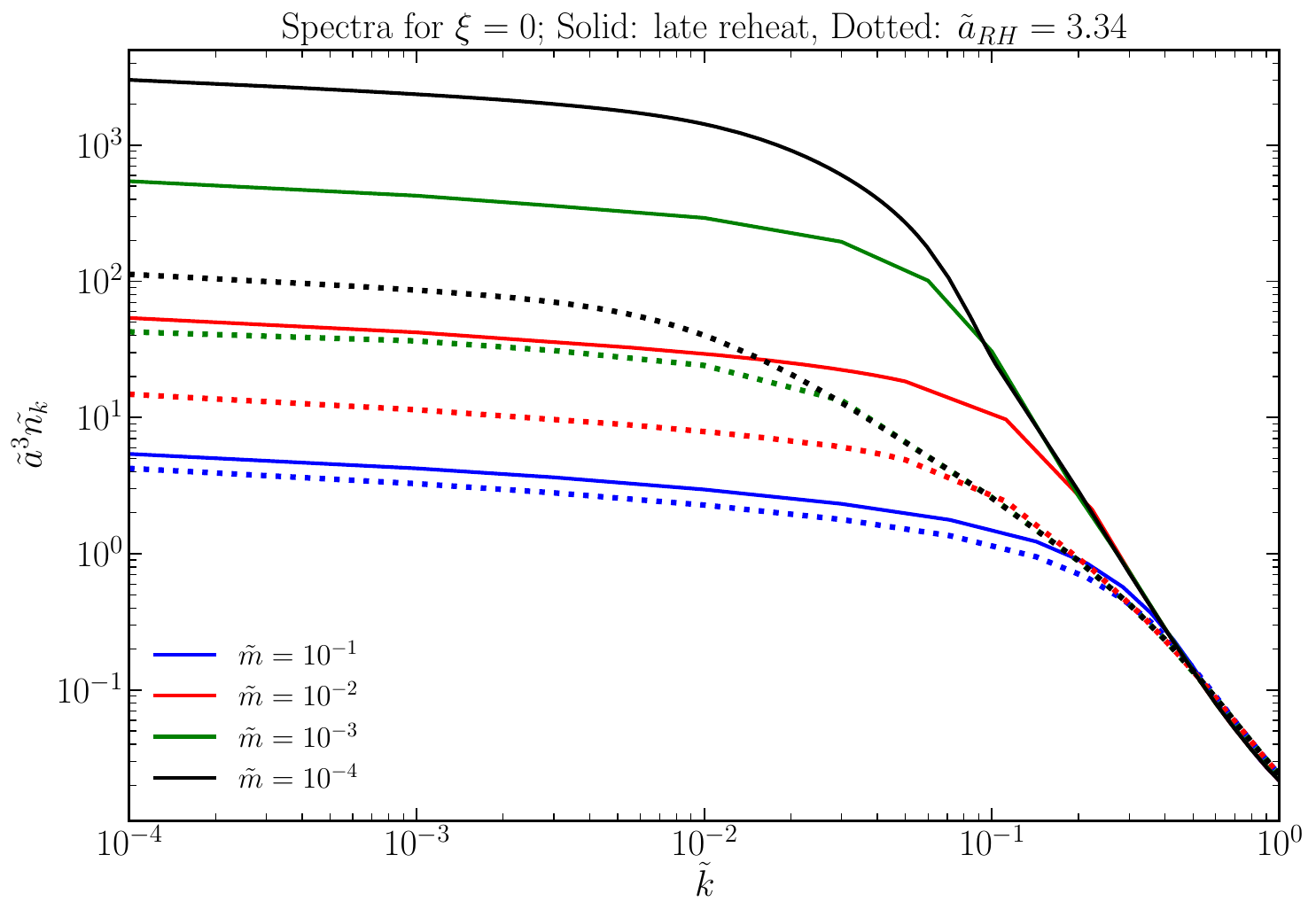}
\caption{Spectra of minimally-coupled scalars, late (solid) and early (dotted) reheating.  }
\label{fig:doubleMinimal}
\end{center}
\end{figure}

Having enumerated the features of the spectrum for conformally-coupled scalars, let us now discuss minimally-coupled scalars. In Fig.~\ref{fig:doubleMinimal} we present numerical results for $\tilde{a}^{3}\tilde{n}_{k}$ as a function of $\tilde{k}$ for minimally-coupled scalars with masses $\tilde{m}=10^{-4}$ to $\tilde{m}=10^{-1}$ for late (solid) and early (dotted) reheating. As we noted before for conformally-coupled scalars, the early and late reheating curves approach each other as $\tilde{m}$ approaches $1$. The $\tilde{k}$ range is chosen from $10^{-4}$ to $1$, so that we can see that the spectra tend toward flatness at low $\tilde{k}$, and fully converge around $\tilde{k}=1$.

\begin{figure}[htb!]
\begin{center}  
\includegraphics[width=0.8\linewidth]{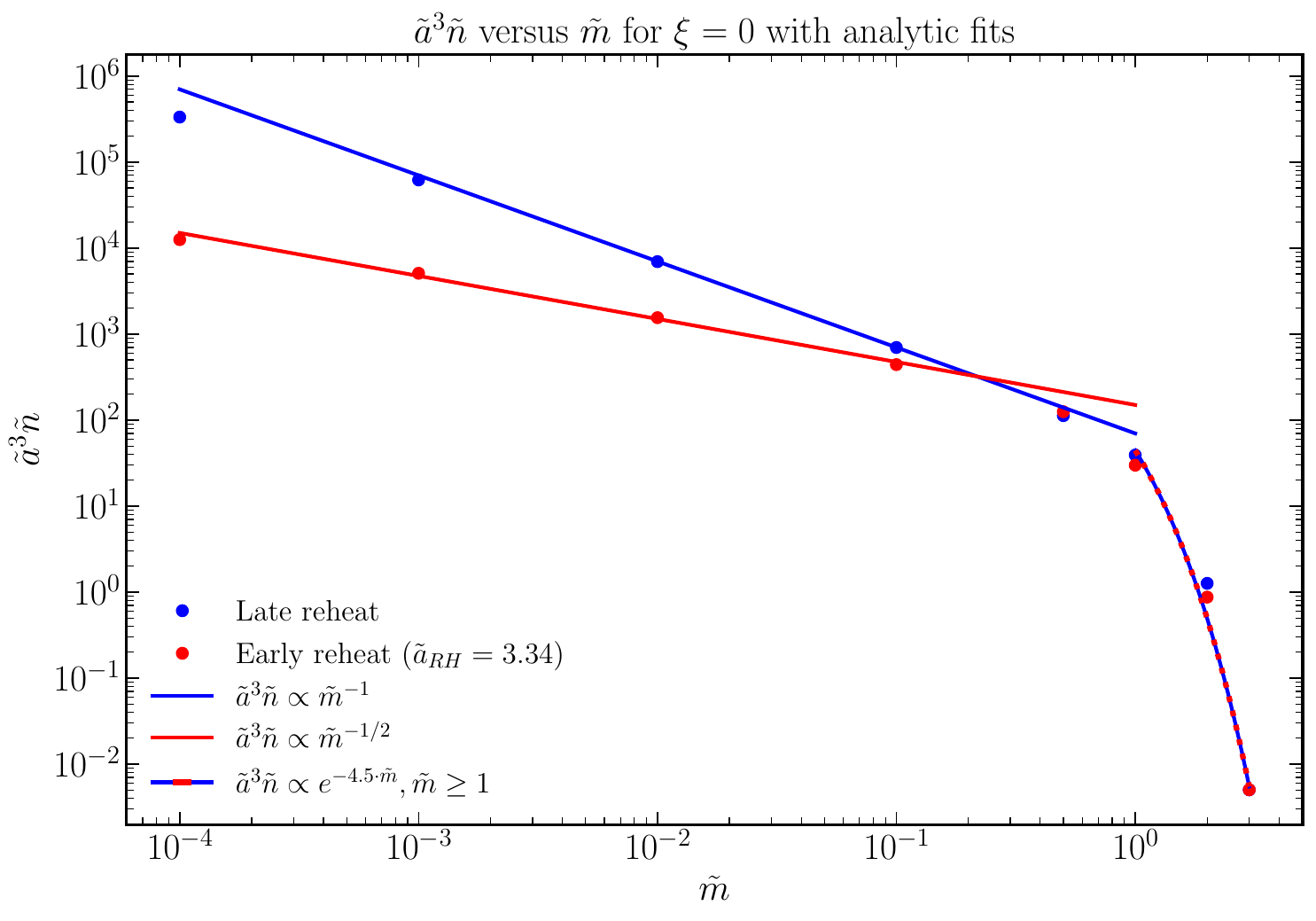}
\caption{Comoving number density ($\tilde{a}^3\tilde{n}$) for a range of masses, with approximate fits to the numerical points. Late reheating (red) and early reheating (blue) agree for $\tilde{m}\geq 0.5$.}
\label{fig:n_vs_m_Minimal}
\end{center}
\end{figure} 

We now again find the comoving number density $\tilde{a}^3\tilde{n}$ from $\tilde{a}^3\tilde{n}_k$. Unlike for conformally-coupled scalars where the spectra are blue-tilted before reaching $\kt_{\mathrm{peak}}$, the spectra of minimally-coupled scalars are red-tilted in the low-$\kt$ region, and continue to decrease as $\kt$ increases. This means that the $\tilde{k}$ range we have calculated numerically only captures the ``tail end'' of the area under the curve, and modes with smaller $\kt$ carry more of the number density. Therefore, before finding $\tilde{a}^3\tilde{n}$ by integrating the $\ant$ spectrum, we first extrapolate the spectrum to $\kt=10^{-20}$ by fitting a line to the low-$\kt$ region. This value of minimum $\kt$ is chosen since CMB observations probe scales of approximately $10^4\kt_0$, where $\kt_0=a_0H_0/a_eH_e$ are the modes that are of a scale equal to the Hubble scale today. Using $a_0/a_\rh=T_\rh/T_0$ and $a_\rh/a_e=(H_e/H_\rh)^{3/2}$, we find $\kt_0 \sim 10^{-24}(10^{9}\GeV/T_\rh)^{1/3}(10^{12}\GeV/H_e)^{1/3}$. Therefore, the spectrum is extended to $\kt=10^4\kt_0=10^{-20}$. We also note that as a dark matter candidate, light, minimally coupled scalars face cosmological challenges, as they induce large isocurvature fluctuations which are stringently constrained by the CMB \cite{PhysRevD.64.023508}. Beyond dark matter, if the produced particles are unstable, they may decay and be unconstrained by CMB isocurvature. Therefore, we include analysis of such models here for completeness.

Figure \ref{fig:n_vs_m_Minimal} shows  $\tilde{a}^3\tilde{n}$ for masses ranging from $10^{-4}$ to $3$. For $\tilde{m}<1$, $\tilde{a}^3\tilde{n}$ decreases as $\tilde{m}^{-1}$ for late reheating, and as $\mt^{-1/2}$ for early reheating. For $\tilde{m}>1$, $\tilde{a}^3\tilde{n}$ exponentially decreases in $\tilde{m}$. After $\tilde{m}\approx 0.5$, the late reheating and early reheating $\tilde{a}^3\tilde{n}$ agree. As we will discuss further in the next section, the analytic and numerical results agree overall.

\section{Analytic Methods and Results}
\label{sec:methods-results}
Having numerically calculated the GPP, we now turn to the analytic approximations. We first provide a brief explanation of the various analytic methods and their regimes of validity in Sec.~\ref{sec:methods}. We then present our results for conformally and minimally-coupled scalars in Sec.~\ref{sec:results-analytic-conf} and Sec.~\ref{sec:results-analytic-min}, respectively, as well as a comparison between the analytic approximations and numerical results.

\subsection{Description of Analytic Methods}
\label{sec:methods}

We now outline the various methods available to analytically approximate particle production: boundary matching, Stokes phenomenon, steepest descent, and inflaton scattering. Here we provide a brief summary and discuss where each method is valid. We will be particularly interested in the $\tilde{k}$ and $\tilde{m}$ scaling of the solutions. Further details of each method are outlined in the Appendices and relevant references.

\begin{enumerate}
    \item \textbf{Boundary Matching:} In the boundary matching method, we find solutions to the mode equation as it evolves through the quasi-de Sitter phase into radiation and/or matter domination, such that the boundary conditions are set by the solution in the previous epoch. The modes and their derivatives are then matched at each boundary. This method is applicable when $\tilde{k}\ll 1$ and $\tilde{m} \ll 1$, and detailed calculations are found in Appendix \ref{app:BoundaryMatching}. 
    
   \item \textbf{Stokes Phenomenon:} The mixing of mode functions leading to particle production can also be understood in terms of the Stokes phenomenon. We follow Ref.~\cite{Hashiba:2021npn} in applying the Stokes phenomenon to GPP. The Stokes phenomenon is applicable when $\kt>\mt^{1/3}$ and $\kt>\mt$.    See Appendix~\ref{app:Stokes} for the calculations.   The Stokes approach will only be used for conformal coupling. 

   \item \textbf{Steepest Descent:} Similar to the Stokes phenomenon, the steepest descent method approximates particle production by evaluating the integral for $\beta_k$ around its poles in the complex $\etat$ plane. This method is described in Ref.~\cite{PhysRevD.67.083514}.   The steepest descent method can be applied when $\tilde{k}>\tilde{m}^{1/3}$ for light particles ($\tilde{m}<1$), where it agrees with the results of the Stokes phenomenon method, and for $\tilde{k}<\tilde{m}$ for heavy particles ($\tilde{m}>1$). See Appendix~\ref{app:steepest} for the calculations.
  
    \item \textbf{Inflaton Scattering:} As the name suggests, this method approximates particle production as a result of inflaton annihilation into two $\chi$ particles. For particles with $m\leq m_\varphi$, the  dominant channel is $\varphi\varphi \rightarrow \chi \chi$, while for particles with $m > m_\varphi$,  one must consider $n\varphi \rightarrow \chi \chi$, where $n$ is the smallest even number such that $m<nm_\varphi/2$. This process is kinematically allowed when $m<m_\varphi$, and is the dominant production mechanism when $k \gtrsim m_\varphi$.     For quadratic inflation, $m_\varphi\approx2H_e$, while for Hilltop inflation with potential $V(\varphi) = (m_\varphi^2v^2/72)(1-\varphi^6/v^6)^2$, $m_\varphi\approx 31H_e$ \cite{Basso_2022}. For the rest of this study, we consider the quadratic inflation model. We find numerically that inflaton scattering becomes effective at $\kt \gtrsim 1$ for $\mt\lesssim1$, and at $\kt \gtrsim 2$ for $\mt\gtrsim1$. See Appendix~\ref{app:scattering} for the calculations  and Refs.\ \cite{Tang_2017,Basso_2022} for further discussions.  
\end{enumerate}

\subsection{Analytic Results: Conformal Coupling}
\label{sec:results-analytic-conf}

\begin{figure}[htb!]
\begin{center}      
\includegraphics[height=1.9in]{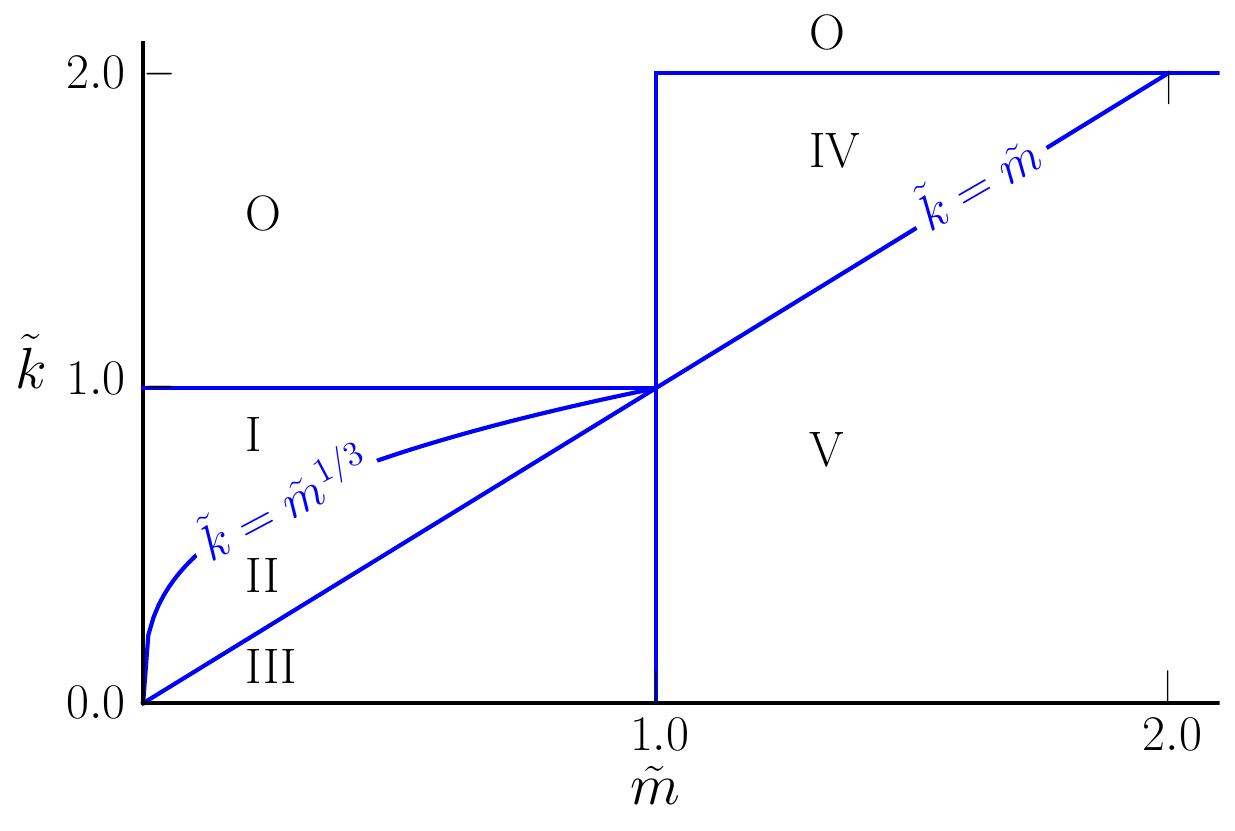}
\includegraphics[height=1.9in]{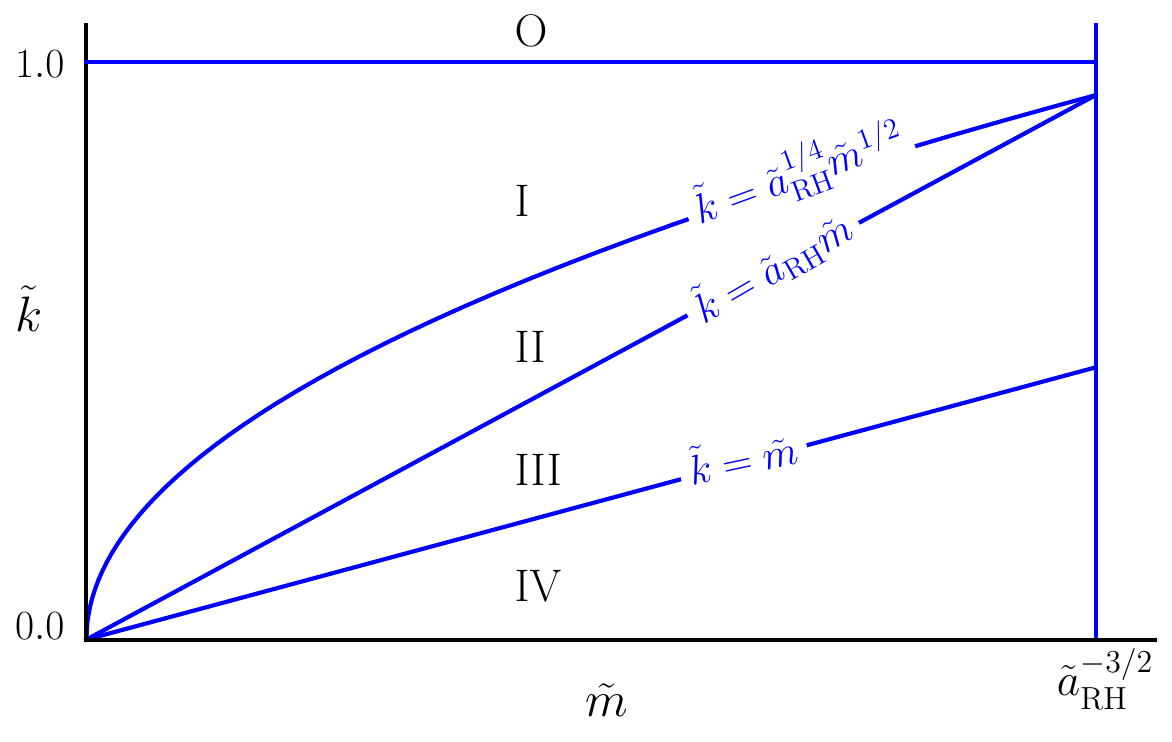}
\caption{Regions of $\kt$ vs.\ $\mt$ for conformally-coupled scalars for late reheating (left) and early reheating (right). The methods for each region of the left panel are as follows: Region O: inflaton scattering, Region I: steepest descent and Stokes phenomenon, Regions II-III: boundary matching, and  Region V: steepest descent. Region IV is a transition region between the behaviors of Regions O and V. The methods for each region of the right panel are as follows - Region O: inflaton scattering, Region I: Stokes phenomenon, and Regions II-IV: steepest descent. }
\label{fig:ConfRegions}
\end{center}
\end{figure}

We now discuss the application of the  above analytic approximations to calculating $\at^3\nt_k$ for conformally coupled scalars for both early and late reheating scenarios. It is convenient to characterize the particle production in terms of the $\tilde{m}-\tilde{k}$ plane, shown in Fig.~\ref{fig:ConfRegions} for late (left) and early (right) reheating. For late reheating, the particle production can be divided into regions delineated by $\kt_\star = \mt^{1/3}$ and $\kt =\mt$, while for early reheating, the regions are delineated by $\kt = \mt$, $\kt = \tilde{a}_\rh \mt$, and $\kt_\star = \tilde{a}_\rh^{1/4}\mt^{1/2}$. 

For each of the numbered regions we analytically approximate the particle production. Tables~\ref{tab:LateConf} and~\ref{tab:InterConf} show the results $\at^3 \tilde{n}_k$ for late and early reheating, respectively. We show the results of each method in the regions where it is applicable, as well as the best fit to the numerical results found in Sec.~\ref{sec:numerical}. We focus primarily on the scaling of $\at^3\nt_k$ with $\kt$ and $\mt$ and collect numerical factors into constants. Each table is further divided into $\mt < 1$ and $\mt > 1$. 

For late reheating, summarized in Table~\ref{tab:LateConf}, we see that for $\mt < 1$, $2\rightarrow 2$ scattering is relevant in Region O, both the Stokes phenomenon method and steepest descent can be applied in Region I, and boundary matching in Regions II and III. In each of these cases, the analytic results reproduce the numerical scaling in $\kt$ and $\mt$ up to numerical factors. For $\mt > 1$, we now have $n \rightarrow 2$  scattering for $\mt<n$ (e.g. $4 \rightarrow 2$ for $2<\mt<4$) in Region O, and the steepest descent method applies in Region V. In this scenario, Region IV only exists from $1 < \kt < 2$ and $1<\mt<2$, and the numerical results show that spectrum of Region IV appears as a transitional region between the $\sim \kt^{3}$ behavior of Region V and the $\sim \kt^{-3/2}$ behavior of Region O.

The early reheating results are shown in Table~\ref{tab:InterConf}. In this scenario, we only consider small mass, $\mt < 1$. We see that in Regions II-IV, the $\kt^2$ scaling of the numerical results is well captured by the boundary matching method. In Region O, we numerically find an exponential suppression, which is well approximated by $2\rightarrow 2$ scattering. In this case, we individually consider the contribution to scattering during the MD and RD eras, demarcated by $\kt=\mt_\varphi\at_\rh$. We define $\Phi_\md$ and $\Phi_\rd$:
\begin{align}
    \Phi_\md &= \exp\left(-\dfrac{2\Gamma_\varphi \tilde{k}^{3/2}}{3H_e\tilde{m}_\varphi^{3/2}}\right) \ & (\kt<\mt_\varphi\at_\rh) \ , \nonumber \\
    \Phi_\rd &=\exp\left(-\dfrac{\Gamma_\varphi\tilde{k}^{2}}{2H_e\tilde{a}_\rh^{1/2}\tilde{m}_{\varphi}^{2}}\right) \ & (\kt>\mt_\varphi\at_\rh) \ .
    \label{eq:phirdmd}
\end{align}
The derivation is discussed in further detail in Appendix~\ref{app:scattering}. Lastly, we note that in Region I, there is a discrepancy between the numerical scaling of $\kt^{-6}$ and the exponentially decaying behavior obtained from applying the Stokes phenomenon. 

\renewcommand*\arraystretch{1.16}
\begin{table}[th]
\begin{center}
\begin{tabular}{|c|c|c|c|c|c|}
\hline 
\multicolumn{6}{|c|}{Late reheating ($\tilde{a}_\rh>\tilde{m}^{-2/3}$);\ Conformal coupling $(\xi=1/6)$} \tabularnewline[1ex]
\hline \hline
\multicolumn{6}{|c|}{$\tilde{m}<1; \ \kt_\star=\mt^{1/3}$}\tabularnewline[1ex]
\hline
\multirow{2}{*}{Region}  & \multirow{2}{*}{Numerical} & Inflaton & \multirow{2}{*}{Stokes} & Steepest & Boundary  \\
                         &                            & Scattering &      & Descent  & Matching \\[1ex] 
\hline 
O & \multirow{2}{*}{$2\rightarrow 2$: $\tilde{k}^{-3/2}$} & \multirow{2}{*}{$c\mt^4\tilde{k}^{-3/2}$} &  &  & \\
$1<\kt$ & & & & & \\
\hline 
I  & \multirow{2}{*}{$\tilde{k}^{3} e^{-\sqrt{\tilde{k}^3/\tilde{m}}}$} &  & \multirow{2}{*}{$\dfrac{\tilde{k}^{3}}{2\pi^2} e^{-c_s\sqrt{\tilde{k}^3/\tilde{m}}}$} & \multirow{2}{*}{$\dfrac{\tilde{k}^{3}}{18} e^{-c_d\sqrt{\tilde{k}^3/\tilde{m}}}$} & \\
$\kt_\star<\tilde{k}<1$ & & & & & \\
\hline 
II  & \multirow{4}{*}{$\tilde{k}^{2}$} & \multirow{2}{*}{} & \multirow{2}{*}{} & \multirow{2}{*}{} & \multirow{4}{*}{$c_m\mt^{1/3} \tilde{k}^{2}$}\\
$\tilde{m}<\tilde{k}<\kt_\star$ & & & & & \\
\cline{1-1} 
III  &  &  &  &  & \\
$0<\tilde{k}<\tilde{m}$ & & & & & \\
\hline 
\multicolumn{6}{|c|}{$\tilde{m}>1$}  \tabularnewline[1ex]
\hline 
O & $2\rightarrow 2$: $ \tilde{k}^{-3/2}$ & $c\mt^4\tilde{k}^{-3/2}$ & \multirow{2}{*}{} & \multirow{2}{*}{} & \\
\cline{2-3} \cline{3-3} 
$2<\kt$  &$4\rightarrow2$: $ \tilde{k}^{-15/2}$ & \multicolumn{1}{c|}{$\tilde{k}^{-15/2}$} &  &  & \tabularnewline 
 \hline
IV  & \multirow{2}{*}{Transition} &  &  &  & \\
$\tilde{m}<\tilde{k}<2$ & & & & & \\
 \hline 
V  & \multirow{2}{*}{$\tilde{k}^{2.7}e^{-2.2\tilde{m}}$} & \multirow{2}{*}{} &  & \multirow{2}{*}{$\dfrac{\tilde{k}^{3}}{18}e^{-2\sqrt{3}\tilde{m}}$} & \\
$0<\tilde{k}<\tilde{m}$ & & & & & \\
\hline 
\end{tabular}
\captionof{table}{Analytic results of $\tilde{a}^3\tilde{n}_k$ for conformally coupled scalars in the late reheating scenario. The constants are defined as $c=9/(512 m_\varphi^{5/2}\pi)$, $c_{s}=2\Gamma^2[1/4]/3\sqrt{\pi}$, $c_d=8\sqrt{6}/5^{3/4}$ and $c_m=3^{2/3}\Gamma^2[5/6]/8\pi^{3}$. Note that in the quadratic inflation model, Region IV only exists for $1 < \kt < 2$. In this region, we find numerically that the spectrum is in transition between the blue-tilted behavior for $\kt<\mt$ and the red-tilted behavior for $\kt>2$, where $2\rightarrow2$ scattering is effective.}
\label{tab:LateConf}
\end{center}
\end{table}

\renewcommand*\arraystretch{1.15}
\begin{table}
\begin{center}
\begin{tabular}{|c|c|c|c|c|}
\hline 
\multicolumn{5}{|c|}{Early reheating ($1<\tilde{a}_\rh<\tilde{m}^{-2/3}$); Conformal coupling $(\xi=1/6)$}\tabularnewline[1ex]
\hline \hline
\multicolumn{5}{|c|}{$\mt<1; \ \kt_\star=\at_\rh^{1/4}\mt^{1/2}$} \\
\hline
\multirow{2}{*}{Region}  & \multirow{2}{*}{Numerical} & Inflaton & \multirow{2}{*}{Stokes} & Boundary   \\
                         &                            & Scattering &      & Matching  \\[1ex] 
\hline 
\hline 
 &  &  &  & \\
O & \multirow{2}{*}{\makecell{exponential \\ decay in $\kt$}} & \multirow{2}{*}{\makecell{
    $\dfrac{c_1\mt^4}{\kt^{3/2}}\Phi_\md^2 \   (\kt<\mt_\varphi\at_\rh)$ \\ [2ex]
    $\dfrac{c_2\mt^4}{\at_{\rh}^{1/2}\kt}\Phi_\rd^2 \   (\kt>\mt_\varphi\at_\rh)$
}} & & \\
$1<\kt$ & & & & \\
  &  &  &  & \\
   &  &  &  & \\
\hline 
I & \multirow{2}{*}{$ \tilde{k}^{-6}$} &  & \multirow{2}{*}{$\dfrac{\tilde{k}^{3}}{2\pi^{2}}e^{-\tilde{k}^{2}\pi/\tilde{a}_\rh^{1/2}\tilde{m}}$}  & \\
$\kt_\star<\kt<1$ & & & & \\
\hline 
II & \multirow{6}{*}{$ \tilde{k}^{2}$} &  &  & \multirow{6}{*}{$c_m\at_\rh^{1/4}\tilde{k}^{2}\tilde{m}^{1/2}$} \\
$\at_\rh\mt<\kt<\kt_\star $ & & & & \\
\cline{1-1} 
III &  &  &  & \\
$\mt<\kt<\at_\rh\mt$ & & & & \\
\cline{1-1} 
IV &  &  &  & \\
$0<\kt<\mt$ & & & & \\
\hline 
\end{tabular}
\captionof{table}{Analytic results for $\tilde{a}^3\tilde{n}_k$ for conformally-coupled scalar in the early reheating model. The constants are defined as $c_1=9/(512 \mt_\varphi^{5/2}\pi)$, $c_2=9/(512 \mt_\varphi^{3}\pi)$, and $c_m= \Gamma^2[3/4]/4\pi^3$. The expressions $\Phi_\rd$ and $\Phi_\md$ are decaying exponentials defined in Eq.~\eqref{eq:phirdmd}. }
\label{tab:InterConf}
\end{center}
\end{table}

\begin{figure}[htb!]
\begin{center}
\includegraphics[width=0.8\linewidth]{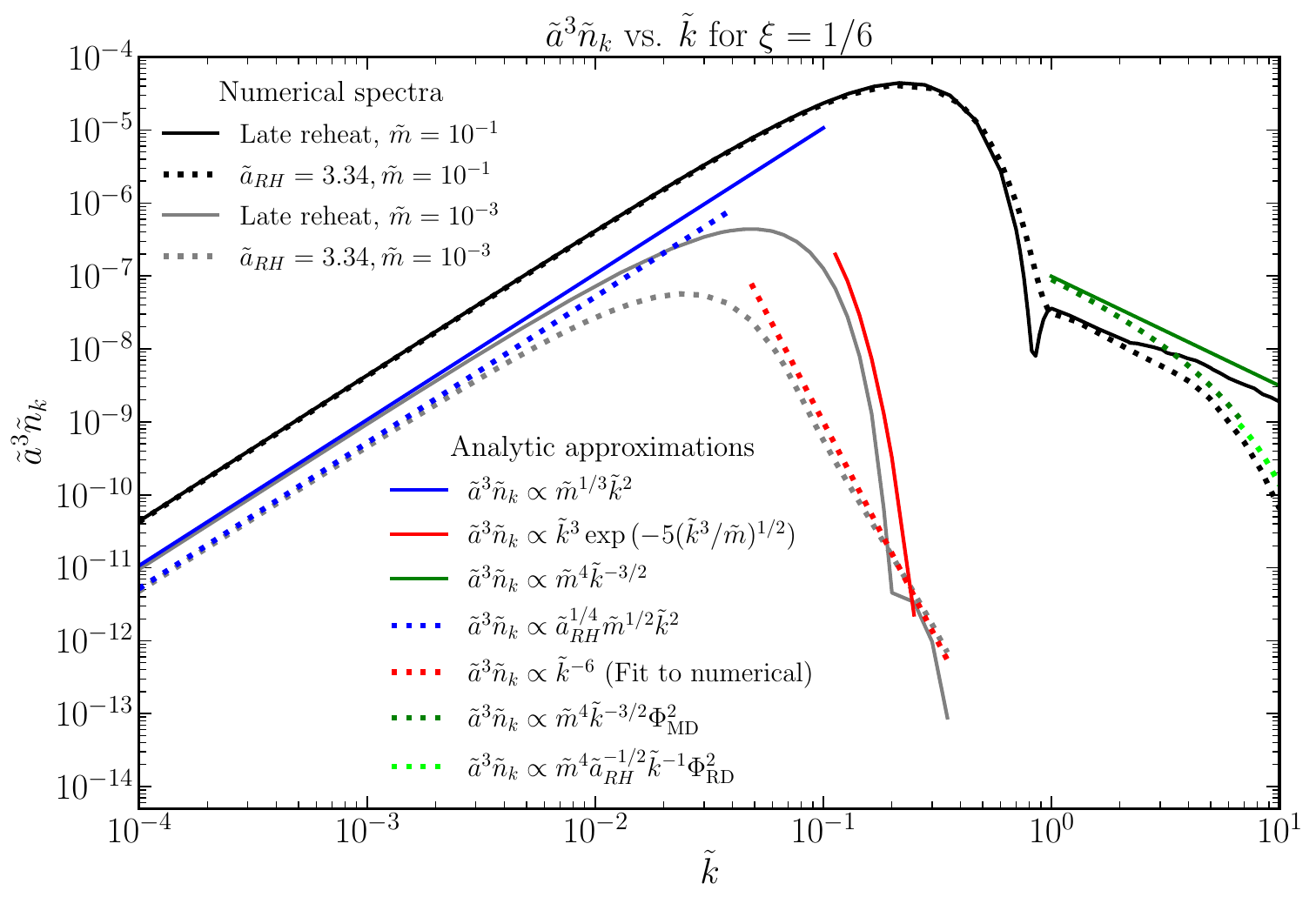}
\caption{Comparison of numerical spectra ($\tilde{m}=10^{-1}$ in black and $\tilde{m}=10^{-3}$ in grey) and analytic approximations (colored) for conformally coupled scalars. The late reheating spectrum and approximations are represented by solid lines, and the early reheating spectrum and approximations are represented by dotted lines. For both early and late reheating, the analytic approximations are divided into ``low-$\tilde{k}$'' regions (blue), ``high-$\tilde{k}$'' regions (red), and ``scattering'' regions (green). The scattering region of early reheating is further divided into two regions: MD (dark green) where $\kt<2\at_\rh$, and RD (light green) where $\kt>2\at_\rh$.}
\label{fig:compareConformal}
\end{center}
\end{figure}

To further visually appreciate the congruence between the analytic and numerical results, Figure~\ref{fig:compareConformal} compares $\tilde{a}^3 \tilde{n}_k$ calculated numerically to the various analytic approximations in the appropriate regimes. We show $\at^3 \nt_k$ for $\tilde{m}=10^{-1}$ (blue) and $\tilde{m}=10^{-3}$ (green) as representative examples in Figure~\ref{fig:doubleConformal}. As expected, in the low-$\kt$ regions of both early and late reheating, the spectrum scales as $\tilde{k}^{2}$. The high-$\kt$ region of late reheating decays as $\exp\left(-\kt^{3/2}/\mt^{1/2}\right)$, and the high-$\kt$ region of early reheating decays as $\kt^{-6}$. We note that this power law behavior is directly approximated from the numerical results, rather than an analytic approximation. Finally, the scattering region of late reheating obeys the $\kt^{-3/2}$ behavior for $2$-to-$2$ scattering, and early reheating exhibits an exponential decay sourced by the decay of the inflaton field.

\begin{figure}[htb!]
\begin{center}
\includegraphics[width=0.8\linewidth]
{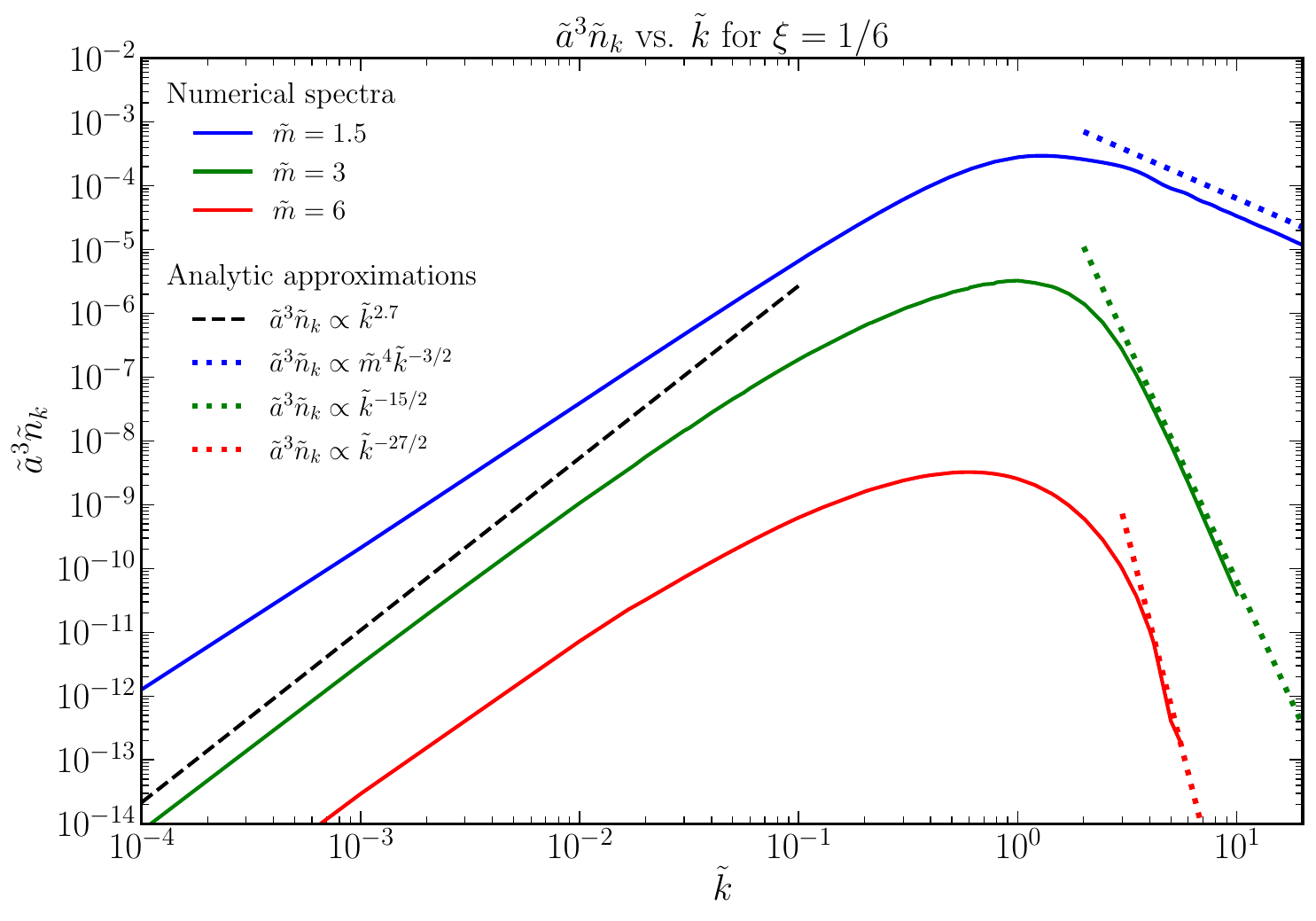}
\caption{Comparison of numerical spectra for conformally coupled heavy scalars with their analytic approximations, for $\mt=1.5$ (in blue), $\mt=3$ (in green), and $\mt=6$ (in red). The black dashed line represents the low-$\kt$ behavior shared by the spectra. The dotted lines represent the inflaton scattering behavior for $\kt\gtrsim 2$ for each of the spectra, where the dominant scattering processes are respectively $2\rightarrow2$ (blue), $4\rightarrow2$ (green), and $6\rightarrow2$ (red).}
\label{fig:compareConformalLargeM}
\end{center}
\end{figure}

We present in Figure~\ref{fig:compareConformalLargeM} the numerical spectra and the corresponding analytic approximations for conformally coupled ``heavy'' scalars, where $\mt>1$. In the $\kt$-range shown in Fig.~\ref{fig:compareConformalLargeM}, the low-$\kt$ behavior scales as $\ant\propto\kt^{2.7}$. As we extend to even lower $\kt$, the spectrum approaches $\ant\propto\kt^3$, which is the result of analytic calculations shown in Region V of Table~\ref{tab:LateConf}. For the $\kt\gtrsim 2$ region where inflaton scattering dominates, the allowed $n\rightarrow2$ processes for a given $\mt$ have even $n$ where $n\geq\mt$. For $\mt=1.5$, we consider $2\rightarrow2$ scattering where $\ant\propto \kt^{-3/2}$; for $\mt=3$, we consider $4\rightarrow2$ scattering where $\ant\propto \kt^{-15/2}$; for $\mt=6$, we consider $6\rightarrow2$ scattering where $\ant\propto \kt^{-27/2}$. 

\subsection{Analytic Results: Minimal Coupling}
\label{sec:results-analytic-min}

\begin{figure}[htb!]
\begin{center}
\includegraphics[height=1.9in]{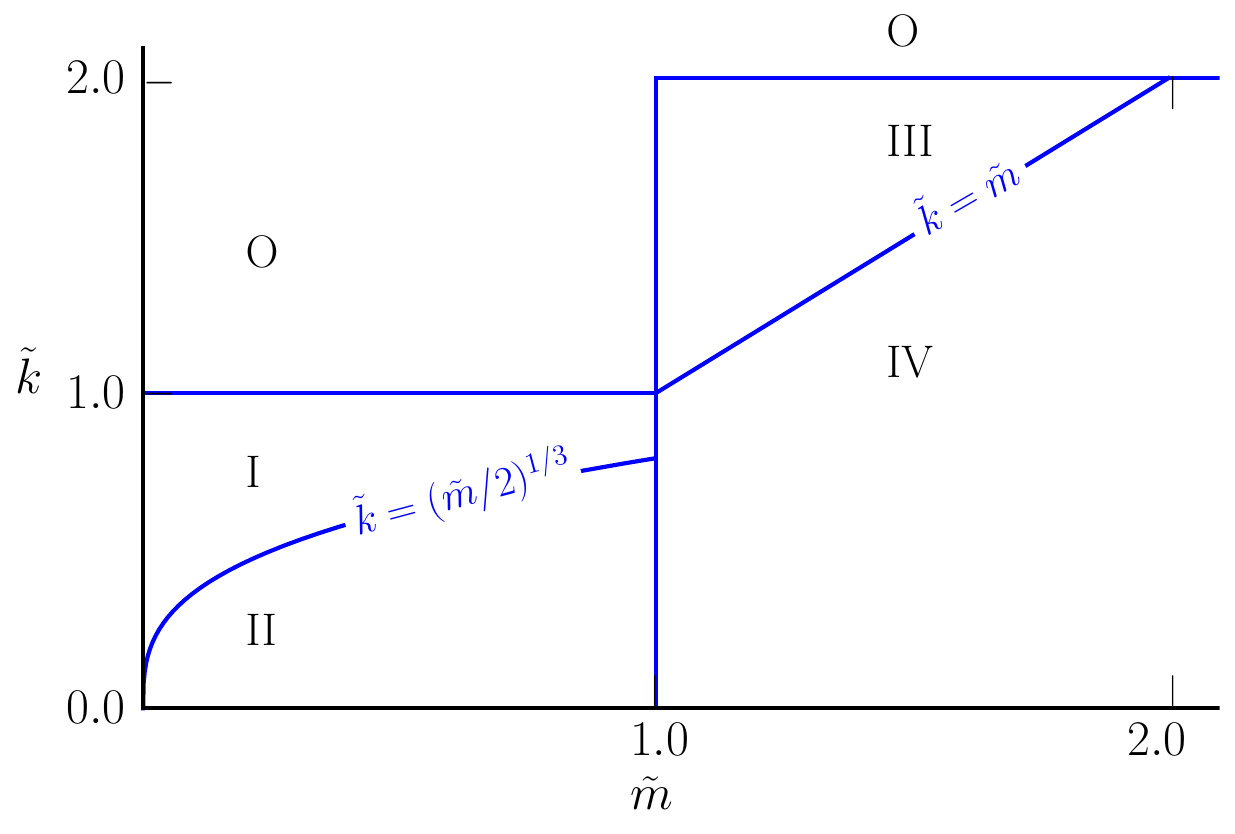} 
\includegraphics[height=1.9in]{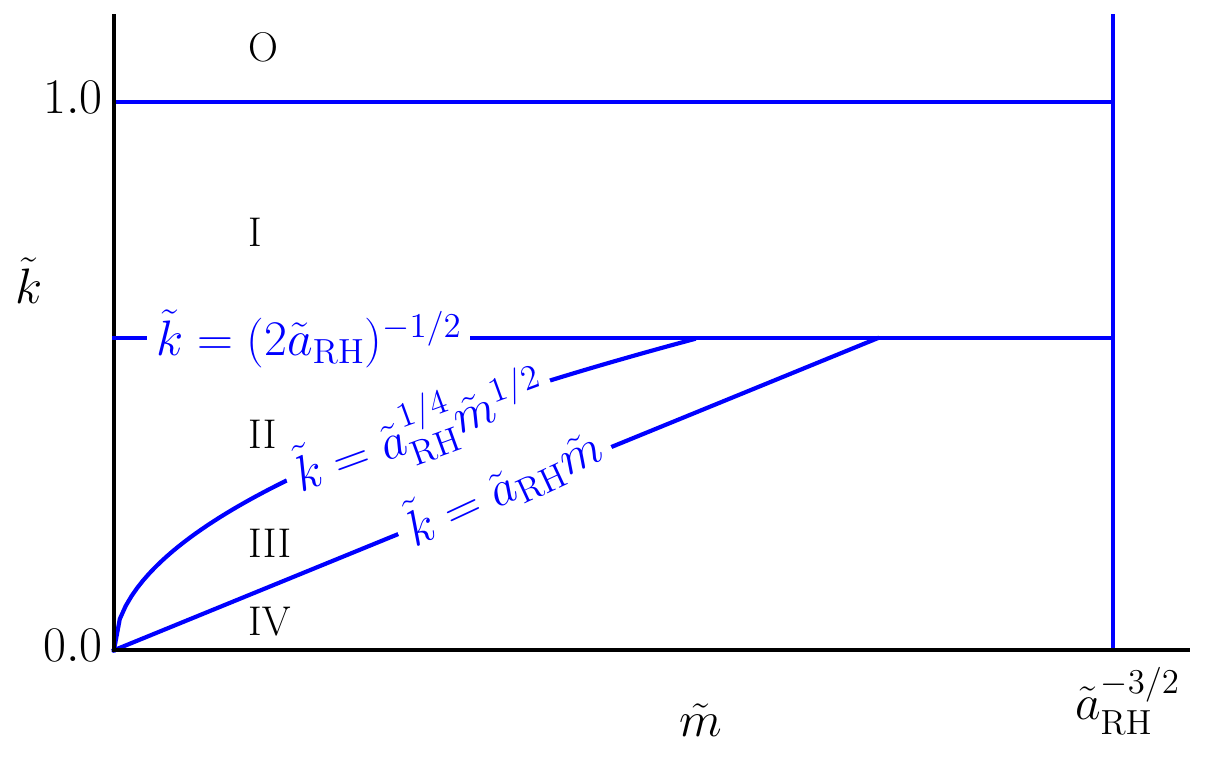}
\caption{Regions of $\tilde{k}$ vs. $\tilde{m}$ for minimal coupling, late reheating (left) and early reheating (right). The methods for each region of the left panel is as follows - Region O: scattering, Region I: matching, Region II: matching, Region III: steepest descent. The methods for each region of the right panel is as follows: Region O: scattering, Regions I through IV: matching.}
\label{fig:MinRegions}
\end{center}
\end{figure}

We now discuss our results for minimally coupled scalars. Figure~\ref{fig:MinRegions} shows the various regions in the $\mt-\kt$ plane for late (left) and early (right) reheating, which has a further delineation at $\kt = 1$ compared to the conformally coupled scenario.
 
As before, we approximate $\at^3\nt_k$ in each numbered region for late and early reheating, shown in Tables~\ref{tab:LateMin} and~\ref{tab:InterMin}, respectively. In Table~\ref{tab:LateMin}, we see that for $\mt < 1$, $2\rightarrow 2$ scattering yields the $k^{-3/2}$ scaling in Region O, boundary matching provides the correct $\kt^{-3}$ scaling in regions I and $\kt^0$ in region II. For $\mt > 1$, we have the usual $2\rightarrow 2$ result, as well as a contribution which scales as $k^{-15/2}$ from $4 \rightarrow 2$ scattering in Region O. As in the conformally coupled scenario, Region III is a small numerical transition region, and we find that steepest descent reproduces the numerical result in Region IV. For early reheating, shown in Table~\ref{tab:InterMin}, the scattering calculation yields an exponential falloff, and we similarly find that boundary matching method reproduces the $\kt^{-3}, \kt^{-1}$, and $\kt^{0}$ scaling in Regions I, II, and III-IV, respectively.

\renewcommand*\arraystretch{1.15}
\begin{table}
\begin{center}
\begin{tabular}{|c|c|c|c|c|c|}
\hline 
\multicolumn{5}{|c|}{Late reheating; Minimal coupling ($\xi=0$)}\\
\hline \hline
\multicolumn{5}{|c|}{$\tilde{m}<1$} \\
\hline
\multirow{2}{*}{Region} & \multirow{2}{*}{Numerical} & Inflaton & Steepest & Boundary \tabularnewline
                        &                            & Scattering &  Descent & Matching \\ [1ex]       
\hline 
O & \multirow{2}{*}{$2\rightarrow 2:  \tilde{k}^{-3/2}$} & \multirow{2}{*}{$c \tilde{k}^{-3/2}$} &  & \\
$1<\kt$ & & & & \\

\hline 
I  & \multirow{2}{*}{$\dfrac{1}{\tilde{k}^{3}}$} &  &  & \multirow{2}{*}{$\dfrac{1}{8\pi^{2}\tilde{k}^{3}}$} \\
$(\mt/2)^{1/3}<\kt<1$ & & & & \\
\hline 
II  & \multirow{2}{*}{$ \tilde{k}^{0}$} &  &  & \multirow{2}{*}{$\dfrac{1.87}{2\pi^2 \tilde{m}}$}\\
$0<\kt<(\mt/2)^{1/3}$ & & & & \\
\cline{1-1} 
\hline 
\multicolumn{5}{|c|}{$\mt>1$} \\
\hline
O & $2\rightarrow2$: $ \tilde{k}^{-3/2}$ & $c\tilde{k}^{-3/2}$ & & \\
$2<\kt$ & $4\rightarrow2$: $ \tilde{k}^{-15/2}$ & $ \tilde{k}^{-15/2}$ & & \\
\hline 
III & \multirow{2}{*}{Transition} & & & \\
$\mt<\kt<2$ & & & & \\
\hline
IV & \multirow{2}{*}{$ \kt^{2.7}e^{-2.2\mt}$} & &  \multirow{2}{*}{$\dfrac{\kt^3}{18}e^{-2\sqrt{3(\mt^2-2)}}$} & \\
$0<\kt<\mt$ & & & & \\
\cline{1-1} 
\hline
\end{tabular}
\captionof{table}{Analytic results of $\tilde{a}^3\tilde{n}_k$ for minimally-coupled light scalar in the late reheating model. The constants $c=9 \mt_\varphi^{3/2}/128\pi$, and $c_{s}=2\Gamma^2[1/4]/3\sqrt{\pi}$. Note that for $\mt>1$, the spectra of the minimally coupled scalar approaches that of the conformally coupled scalar. We therefore apply the result in Region O of Table~\ref{tab:LateConf} to this table's Region O, and we apply the conclusions about Region IV of Table~\ref{tab:LateConf} to this table's Region III. For large $\mt$, the result in Region IV approaches that of Region V of Table~\ref{tab:LateConf}.}
\label{tab:LateMin}
\end{center}
\end{table}

\renewcommand*\arraystretch{1.15}
\begin{table}
\begin{center}
\begin{tabular}{|c|c|c|c|}
\hline 
\multicolumn{4}{|c|}{Early reheating ($1\leq\tilde{a}_\rh<\tilde{m}^{-2/3}$); Minimal coupling $(\xi=0)$}\tabularnewline[1ex]
\hline \hline
\multirow{2}{*}{Region} & \multirow{2}{*}{Numerical} & Inflaton & Boundary \\
                        &                            & Scattering & Matching \\
\hline 
 &  &   & \\
O & \multirow{2}{*}{\makecell{exponential \\ decay in $\kt$}} & \multirow{2}{*}{\makecell{$\dfrac{c_1}{\kt^{3/2}}\Phi_\md^2 \ (\kt>\mt_\varphi\at_\rh)$ \\ [2ex]
$\dfrac{c_2}{\at_{\rh}^{1/2}\kt}\Phi_\rd^2 \ (\kt>\mt_\varphi\at_\rh)$}} & \\
$1<\kt$ & & & \\
 &  &   & \\
  &  &   & \\
\hline 
I & \multirow{2}{*}{$ \tilde{k}^{-3}$} &  & \multirow{2}{*}{$\dfrac{1}{8\pi^{2}\tilde{k}^{3}}$} \\
$(2\at_\rh)^{-1/2}<\kt<1$ & & & \\
\hline 
II & \multirow{2}{*}{$ \tilde{k}^{-1}$} &  & \multirow{2}{*}{$\dfrac{\tilde{a}_\rh}{6\pi^{2}\tilde{k}}$} \\
$\at_\rh^{1/4}\mt^{1/2}<\kt<(2\at_\rh)^{-1/2}$ & & & \\
\hline 
III & \multirow{4}{*}{$ \tilde{k}^{0}$} & \multirow{4}{*}{} & \\
$\at_\rh\mt<\kt<\at_\rh^{1/4}\mt^{1/2}$ & & & $\dfrac{c_m\at_\rh^{3/4}}{\mt^{1/2}}$\\
\cline{1-1} 
IV &  &  & \\
$0<\kt<\at_\rh\mt$ & & & \\
\hline 
\end{tabular}
\caption{Analytic results of $\tilde{a}^3\tilde{n}_k$ for minimally-coupled scalars in the early reheating model. The numerical constants are given by $c_1=9 \mt_\varphi^{3/2}/128\pi$, $c_2=9 \mt_\varphi/128\pi$ and $c_m=\Gamma^2[1/4]/(12\pi^{3})$. The expressions $\Phi_\rd$ and $\Phi_\md$ are decaying exponentials defined in Eq.~\eqref{eq:phirdmd}.}
\label{tab:InterMin}
\end{center}
\end{table}

\begin{figure}[htb!]
\begin{center}  
\includegraphics[width=0.8\linewidth]{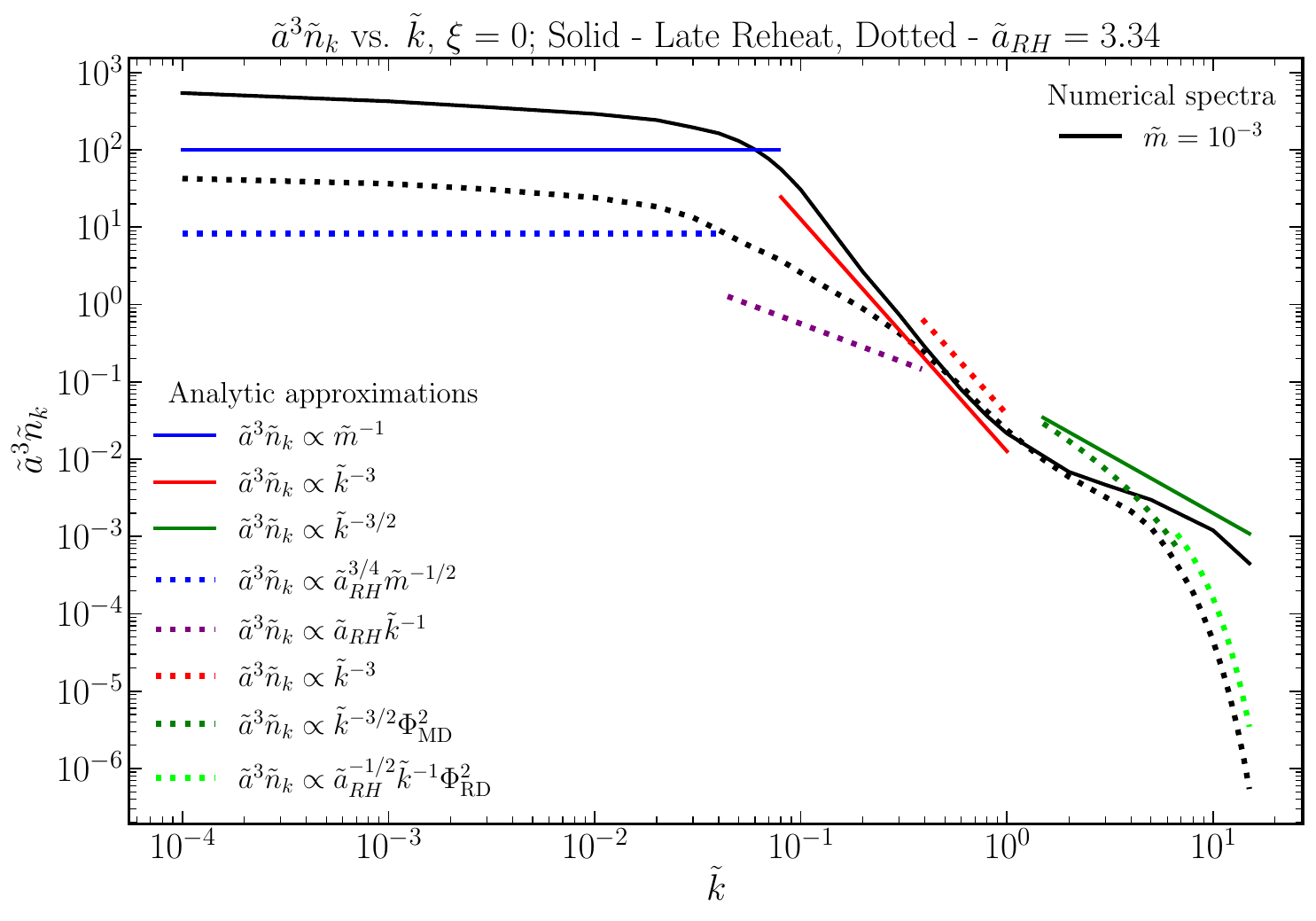}
\caption{Comparison of numerical spectra (black) and analytic approximations (colored) for a minimally-coupled scalar of mass $\tilde{m}=10^{-3}$. The late reheating spectrum and approximations are represented by solid lines, and the early reheating spectrum and approximations are represented by dotted lines. The analytic approximation of both the late and early reheating spectra are divided into ``low-$\tilde{k}$'' (blue), ``high-$\tilde{k}$'' (red), and ``scattering'' (green) regions, while the early reheating spectrum has an additional ``medium-$\tilde{k}$'' (purple) region, and its scattering region is further divided into two regions: MD (dark green) where $\kt<2\at_\rh$, and RD (light green) where $\kt>2\at_\rh$. The ``high-$\tilde{k}$'' regions of late and early reheating overlap.}
\label{fig:compareMinimal}
\end{center}
\end{figure}

 In Fig.\ \ref{fig:compareMinimal} we now compare a numerically calculated spectrum to its analytic approximations. In the ``low-$\tilde{k}$'' region of the spectra, the analytic calculations approximate the spectra as flat, whereas the numerical spectra are slightly red-tilted. The tilt of the spectra is dependent on the specific inflationary model and profile of $\Ht$. Our analytic approximation assumes an initially constant $\tilde{H}$ during the dS phase, which corresponds to a potential that begins in a flat region. This produces a flat low-$\kt$ spectrum, which is a reasonable assumption for a wide range of inflationary models. The numerical calculation, however, uses the quadratic inflation model, which begins in a descent. Using instead an inflation model that begins in a less steeply descending region (e.g. T-inflation model) generates numerical spectra that are closer to flatness in this region. In the high-$\kt$ region of late reheating, the spectrum scales as $\kt^{-3}$. For early reheating, there exists an extra ``medium''-$\kt$ region where the spectrum approximately scales as $\kt^{-1}$. This region is followed by a high-$\kt$ region whose behavior is the same as that of the high-$\kt$ region for late reheating. The scattering region of both the early and late reheating spectra have the same overall scaling as that of conformal coupling - the late reheating spectrum exhibits the $\kt^{3/2}$ behavior of 2-to-2 scattering process that is dominant for low mass scalars, and the early reheating spectrum exhibits an exponential decay. 

Not shown here are the late reheating spectra for $\mt\geq 1$, which gradually resemble that of the conformally-coupled scalars in Fig.~\ref{fig:compareConformalLargeM} as $\mt$ increases. This is because the the dispersion relation for minimal coupling (with $\xi=0$ in Eq.~\eqref{eq:dispersion}) differs from that for conformal coupling (with $\xi=1/6$ in Eq.~\eqref{eq:dispersion}) only in the $R$ term, and this term becomes subdominant if the mass is sufficiently large.

\section{Discussion and Conclusions}
\label{sec:conclusions}

In this work, we have characterized the cosmological gravitational particle production for massive scalars by carefully understanding the features of the spectra. We have considered both conformally and minimally-coupled fields, in both early and late reheating models in order to present as full a picture as possible. In particular, we have compared numerical results for the GPP to various analytic approximations, presenting novel numerical results for early reheating and analytic details for boundary matching calculations. We have considered four different analytic methods: boundary matching, the Stokes phenomenon, steepest descent, and inflaton scattering. For each method, we have outlined the regions of validity in the $\mt-\kt$ plane and calculated the squared Bogoliubov coefficient, $|\beta_k|^2$ and the subsequent number density, $\tilde{a}^3\tilde{n}_k$ for the relevant regions. We then compared these analytic approximations to the scaling of the numerically calculated $\tilde{a}^3\tilde{n}_k$ spectra, and found overall good agreement between the results.

We summarize here the behavior of the number density spectra $\ant$ in terms of their $\kt$ and $\mt$ dependence. The full results with all numerical prefactors were presented in Section~\ref{sec:methods-results}. For the late reheating scenario, GPP depends on $\mt$, in particular whether $\mt < 1$ (small mass) or $\mt > 1$ (large mass). The early reheating scenario is only effective for small masses, and we thus only consider the case where $\mt < 1$. 

We first considered conformally-coupled scalars, comparing the numerical results to analytic approximations. For scalars with $\mt<1$ undergoing late reheating, we found that the $\ant$ spectrum scales like $\kt^2$  for $\kt < \kt_\star$, then exponentially drops as $\kt^{3} e^{-\sqrt{\kt^3/\mt}}$ from $\kt < \kt_\star < 1$, and finally scales as $\kt^{-3/2}$ for $\kt> 1$. For $\mt > 1$, $\at^3 \tilde{n}_k$ scales as $\kt^3 e^{-\mt}$ for $0 < \kt < \mt$ or $\kt<2$, and as $\kt^{-n}$ for $\kt > 2$ or $\kt>\mt$, where $n = 3/2$ for $2\rightarrow 2$ scattering and $n = 15/2$ for $4 \rightarrow 2$ scattering. For early reheating, the spectrum scales as $\kt^2$ for $0 < \kt < \kt_\star$, $\kt^{-6}$ for $\kt_\star < \kt < 1$ and exponentially drops in $\kt$ for $\kt > 1$. The analytic approximations reproduce the numerical results barring the $\kt^{-6}$ scaling for $\kt_* < \kt < 1$ for early reheating. We defer a closer analysis of this particular case to future work, but emphasize that the analytic approximations reproduce all other regions of the spectrum. Furthermore, for the conformally-coupled scalars, we have shown that the comoving number density $\at^3\nt$ is well approximated by the dominant modes of the spectrum.

We performed the same analysis for minimally-coupled scalars. For the late reheating scenario with $\mt< 1$, the spectrum is $\kt$-independent for $0 < \kt < \mt$, then scales as $\kt^{-3}$ for $\mt^{1/3} < \kt < 1$, and scales as $\kt^{-3/2}$ for $\kt > 1$. For larger masses with $\mt > 1$, the spectrum scales as $\kt^3 e^{-\mt}$ for $0 < \kt < \mt$ and as $\kt^{-n}$ for $\kt > \mt$, where $n=3/2$ or $n=15/2$ for $2\rightarrow 2$ and $4\rightarrow 2$ scattering, respectively. Finally, for minimally-coupled scalars in the early reheating model, $\at^3 \nt_k$ is $\kt$-independent for $0 < \kt < \at_\rh^{1/4}\mt^{1/2}$, and scales as $\kt^{-1}$ for $\at_\rh^{1/4}\mt^{1/2}<\kt<(2\at_\rh)^{-1/2}$, then scales as $\kt^{-3}$ for $(2\at_\rh)^{-1/2} < \kt < 1$, and finally exponentially drops in $\kt$ for $\kt > 1$. If the minimally-coupled scalar survives as dark matter, perturbations in the present dark matter density would arise from a quantum process uncorrelated with the quantum origin of the radiation (on cosmological scales).  This will lead to an unacceptably large isocurvature signature in the CMB \cite{Chung_2005}.

There are many directions forward. In particular, it would be interesting to apply our numerical early reheating analysis to particles of spin $s > 0$ to e.g., the vector, tensor, or fermion scenarios. It would also be enlightening to carry out a similar broad analytic analysis for such scenarios, as well as considering further phenomenological implications of the particle production. Additionally, recent work \cite{Chung:2024lky} has discussed an alternative topological method for calculating the Bogoliubov coefficients, which may be useful to compare with our study. We leave these analyses and others to future work.

\acknowledgments

We thank Andrew Long for useful discussions and suggested improvements of this manuscript. The work of L.J.\ was supported by the Kavli Institute for Cosmological Physics at the University of Chicago. The work of E.W.K.\ was supported in part by the US Department of Energy contract DE-FG02-13ER41958 and the Kavli Institute for Cosmological Physics at the University of Chicago. The work of K.T. was supported by the Kavli Institute for Cosmological Physics at the University of Chicago.

\appendix

\section{Boundary Matching}
\label{app:BoundaryMatching}

We now provide details of the boundary matching results presented in Sec.~\ref{sec:methods-results}. Recall that in this method, we find solutions to the mode equation as it evolves through de Sitter (dS), matter-dominated (MD) and (possibly) radiation dominated (RD) epochs and match the solutions at each boundary, which yields an approximate analytic solution for the late-time behavior of the mode functions. For convenience, we will simplify the calculations solving the mode equation by considering sub-regions of the epochs, in which only one term dominates the expression of $\omega_k$.

In order for the boundary matching method to work well, we cannot have too rapid oscillations in the mode function $\beta_k(\eta)$.  To see where this might occur, consider the differential equation for $\beta_k(\eta)$:
\begin{equation}
\beta_{k}^\prime(\eta)=-\frac{1}{2}A_{k}(\eta)\omega_{k}(\eta)\alpha_{k}(\eta)e^{2\ii\int^\eta \dd\eta_1 \: \omega_{k}(\eta_1)} \ , 
\label{eq:betaprime}
\end{equation}
where $A_k$ is the adiabaticity parameter $A_k=\omega_k^\prime/\omega_k^2$. If $|\beta_k| <1$, we can set $\alpha_k=1$ in Eq.~\ref{eq:betaprime}. Generally, $A_k(\eta)\omega_k(\eta)$ is slowly varying in $\eta$ (varying on a timescale $\mathcal{O}(H)$), so if the exponent in Eq.~\ref{eq:betaprime} is rapidly varying, there will be no growth in $\beta_k$.  Let $\Pi=\int^\eta d\eta_1 \, \omega_{k}(\eta_1)$. The oscillations will be ``too large" for boundary matching if $\eta\, d\Pi/d\eta>1$, or $\eta\,\omega_k(\eta)>1$. So the requirement for boundary matching can be stated as $\etat^2 \times \left(\kt^2+\mt^2\at^2\right)<1$. In MD $\etat^2\sim \at$, and in RD $\etat^2\sim \at^2$; in either case $\etat^2>1$.   Since in MD and RD, $\etat>1$ and $\at>1$, we must have  we have  $\kt\ll 1$ and $\mt\ll 1$, in order to avoid rapid oscillations in $\beta_k(\eta)$ which would invalidate the boundary matching method. Thus, the boundary matching method works in regions where both $\kt$ and $\mt$ are less than unity. 

For boundary matching it proves convenient to rescale the mode functions to a dimensionless variable $\chitk=\sqrt{2k}\,\chi_k$ and seek solutions to the mode equation $\partial_\etat^2\,\chitk(\etat) + \omegat_k\,\chitk(\etat) = 0$.  The dispersion relation for $\omegat_k$ is $\omegat_{k}^{2}(\eta)=\tilde{k}^{2}+\tilde{a}^{2}(\eta)\tilde{m}^{2}+\Rt\tilde{a}^{2}(\eta)$, and the Bunch-Davies boundary condition for $\chitk$ reads $\chitk(\etat\to-\infty)=e^{-\ii\kt\etat}$.

In terms of the rescaled mode function
\begin{equation}
|\beta_{k}|^{2}=\underset{\etat\rightarrow\infty}{\text{lim}}\left(\frac{1}{4\kt\omegat_k}|\partial_{\etat}\chit(\etat)|^{2}
                                                               +\frac{\omegat_\kt}{4\kt}|\chit(\etat)|^{2}-\frac{1}{2}\right) \ ,
\end{equation}
and $\at^3\nt_k$ can be found from $|\beta_{k}|^{2}$ in the usual way.

We now calculate the Bogoliubov coefficients, first for conformal coupling in Sec.\ \ref{ssec:bdryConfApp}, and then for minimal coupling in Sec\ \ref{ssec:bdryMinApp}.  For late reheating we need only consider evolution through dS and MD.  For early reheating we must also consider evolution through RD. We will consider late and early reheating in each of these cases.

\subsection{Conformal Coupling}
\label{ssec:bdryConfApp}

For conformally-coupled scalars, the dispersion relation is 
\begin{equation}
\omegat_{k}^{2}(\etat)=\kt^{2}+\at^{2}(\etat)\mt^{2}.
\end{equation}
Using Table \ref{tab:etaRelation}, 
\begin{equation}
    \omegat^2(\etat) = \left\{ \begin{array}{ll}
\kt^2 + \mt^2/\etat^2 & \quad \mathrm{dS} \\[1ex]
\kt^2 + \mt^2 \left(\dfrac{3}{2} + \dfrac{1}{2}\etat\right)^4 & \quad \mathrm{MD} \\[2ex]
\kt^2 + \mt^2\at_\rh\etat^2 & \quad \mathrm{RD}\ (\etat>\etat_\rh) \ .
\end{array} \right.
\end{equation}
For conformal coupling we will be interested in boundary matching for $\kt<\kt_\star$, where $\kt_\star$ is the dominant mode.

\begin{figure}[htb!]
\begin{center}  
\includegraphics[width=0.48\linewidth]{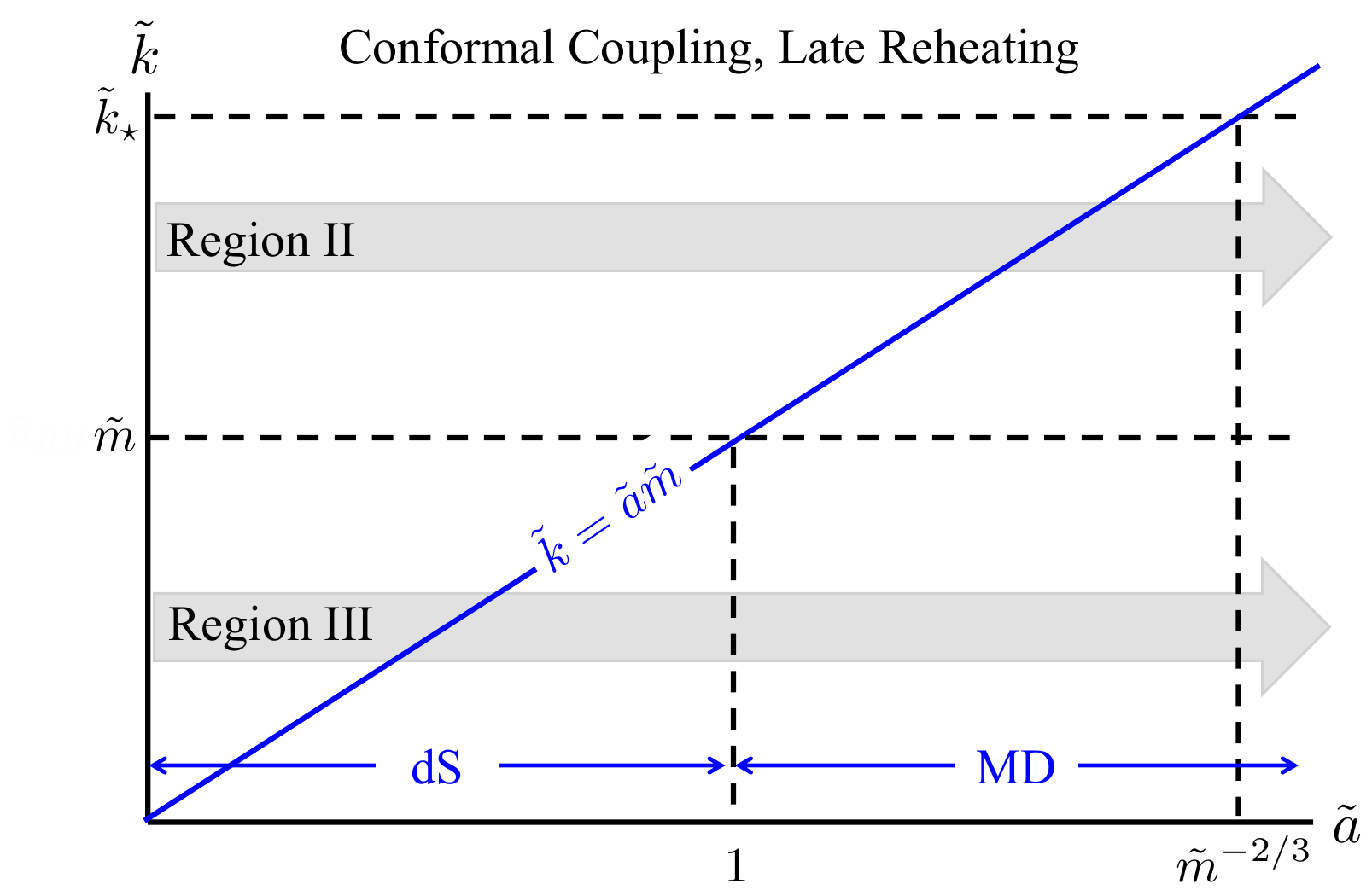}
\includegraphics[width=0.48\linewidth]{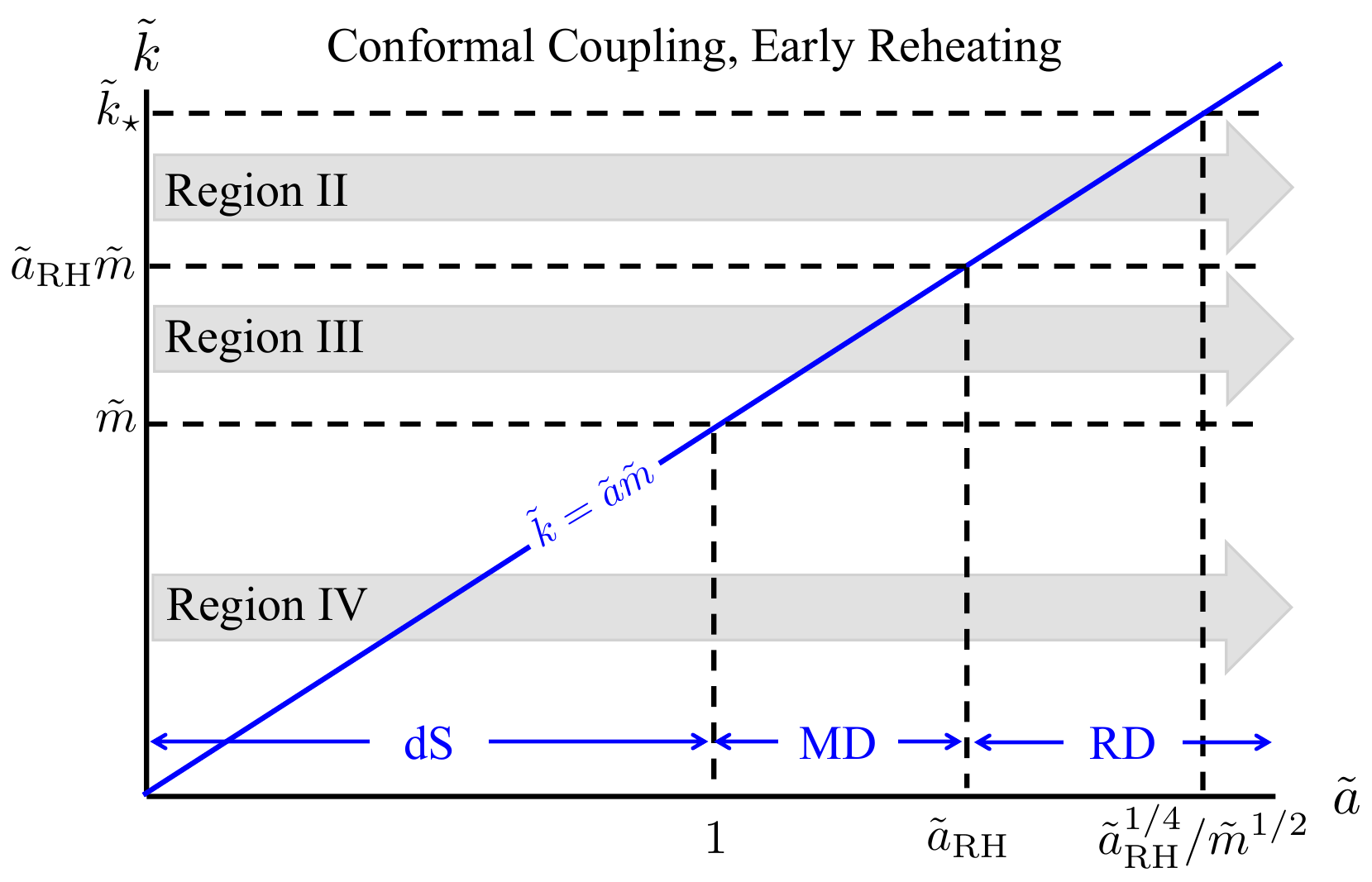}
\caption{Boundary matching method schematics for conformally-coupled scalars. Left panel: $\tilde{k}$ versus $\tilde{a}$ schematics showing the evolution of modes through the epochs of dS and MD in late reheating. Right panel: $\tilde{k}$ versus $\tilde{a}$ schematics showing the evolution of modes through the epochs of dS, MD, and RD in early reheating. The horizontal dotted lines demarcate the $\kt$ versus $\mt$ regions (see Fig.~\ref{fig:ConfRegions}). The vertical dotted lines demarcate the boundaries between epochs, as well as the transition to the sub-Hubble region ($\at=\mt^{-2/3}$ for late reheating, and $\at=\at_\rh^{1/4}\mt^{-1/2}$ for early reheating). The solid line $\kt=\at\mt$ demarcates the boundary between relativistic versus nonrelativistic regions.  The $\kt_\star$ line is the dominant mode: $\kt_\star=\mt^{1/3}$ for late reheating, and $\kt_\star=\at_{\rh}^{1/4}\mt^{1/2}$ for early reheating. }
\label{fig:ConformalModes}
\end{center}
\end{figure} 

Figure~\ref{fig:ConformalModes} illustrates the evolution of conformally-coupled $\kt$-modes through the different epochs of the early universe, with the left panel for late reheating (dS and MD), and the right panel for early reheating (dS, MD, and RD). 

Each mode evolves through different $\etat$-intervals, characterized by the term that dominates $\omegat_k$ in that interval. Boundary crossing occurs each time the mode crosses a solid line or a dotted line. 

We first consider late reheating. We follow the same conventions as the $\kt$ versus $\mt$ parameter space as the left panel of Fig.~\ref{fig:ConfRegions}.  Region II corresponds to modes that become nonrelativistic during MD ($\mt<\kt<\mt^{1/3}$). In Region III, modes become nonrelativistic during dS.

\subsubsection{Late Reheat, Region II, $\mt \leq \kt \leq \kt_\star$}
\label{sssec:bdryLateRHConfII}

In Region II we can separate $\omegat_k^2$ into two intervals in $\etat$:
\begin{equation}
\omegat_k^{2}=\left\{ 
\begin{array}{llll}
\kt^{2}                                              & \quad\etat<\etat_1      & \quad \mathrm{relativistic\ in\ dS\ and\ MD}    & \quad \mathrm{Interval\ 1} \\
\mt^{2}\left(\frac{3}{2}+\frac{1}{2}\etat\right)^{4} & \quad \etat_1<\etat  & \quad \mathrm{nonrelativistic\ in\ MD} & \quad \mathrm{Interval\ 2} \ .\\
\end{array}
\right.
\end{equation}
The boundary $\etat_1$ occurs at $\kt^{2}=\mt^{2}(3/2+\etat/2)^{4}$. Since $\etat_1>-1$,
we choose the positive root: $\etat_1=2(\kt/\mt)^{1/2}-3$. 

\begin{center}\textbf{Interval 1 -- relativistic in dS and MD}\end{center}

The mode equation in this region is 
$\partial_{\etat}^{2}\chit=-k^{2}\chit$, with solution in accordance with Bunch-Davies initial condition to be $\chit_1(\etat)=e^{-\ii\kt\etat} \, \mathrm{and} \ \partial_\etat\chit_1(\etat)=-\ii \kt \, e^{-\ii\kt\etat}$ where $\etat\leq \etat_1$.

\begin{center}\textbf{Interval 2 -- nonrelativistic in MD}\end{center}

After the mode evolves into the region of non-relativistic matter domination, it obeys the mode equation
\begin{equation}
   \partial_{\etat}^{2}\chit=-\mt^{2}\left(\frac{3}{2}+\frac{1}{2}\etat\right)^{4}\chit, 
   \label{eq:NRMD}
\end{equation}
where $\etat_1<\etat$.  The general solution in Interval 2 is 
\begin{equation}
\chit_{2}=A\chit_{A}+B\chit_{B},
\end{equation}
where $\chit_{A}$ and $\chit_{B}$ are expressed in terms of Bessel functions of the first kind:
\begin{align}
\chit_{A}(\etat) & =\frac{\mt^{1/6}}{3^{1/6}}\left(\frac{3}{2}+\frac{1}{2}\etat\right)^{1/2}
      + \Gamma[5/6] \, J_{-1/6}\left[\frac{2}{3}\mt\left(\frac{3}{2}+\frac{1}{2}\etat\right)^{3}\right] \ , \nonumber \\
\chit_{B}(\etat) & =\frac{\mt^{1/6}}{3^{1/6}}\left(\frac{3}{2}+\frac{1}{2}\etat\right)^{1/2}
      + \Gamma[7/6] \, J_{1/6}\left[\frac{2}{3}\mt\left(\frac{3}{2}+\frac{1}{2}\etat\right)^{3}\right] \ .
\label{eq:BesselJ}
\end{align}
We may find the coefficients by matching $\chit_{1}$ and $\chit_{2}$ and their derivatives at the boundary $\etat=\etat_1$:
\begin{align}
A &=\frac{\partial_{\etat}\chit_{B}(\etat)+ik\chit_{B}(\etat)}{\chit_{A}(\etat)\partial_{\etat}\chit_{B}(\etat)-\chit_{B}(\etat)\partial_{\etat}\chit_{A}(\etat)}e^{-ik\etat}\Big|_{\etat=\etat_1},
\nonumber
\\
B &=\frac{\partial_{\etat}\chit_{A}(\etat)+ik\chit_{A}(\etat)}{\chit_{B}(\etat)\partial_{\etat}\chit_{A}(\etat)-\chit_{A}(\etat)\partial_{\etat}\chit_{B}(\etat)}e^{-ik\etat}\Big|_{\etat=\etat_1}.
\label{eq:LateIICoeff2}
\end{align}

Once evaluated at the boundary, the coefficients are expressed in terms of $J_{\pm 1/6}$ with arguments  $[2/3(\kt/\mt^{1/3})^{3/2}]$. Thus, their behavior depends upon whether $\kt\ll \mt^{1/3}$ or $\kt\gg \mt^{1/3}$. In this Region $\kt\ll \mt^{1/3}$, and in this limit the coefficients are
\begin{align}
A  \approx e^{3\ii\kt} \ , \qquad
B  \approx -2\ii\,3^{1/3}e^{3\ii\kt}\frac{\kt}{\mt^{1/3}} \ .
\label{eq:AB}
\end{align}

Expanding Eqs.\ \ref{eq:BesselJ} and \ref{eq:LateIICoeff2} to first order in $\etat$ as $\etat\rightarrow\infty$, we find 
\begin{equation}
|\beta_{k}|^{2}=\frac{3^{2/3}\Gamma^2[5/6]\mt^{1/3}}{4\pi \kt}|A|^{2} + \frac{3^{2/3}\Gamma^2[7/6]\mt^{1/3}}{4\pi \kt}|B|^{2}+\frac{3^{1/6}\mt^{1/3}}{8\kt}(A^{*}B+B^{*}A) \ .
\label{eq:bogoMD}
\end{equation}
Substituting \eqref{eq:AB} into \eqref{eq:bogoMD}, we find the leading term for $\ant$ in the limit $\etat\to\infty$ to be 
\be
\ant = \frac{1}{2\pi^2}\kt^3  |\beta_{k}|^{2} = \frac{3^{2/3}\Gamma^2[5/6]}{8\pi^3}\mt^{1/3}\kt^2 \ .
\label{eq:LateIIbeta}
\ee
This is the result for late reheating, conformal coupling, Region II, as listed in Table \ref{tab:LateConf}.

\subsubsection{Late Reheat, Region III, $0\leq \kt \leq \mt$}
\label{sssec:bdryLateRHConfIII}

We now consider the modes for which $\kt<\mt$ corresponding to Region III in the left panel of Fig.\ \ref{fig:ConformalModes}. These modes transition to the nonrelativistic regime during dS. Since the modes are nonrelativistic before the end of inflation, we can separate $\omegat_k^2$ into three intervals:
\begin{equation}
\omegat_k^{2}(\etat) = \left\{ 
\begin{array}{llll}
\kt^{2}                                              & \quad\etat<\etat_1      & \quad \mathrm{relativistic\ in\ dS}    & \quad \mathrm{Interval\ 1} \\
\mt^{2}\etat^{-2}                                    & \quad\etat_1<\etat<\etat_e   & \quad \mathrm{nonrelativistic\ in\ dS} & \quad \mathrm{Interval\ 2} \\
\mt^{2}\left(\frac{3}{2}+\frac{1}{2}\etat\right)^{4} & \quad \etat_e<\etat  & \quad \mathrm{nonrelativistic\ in\ MD} & \quad \mathrm{Interval\ 3} \\
\end{array}
\right.
\end{equation}
The boundary $\etat_1$ occurs at $\kt^{2}=\mt^{2}/\etat^{2}$. Since $\etat_1<-1$, we choose the negative root: $\etat_1=-\mt/\kt$. 

\begin{center}\textbf{Interval 1 -- relativistic in dS}\end{center}

The solution to the mode equation in Interval 1 is again $\chit_{1}(\etat)=e^{-\ii k\eta}$, so at the end of Interval 1, $\chit_1(\eta_1)=e^{-\ii\mt}$, and $\partial_{\etat}\chit_{1}(\etat_1)=-\ii\kt e^{\ii\mt}$.

\begin{center}\textbf{Interval 2 -- nonrelativistic in dS}\end{center}

The mode equation in this interval is $\partial_{\etat}^{2}\,\chit=-\mt^{2}\etat^{-2} \, \chit$, with general solution  
\begin{equation}
\chit_{2}(\etat)=C\,\etat^{(1+\sqrt{1-4\mt^2})/2} + D\,\etat^{(1-\sqrt{1-4\mt^2})/2} \ ,
\end{equation}
where $\etat_1<\etat<-1$. In order to ensure that the power of $\etat$ is real, we make the restriction that $\mt<1/2$.  We find expressions for $C$ and $D$ by solving the equations $\chit_{2}(\etat_1)=\chit_{1}(\etat_1)$ and $\partial_{\etat}\chit_{2}(\etat_1) = \partial_{\etat}\chit_{1}(\etat_1)$:
\begin{align}
C & =\frac{e^{-\ii\kt\etat_1}\,\etat_1^{-(1+\sqrt{1-4\mt^{2})}/2} (-1+\sqrt{1-4\mt^{2}}-2\ii \kt\etat_1)}{2\sqrt{1-4\mt^{2}}}, \nonumber \\   
D & =\frac{e^{-\ii\kt\etat_1}\,\etat_1^{-(1-\sqrt{1-4\mt^{2})}/2} (-1+\sqrt{1-4\mt^{2}}+2\ii \kt\etat_1)}{2\sqrt{1-4\mt^{2}}}   \ .
\end{align}
Since $\mt<1$, we can simplify $\chit_{2}$ by series expanding it to first order in $\mt\rightarrow0$ to find 
\begin{align}
\chit_{2}(\etat) & =e^{-\ii \kt\etat_1}\,(1-\ii \kt\etat + \ii \kt\etat_1)= e^{\ii \mt}(1-\ii\kt\etat-\ii\mt), \nonumber \\
\partial_{\etat}\chit_{2}(\etat) & = -\ii \kt e^{-\ii \kt\mt}\ .    
\label{eq:sim}
\end{align}

\begin{center}\textbf{Interval 3  -- nonrelativistic in MD}\end{center}

In Interval 3, the mode equation is the same as \ref{eq:NRMD}, so we write the general solution as
\begin{equation}
\chit_{3}(\etat) = E\,\chit_{3E}(\etat) + F\,\chit_{3F}(\etat) \ ,
\label{eq:bdryconflate3chi3}
\end{equation}
where $\chit_{3E}$ and $\chit_{3F}$ are defined in Eq.~\eqref{eq:BesselJ}.  We find the coefficients $E$ and $F$ by matching $\chit_{2}$ and $\chit_{3}$, as well as $\partial_\etat\chit_{2}$ and $\partial_\etat\chit_{3}$, at the boundary $\etat=-1$. In the limit where $\mt\ll 1$,
\begin{align}
E \approx 1+3\ii\kt -3\kt\mt \ , \qquad
F \approx \frac{2\times 3^{1/3}\kt(-\ii+\mt)}{\mt^{1/3}}.
\label{eq:bdryconflate3coeff3}
\end{align}

Substituting the coefficients $E$ and $F$ for $A$ and $B$ in \eqref{eq:bogoMD}, we find the leading term for $\kt^{1/3}/\mt\ll 1$ to be 
\be
\ant = \frac{1}{2\pi^2}\kt^3  |\beta_{k}|^{2} = \frac{3^{2/3}\Gamma^2[5/6]}{8\pi^3}\mt^{1/3}\kt^2 \ ,
\label{eq:LateIIIbeta}
\ee
which is the same result as that of Region II in Eq.~\eqref{eq:LateIIbeta}.  The result for late reheating, conformal coupling, Region III is listed in Table \ref{tab:LateConf}.

\subsubsection{Early Reheat, Region II, $\at_\rh\mt<\kt<\kt_\star$}

The calculation for early reheating follows the same conventions as the $\kt$ versus $\mt$ parameter space as the right panel of Fig.~\ref{fig:ConfRegions}. Region IV corresponds to the modes for which $\kt<\mt$, and become nonrelativistic during dS. Region III corresponds to modes for which $\mt<\kt<\at_\rh\mt$, and become nonrelativistic during MD. Region II corresponds to modes for which $\at_\rh\mt<\kt<\at_\rh^{1/4}\mt^{1/2}$, and become nonrelativistic during RD.

Region II modes of early reheating (see right panel of Fig.~\ref{fig:ConformalModes}) transition to the nonrelativistic regime during RD era, corresponding to $\at_{\rh}\mt<\kt<\kt_\star$. We therefore consider two $\etat$-intervals for Reion II:
\begin{equation}
\omegat_k^{2}=\left\{ 
\begin{array}{llll}
\kt^{2}                                              & \quad\etat<\etat_1      & \quad \mathrm{relativistic\ in\ dS, MD, RD}    & \quad \mathrm{Interval\ 1} \\
\at_\rh\mt^2\etat^2 & \quad \etat_1<\etat &
\quad\mathrm{nonrelativistic\ in\ RD} & \quad \mathrm{Interval\ 2} \ .
\end{array}
\right.
\end{equation}
Now the boundary $\etat_1$ occurs at $\kt^{2}=\at_\rh\mt^{2}\etat^{2}$.  Since $\etat_1>-1$, we choose the positive root: $\etat_1=\kt/(\at_\rh^{1/2}\mt)$.

\begin{center}\textbf{Interval 1  -- relativistic in dS, MD, and RD}\end{center}

The mode equation in this region is again $\partial_{\etat}^{2}\chit=-\kt^{2}\chit$, with values at $\etat_1$ of $\chit_{1}(\etat_1)=e^{-\ii\kt\etat_1}$, and $\partial_\etat\chit_{1}(\etat_1) = -\ii\,ke^{-\ii\kt\etat_1}$. 

\begin{center}\textbf{Interval 2  -- nonrelativistic in RD}\end{center}

The mode equation in this interval is $\partial_{\etat}^{2}\chit(\etat) = -\at_\rh\etat^{2}\mt^{2}\etat^{2}\chit(\etat)$, with the general solution 
\begin{equation}
\chit_{2}(\etat)=G\chit_{2G}(\etat)+H\chit_{2H}(\etat),    
\end{equation}
where $\chit_{2G}(\etat)$ and $\chit_{2H}(\etat)$ are parabolic cylinder functions $D_{-1/2}$:
\begin{align}
\chit_{2G}(\etat)=D_{-1/2}[(1+\ii)\mt^{1/2}\at_\rh^{1/4}\etat]\ , \qquad
\chit_{2H}(\etat)=D_{-1/2}[(-1+\ii)\mt^{1/2}\at_\rh^{1/4}\etat] \ .
\label{eq:parabolicCylinder}
\end{align}
We series expand the expression to first order in $\etat$ as $\etat\rightarrow\infty$, to find 
\begin{equation}
|\beta_{k}|^{2}=\frac{\mt^{1/2}\at_\rh^{1/4}}{2\sqrt{2}\kt}|G|^{2}+\frac{3\mt^{1/2} \at_\rh^{1/4}}{2\sqrt{2}\kt}|H|^{2}+\frac{(1+i)\mt^{1/2} \at_\rh^{1/4}}{2\sqrt{2}\kt}G^{*}H+\frac{(1-i)\mt^{1/2} \at_\rh^{1/4}}{2\sqrt{2}\kt}H^{*}G-\frac{1}{2}
\label{eq:bogoRD}
\end{equation}

We can find the coefficients $G$ and $H$ by matching the solution to that of Interval 1 at $\etat_1=\kt/(\mt\at_\rh^{1/2})$:
\begin{align}
   G = \frac{\partial_{\etat}\chit_{2H}(\etat)+\ii\kt\chit_{2H}(\etat)}{\chit_{2G}(\etat)\partial_{\etat}\chit_{2H}-\chit_{2H}(\etat)\partial_{\etat}\chit_{2G}(\etat)}e^{-\ii\kt\etat}\Big\rvert_{\etat=\eta_1},
\nonumber
\\
    H = \frac{\partial_{\etat}\chit_{2G}(\etat)+ik\chit_{2G}(\etat)}{\chit_{2H}(\etat)\partial_{\etat}\chit_{2G}-\chit_{2G}(\etat)\partial_{\etat}\chit_{2H}(\etat)}e^{-\ii\kt\etat}\Big\rvert_{\etat=\etat_1}.
\end{align}

Since the arguments of $\chit_{2G}(\etat_1)$ and $\chit_{2H}(\etat_1)$ scale with $\kt/\at_\rh \mt^{1/2}$, their behavior depend on whether $\kt\gg \at_\rh \mt^{1/2}$ or $\kt\ll \at_\rh \mt^{1/2}$.  We will only consider the case where $\kt\ll \mt^{1/2}\at_\rh^{1/4}$. The case where $\kt\gg \mt^{1/2}\at_\rh^{1/4}$ is not in the boundary matching method's region of applicability. 

Expanding to first order in $\kt/(\mt^{1/2}\at_\rh^{1/4})\rightarrow 0$, we find 
\begin{align}
    G  \approx \frac{(1-\ii)\Gamma[3/4]}{2^{3/4}\sqrt{\pi}} \ , \qquad
    H  \approx \frac{(1+\ii)\Gamma[3/4]}{2^{3/4}\sqrt{\pi}} \ .
\end{align}
Substituting the expressions for $G$ and $H$ into Eq.~\eqref{eq:bogoRD}, we find that
\begin{equation}
\ant = \frac{\kt^3}{2\pi^2}|\beta_k|^2 = \frac{\at_\rh^{1/4}\Gamma^2[3/4]}{4\pi^3}\kt^2\mt^{1/2}
\label{eq:EarlyIIbeta}
\end{equation}
as shown in Region II of Table \ref{tab:InterConf}.

\subsubsection{Early Reheat, Region III, $\mt<\kt<\at_\rh\mt$}

These modes transition to the nonrelativistic regime during the MD era. For early reheating, this corresponds to $\mt<\kt<\at_\rh\mt$, and Region III in the right panel of Fig.~\ref{fig:ConformalModes}. 
The calculations in this region differ from those of Region II of late reheating (Sec.~\ref{sssec:bdryLateRHConfII}) in an additional nonrelativistic RD interval, Interval 3. 
\begin{equation}
\omegat_k^{2}=\left\{ 
\begin{array}{llll}
\kt^{2}                                              & \quad\etat<\etat_1      & \quad \mathrm{relativistic\ in\ dS, MD}    & \quad \mathrm{Interval\ 1} \\
\mt^{2}\left(\frac{3}{2}+\frac{1}{2}\etat\right)^{4} & \quad \etat_1<\etat<\etat_\rh  & \quad \mathrm{nonrelativistic\ in\ MD} & \quad \mathrm{Interval\ 2} \\
\at_\rh\etat^{2}\mt^{2} & \quad \etat_\rh<\etat &
\quad\mathrm{nonrelativistic\ in\ RD} & \quad \mathrm{Interval\ 3} \\
\end{array}
\right.
\end{equation}

The boundary $\etat_1=2\sqrt{\kt/\mt}-3$. The boundary $\etat_\rh$ occurs at $\etat_\rh=2\at_\rh^{1/2}-3$, since reheating happens at the end of MD, where $\at_\rh=(3/2+\etat_\rh/2)^2$. 

We begin our calculations in Interval 3, since Intervals 1 and 2 have been discussed in Sec.~\ref{sssec:bdryLateRHConfII}. At the end of Eq.~\eqref{eq:LateIICoeff2}, we discussed choosing the limit where $\kt\ll \mt^{1/3}$ during the matter-dominated MD epoch (Interval 2). The same choice applies here, since in early reheating, the MD epoch lasts only until $\at=\at_\rh<\mt^{-2/3}$. We can therefore continue our calculations using the results for $\chit_2$ found in Eq.~\eqref{eq:BesselJ} and Eq.~\eqref{eq:LateIICoeff2}.

\begin{center}\textbf{Interval 3  -- nonrelativistic in RD}\end{center}

Interval 3 is the non-relativistic region of RD, where $\partial_{\etat}^{2}\chit_3(\etat)=-\at_\rh\mt^{2}\etat^{2}\chit_3(\etat)$ with general solution  
\begin{equation}
\chit_{3}(\etat)=K\chit_{3K}(\etat) + L\chit_{3L}(\etat)    \ ,
\end{equation}
where $\chit_{3K}(\etat)$ and $\chit_{3L}(\etat)$ are the parabolic cylinder functions in Eq.~\eqref{eq:parabolicCylinder}.

We may find the $K$, $L$ coefficients by setting $\chit_{2}(\etat_\rh)=\chit_{3}(\etat_\rh)$, and $\partial_\etat\chit_{2}(\etat_\rh)=\partial_\etat\chit_{3}(\etat_\rh)$. Series expand to leading order in $\mt$ as $\mt\rightarrow 0$, we find that 
\begin{align}
K \approx\frac{(1-\ii)e^{3\ii\kt}\Gamma[3/4]}{2^{3/4}\sqrt{\pi}} \ , \qquad
L \approx\frac{(1+\ii)e^{3\ii\kt}\Gamma[3/4]}{2^{3/4}\sqrt{\pi}} \ .
\end{align}

Substituting $G\to K$ and $H\to L$ into Eq.~\eqref{eq:bogoRD} yields 
\begin{equation}    
\ant = \frac{\kt^3}{2\pi^2}|\beta_k|^2 = \frac{\at_\rh^{1/4}\Gamma^2[3/4]}{4\pi^3}\kt^2\mt^{1/2},
\label{eq:EarlyIIIbeta}
\end{equation}
which is the same result as that of Region II of early reheating in Eq.~\eqref{eq:EarlyIIbeta} and  shown in Region III of Table \ref{tab:InterConf}.

\subsubsection{Early Reheat, Region IV, $0<\kt<\mt$}

From Fig.~\ref{fig:ConfRegions}, we see that these modes with $\kt<\mt^{1/3}$ will become nonrelativistic in RD if $\at_\rh<\mt^{-2/3}$.  If $\at_\rh$ is greater than that value, the mode will become nonrelativistic in Interval 3, and effectively we are in the late reheating regime where $\at_\rh\to \infty$.

The calculations for Region IV of early reheating (see Fig.~\ref{fig:ConformalModes}) differs from that of Region III of late reheating (Sec.~\ref{sssec:bdryLateRHConfIII}) only in an additional nonrelativistic RD interval. The four $\etat$-intervals are as follows:
\begin{equation}
\omegat_k^{2}=\left\{ 
\begin{array}{llll}
\kt^{2}                                              & \quad\etat<\etat_1      & \quad \mathrm{relativistic\ in\ dS}    & \quad \mathrm{Interval\ 1} \\
\mt^{2}\etat^{-2}                                    & \quad\etat_1<\etat<\etat_e   & \quad \mathrm{nonrelativistic\ in\ dS} & \quad \mathrm{Interval\ 2} \\
\mt^{2}\left(\frac{3}{2}+\frac{1}{2}\etat\right)^{4} & \quad \etat_e<\etat<\etat_\rh  & \quad \mathrm{nonrelativistic\ in\ MD} & \quad \mathrm{Interval\ 3} \\
\at_\rh\etat^{2}\mt^{2}                              & \quad \etat_\rh<\etat     & \quad \mathrm{nonrelativistic\ in\ RD} & \quad \mathrm{Interval\ 4} \ . \\    
\end{array}
\right.
\end{equation}
The boundaries $\etat_1=-\mt/\kt$, $\etat_e=-1$, and $\etat_\rh=2\at_\rh^{1/2}-3$.

We begin our calculations in Interval 4, which follows after the results of Interval 3 of Late Reheat Region III.

\begin{center}\textbf{Interval 4  -- nonrelativistic in RD}\end{center}

The mode equation in this time interval is $\partial_{\etat}^{2}\chit_4(\etat)=-\at_\rh\etat^{2}\mt^{2}\chit_4(\etat)$, with general solution  
\begin{equation}
\chit_{4}(\etat)=M\chit_{4M}(\etat)+N\chit_{4N}(\etat)    
\end{equation}
where $\chit_{4M}(\etat)$ and $\chit_{4N}(\etat)$ are the parabolic cylinder functions of \eqref{eq:parabolicCylinder}.

We find $M$ and $N$ by matching $\chi_3$ and $\chi_4$ and $\partial_\etat\chi_3$ and $\partial_\etat\chi_4$ at $\eta=\eta_\rh$, where the results of Interval 3 are found in Eq.~\eqref{eq:bdryconflate3chi3} and Eq.~\eqref{eq:bdryconflate3coeff3}. These coefficients may be simplified by series expanding them to first order in $\mt$ around $\mt=0$:
\begin{align}
    M & \approx \frac{-\at_\rh^{-1/4}\mt^{-1/2}\kt(\mt-\ii)\pi+(1-\ii)\Gamma^2[3/4]}{2^{3/4}\sqrt{\pi}\Gamma[3/4]},
\nonumber
\\
    N & \approx \frac{\at_\rh^{-1/4}\mt^{-1/2}\kt(\mt-\ii)\pi+(1+\ii)\Gamma^2[3/4]}{2^{3/4}\sqrt{\pi}\Gamma[3/4]},
\end{align}
Substituting these coefficients into Eq.\ \eqref{eq:bogoRD} and taking the limit where $\kt<\mt$, the leading order term yields 
\begin{equation}
\ant = \frac{\kt^3}{2\pi^2}|\beta_k|^2 = \frac{\at_\rh^{1/4}\Gamma^2[3/4]}{4\pi^3}\kt^2\mt^{1/2},\label{eq:EarlyIVbeta}
\end{equation}
which is the same result as that of Region II \eqref{eq:EarlyIIbeta} and Region III \eqref{eq:EarlyIIIbeta}.  Therefore, the comoving number density for modes in Regions II, III, and IV ($0<\kt<\at_\rh^{1/4}\mt^{1/2}$) undergoing early reheating is 
\begin{equation}
\ant = \frac{\at_\rh^{1/4}\Gamma^2[3/4]}{4\pi^3}\kt^2\mt^{1/2}\ 
\end{equation}
as shown in Table \ref{tab:InterConf}.

\subsection{Minimal Coupling}
\label{ssec:bdryMinApp}

For minimal coupling, the dispersion relation is $\omegat_{k}^{2}(\eta) = \tilde{k}^{2} + \mt^2\at^2 + \Rt\tilde{a}^{2}(\eta)$.  From Table \ref{tab:etaRelation}, in dS $\Rt=-2$, in MD $\Rt=-1/2\at^3$, and in RD $\Rt=0$.  With these simplifications
\begin{equation}
    \omegat_k^2 \simeq \left\{ \begin{array}{lll}
\kt^2 - 2\at^2 & =  \kt^{2} - 2\etat^{-2} & \quad \mathrm{dS} \\[1ex]
\kt^2 + \mt^2\at^2 - \frac{1}{2} \at^{-1} & = \kt^{2}+\mt^{2}\left(\frac{3}{2}+\frac{1}{2}\etat\right)^{4} - \frac{1}{2}\left(\frac{3}{2} + \frac{1}{2}\etat\right)^{-2} & \quad \mathrm{MD} \\[2ex]
\kt^2+\mt^2\at^2 & = \kt^{2}+\mt^{2}\at_\rh\etat^{2} & \quad \mathrm{RD} \ .
\end{array} \right.
\end{equation}

\begin{center}
\begin{figure}[htb!]
\includegraphics[width=0.48\linewidth]{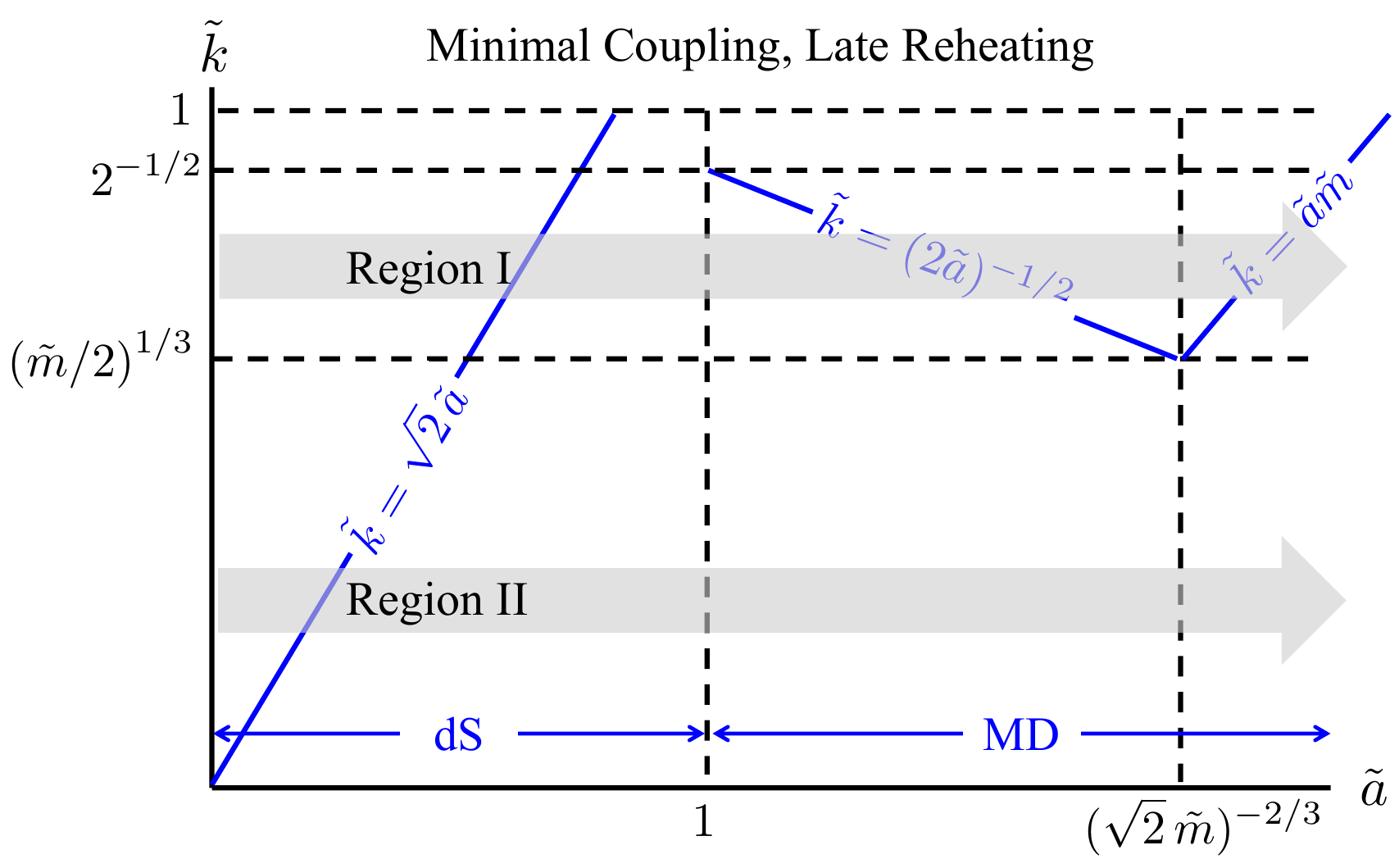}
\includegraphics[width=0.48\linewidth]{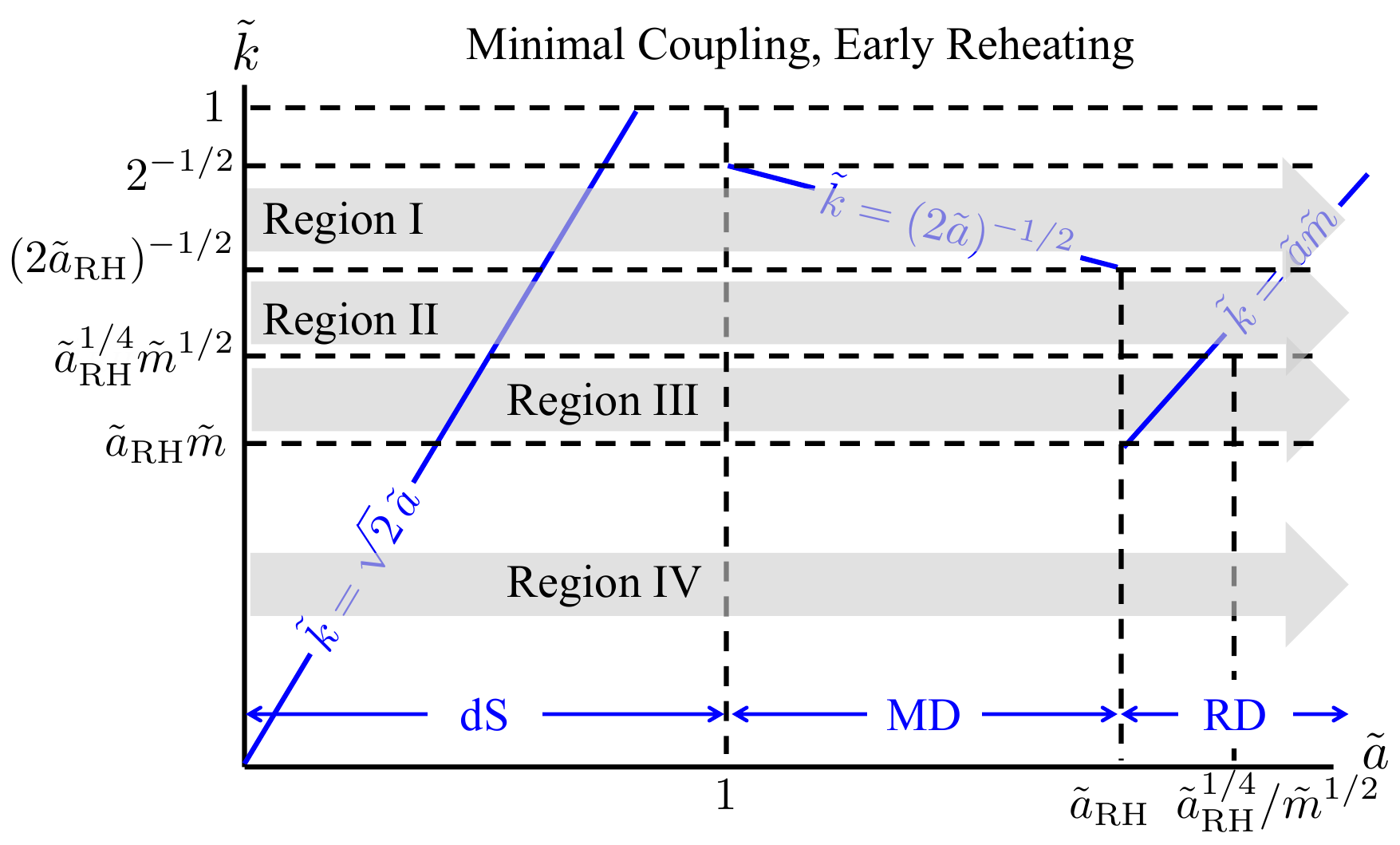}
\caption{Boundary matching schematics for minimally-coupled scalars. Left panel: Evolution of modes through the epochs of dS and MD under late reheating model. Right panel: Evolution of modes through the epochs of dS, MD and RD under early reheating model. }
\label{fig:MinimalRegions}
\end{figure}
\end{center}

Fig.~\ref{fig:MinimalRegions} shows the schematics of minimally-coupled modes evolving through the epochs of the early universe. Note that although the region of validity for the Region I modes have an upper bound of $\kt=1/\sqrt{2}$, we relax the bound to $\kt=1$ and apply the results to all of Region I for both late and early reheating. 

Just as in Sec.~\ref{ssec:bdryConfApp}, the modes of each Region traverse through a different Interval each time it crosses a blue line or an epoch boundary. Since there are now three terms in the dispersion relation, instead of categorizing the Intervals as ``relativistic'' or ``nonrelativistic,'' we use the term that dominates the dispersion relation in each epoch. 

\subsubsection{Late Reheat, Region I, $(\mt/2)^{1/3}<\kt<1$}
\label{sssec:bdryLateRHMinI}

During dS, all modes start with $\omegat_k^2=\kt^2$ (Bunch Davies initial conditions); this is Interval 1.  Then, since $\kt<1/\sqrt{2}$ and inflation ends at $\at=1$, before the end of inflation there will be a transition to $\omegat_k^2=-2\at^2$ at $\at=\kt/\sqrt{2}$; this is Interval 2.  When MD commences at $\at=1$, since $\kt$ and $\mt$ are both less than unity, the dispersion relation will be $\omegat_k^2=1/2\at$; this is Interval 3.  This interval will continue until $\at>1/2\kt^2$ with dispersion relation $\omegat_k^2=\kt^2$ (Interval 4).  Finally, when $\at<\kt/\mt$ the mode enters the nonrelativistic RD region with dispersion relation $\omegat^2=\mt^2\at^2$ (Interval 5). Expressed in terms of $\etat$, 
\begin{equation}
\omegat_k^{2}(\etat) = \left\{ 
\begin{array}{llll}
\kt^{2}                                              & \quad\etat<\etat_1      & \quad\kt^2\mathrm{-dominant\ in\ dS}    & \quad \mathrm{Interval\ 1} \\
-2\etat^{-2} & \quad \etat_1<\etat<\etat_e  & \quad \at^2\Rt\mathrm{-dominant\ in\ dS} & \quad \mathrm{Interval\ 2} \\
- \frac{1}{2}\left(\frac{3}{2}+\frac{1}{2}\etat\right)^{-2} & \quad \etat_e<\etat<\etat_2 &
\quad\at^2\Rt\mathrm{-dominant\ in\ MD} & \quad \mathrm{Interval\ 3} \\
\kt^2 & \quad\etat_2<\etat<\etat_3 & \quad\kt^2\mathrm{-dominant\ in\ MD} & \quad \mathrm{Interval\ 4} \\
\mt^{2}\left(\frac{3}{2}+\frac{1}{2}\etat\right)^{4} & \quad \etat_3<\etat & \quad \at^2\mt^2\mathrm{-dominant\ in\ MD} & \quad \mathrm{Interval\ 5} \ , \\
\end{array}
\right.
\end{equation}
where $\etat_1=-\sqrt{2}\,\kt^{-1}$, $\etat_e=-1$, $\etat_2=\sqrt{2}\,\kt^{-1}-3$, and $\etat_3= 2(\kt/\mt)^{1/2}-3$. 

\begin{center}\textbf{Interval 1  -- $\kt^2$-dominant in dS}\end{center}

The mode equation in Interval 1 is $\partial_{\etat}^{2}\chit_{1}=-\kt^{2}\chit_{1}$, with solution in accordance with Bunch-Davies initial conditions $\chit_{1}(\etat)=e^{-\ii\kt\etat}$, where $\etat<\etat_1$.  At $\etat=\etat_1$, $\chit_1(\etat_1)=e^{\ii\sqrt{2}}$ and $\partial_\etat\chit_k=-\ii\,\,\kt\, e^{\ii\sqrt{2}}$.

\begin{center}\textbf{Interval 2  -- $\at^2\Rt$-dominant in dS}\end{center}

The mode equation in Interval 2 is $\partial_{\etat}^{2}\chit_{2}(\etat)=2\etat^{-2}\chit_{2}(\etat)$, which has solution in the interval $\etat_1<\etat<-1$ given by $\chit_2(\etat) = A\,\etat^{-1} + B\,\etat^2$.  By setting $\chit_2(\etat_1)=\chit_1(\etat_1)$ and $\partial_\etat\chit_2(\etat_1) = \partial_\etat\chit_1(\etat_1)$, we find 
\begin{align}
A = \frac{2\ii-2\sqrt{2}}{3\kt}\,\,e^{\ii\,\sqrt{2}} \ ; \qquad    
B = \frac{\kt^2}{6} \left(1+\ii\sqrt{2}\right)\,e^{\ii\,\sqrt{2}} \ .
\end{align}

\begin{center}\textbf{Interval 3 -- $\at^2R$-dominant in MD}\end{center}

The mode equation in Interval 3 is 
\begin{equation}
\partial_{\etat}^{2}\chit_{3}(\etat) = \frac{1}{2}\left(\frac{3}{2} + \frac{1}{2}\etat\right)^{-2}\chit_{3}(\etat) \ ,
\end{equation}
with solution when $-1<\etat<\etat_2$ given by
\begin{equation}
\chit_{3}(\etat)= C \frac{1}{3+\etat} + D (3+\etat)^2 \ .
\label{eq:bdryminlate2chi3}
\end{equation}
By setting $\chit_{2}(-1)=\chit_{3}(-1)$, and $\partial_\etat\chit_{2}(-1) = \partial_\etat\chit_3(-1)$, we find that 
\begin{align}
C = \frac{2}{3}\kt^2\left(1+\ii\,\sqrt{2}\right)\,e^{\ii\,\sqrt{2}}  \ ; \qquad 
D = - \frac{(1+\ii\sqrt{2})(4\ii+\kt^3)e^{\ii\,\sqrt{2}}}{24\kt} \ .
\label{eq:bdryminlate2coeff3}
\end{align}

\begin{center}\textbf{Interval 4 -- $\kt^2$ dominant in MD}\end{center}

The mode equation in Interval 4 is $\partial_{\etat}^{2}\chit_{4}(\etat)=-\kt^{2}\chit_{4}(\etat)$, with solution $\chit_{4}(\etat)=Ee^{\ii\kt\etat}+Fe^{-\ii\kt\etat}$, where $\etat_2<\etat<\etat_3$.  By setting $\chit_{3}(\etat_2)=\chit_{4}(\etat_2)$ and $\partial_\etat\chit_{3}(\etat_2) = \partial_\etat\chit_{4}(\etat_2)$, with $\etat_2 = \sqrt{2}\,\kt^{-1}-3$, we find that
\begin{align}
E = \frac{4\ii\,\kt^6 -4\ii - \kt^3}{8\kt^3}e^{3\ii\,\kt}   \ ; \qquad 
F = \frac{(2\sqrt{2}+\ii)(4\kt^6-\ii\,\kt^3+4)}{24\kt^3}e^{2\ii\sqrt{2}-3\ii\,\kt} \ .
\label{eq:bdryminlate1coeff4}
\end{align}

\begin{center}\textbf{Interval 5 -- $\mt^2\at^2$ dominant in MD}\end{center}

The mode equation in Interval 5 is 
\begin{equation}
\partial_{\etat}^{2}\chit_{5}(\etat)=-\mt^{2}\left(\frac{3}{2}+\frac{1}{2}\etat\right)^{4}\chit_{5}(\etat) \ ,
\end{equation}
where $\etat_3<\etat$. The solution is $\chit_{5}(\etat)=G\,\chit_{5G}(\etat)+H\,\chit_{5H}(\etat)$, where $\chit_{5G}$ and $\chit_{5H}$ are defined in Eq.~\eqref{eq:BesselJ}. We find expressions for $G$ and $H$ by setting $\chit_{4}(\etat_3)=\chit_{5}(\etat_3)$, and $\partial_\etat\chit_{4}(\etat_3)=\partial_\etat\chit_{5}(\etat_3)$.  The coefficients $G$ and $H$ are expressed in terms of Bessel functions with arguments of $2\sqrt{\kt^3/\mt}/3$.  For these modes, $\kt\gg \mt^{1/3}$. We therefore define a variable $x=\kt/\mt^{1/3}$ and series expand to first order around $x\rightarrow \infty$. 
We then substitute the resulting $G$, $H$ and their complex conjugates into Eq.~\eqref{eq:bogoMD}. The leading order term of $|\beta_{k}|^{2}$ for small $\kt$ is $1/4\kt^6$. The comoving number density is therefore
\begin{equation}
\ant=\frac{\kt^{3}}{2\pi^{2}}|\beta_k|^{2}=\frac{1}{8\pi^{2}}\frac{1}{\kt^{3}} \ ,
\end{equation}
as shown as Region I in Table \ref{tab:LateMin}.

\subsubsection{Late Reheat, Region II, $0<k<(\mt/2)^{1/3}$}
\label{sssec:bdryLateRHMinII}
Compared to the modes of Late Reheat Region I discussed previously in Sec.~\ref{sssec:bdryLateRHMinI}, these modes do not traverse the $\etat$-interval (Interval IV in Sec.~\ref{sssec:bdryLateRHMinI}) where $\omegat_k^2=\kt^2$. 

We begin with Interval 4, which matches to the end of Interval 3 of Region I. The calculations of the previous intervals are found in~\ref{sssec:bdryLateRHMinI}. 

\begin{equation}
\omegat_k^{2}=\left\{ 
\begin{array}{llll}
\kt^{2}                                              & \quad\etat<\etat_1      & \quad\kt^2\mathrm{-dominant\ in\ dS}    & \quad \mathrm{Interval\ 1} \\
-2/\etat^2 & \quad \etat_1<\etat<\etat_e  & \quad \at^2\Rt\mathrm{-dominant\ in\ dS} & \quad \mathrm{Interval\ 2} \\
-\frac{1}{2}(\frac{3}{2}+\frac{1}{2}\etat)^{-2} & \quad \etat_e<\etat<\etat_2 &
\quad\at^2\Rt\mathrm{-dominant\ in\ MD} & \quad \mathrm{Interval\ 3} \\
\mt^{2}\left(\frac{3}{2}+\frac{1}{2}\etat\right)^{4} & \quad \etat_2<\etat & \quad \at^2\mt^2\mathrm{-dominant\ in\ MD} & \quad \mathrm{Interval\ 4} \\
\end{array}
\right.
\end{equation}
where $\etat_1=-\sqrt{2}/\kt$, $\etat_e=-1$, and $\etat_2=2^{5/6}/\mt^{1/3}-3$.

\begin{center}\textbf{Interval 4 -- $\at^2\mt^2$-dominant in MD}\end{center}

The mode equation in this region is 
\begin{equation}
\partial_{\eta}^{2}\chit_{4}(\etat)=-\mt^{2}\left(\frac{3}{2}+\frac{1}{2}\etat\right)^{4}\chit_{4}(\etat),
\end{equation}
where $\etat_2<\eta$. The solution is $\chit_{4}(\etat)=K\,\chit_{4K}(\etat)+L\,\chit_{4L}(\etat)$, with $\chit_{4K}$ and $\chit_{4L}$ defined as in Eq.~\eqref{eq:BesselJ}.

We find expressions for $K$ and $L$ by setting $\chit_{3}(\etat_2)=\chit_{4}(\eta_2)$, and $\partial_\etat\chit_{3}(\etat_2)=\partial_\etat\chit_{4}(\etat_2)$ where $\etat_2=2^{5/6}/\mt^{1/3}-3$. For $\mt\ll 1$,
\begin{align}
    K&\approx\frac{(\sqrt{2}-\ii)e^{\ii\sqrt{2}}(4-\ii \kt^{3})\left(\sqrt{2}J_{-5/6}[\sqrt{2}/3]-2J_{1/6}[\sqrt{2}/3]\right)\Gamma[7/6]}{2\times2^{1/4}\times3^{5/6}\times \kt\times \mt^{2/3}},
    \nonumber
    \\
    L&\approx\frac{(\sqrt{2}-\ii)e^{\ii\sqrt{2}}(4-\ii \kt^{3})\left(-2J_{-7/6}[\sqrt{2}/3]+\sqrt{2}J_{-1/6}[\sqrt{2}/3]\right)\Gamma[5/6]}{2\times2^{3/4}\times3^{5/6}\times \kt\times \mt^{2/3}}.
\end{align}

Putting the coefficients $K$ and $L$ into~\eqref{eq:bogoMD} and using the condition that $\kt\ll 1$, we find the leading term of $|\beta_{k}|^{2}$ to be approximately $1.87/\kt^{3}\mt$.  The comoving number density is then
\begin{equation}
\ant \approx \frac{1.87}{2\pi^2\mt} \ , 
\end{equation}
as shown in Table \ref{tab:LateMin}.

\subsubsection{Early Reheat, Region I, $(2\at_{\rh})^{-1/2}<\kt<1$}
\label{sssec:bdryEarlyRHMinI}

Notice that the Region I modes of early reheating differ from that of late reheating (discussed in Sec.~\ref{sssec:bdryLateRHMinI}) only in Interval 5. 

\begin{equation}
\omegat_k^{2}=\left\{ 
\begin{array}{llll}
\kt^{2}                                              & \quad\etat<\etat_1      & \quad\kt^2\mathrm{-dominant\ in\ dS}    & \quad \mathrm{Interval\ 1} \\
-2/\etat^2 & \quad \etat_1<\etat<\etat_e  & \quad \at^2\Rt\mathrm{-dominant\ in\ dS} & \quad \mathrm{Interval\ 2} \\
-\frac{1}{2}(\frac{3}{2}+\frac{1}{2}\eta)^{-2} & \quad \etat_e<\etat<\etat_2 &
\quad\at^2\Rt\mathrm{-dominant\ in\ MD} & \quad \mathrm{Interval\ 3} \\
\kt^2 & \quad\etat_2<\etat<\etat_3 & \quad\kt^2\mathrm{-dominant\ in\ MD, RD} & \quad \mathrm{Interval\ 4} \\
\at_\rh\etat^2 \mt^2 & \quad \etat_3<\etat & \quad \at^2\mt^2\mathrm{-dominant\ in\ RD} & \quad \mathrm{Interval\ 5} \\
\end{array}
\right.
\end{equation}
where $\etat_1=-\sqrt{2}/\kt$, $\etat_2=-3+\sqrt{2}/\kt$, and $\etat_3=\kt/(\at_\rh^{1/2} \mt)$. We begin with Interval 5. 

\begin{center} \textbf{Interval 5 -- $\at^2\mt^2$-dominant in RD} \end{center}

The mode equation in this time interval is $\partial_{\etat}^{2}\chit_5(\etat)=-\at_\rh\etat^{2}\mt^{2}\chit_5(\etat)$.  The general solution is $\chit_{5}(\etat)=M\chit_{5M}(\etat)+N\chit_{5N}(\etat)$,  where $\chit_{5M}(\etat)$ and $\chit_{5N}(\etat)$ are the parabolic cylinder functions defined in Eq.~\eqref{eq:parabolicCylinder}.

We find $M$ and $N$ by matching the solution to that of Interval 4, found in Eq.~\eqref{eq:bdryminlate1coeff4}, at $\etat_3=\kt/(\at_\rh^{1/2} \mt)$.  Their arguments are in terms of $(\ii\pm1)\kt/(\at_\rh^{1/4}\mt^{1/2})$. Defining $x=\kt/(\at_\rh^{1/4}\mt^{1/2})$ and expanding in the limit of $x\rightarrow\infty$, 
\begin{align}
    M & \approx \frac{e^{-\ii(x^{2}+6\kt)/2}\sqrt{(-1 -\ii)x}} {48\kt^{3}} \nonumber \\
    & \times \left[(4\sqrt{2}+2\ii)e^{2\ii\sqrt{2}}(4\ii\kt^{6}+\kt^{3}+4\ii)+6\ii\sqrt{2}e^{\ii(x^{2}+6\kt)}(4\kt^{6}+\ii\kt^{3}-4)\right],
\nonumber
\\
    N&\approx\frac{e^{\ii(x^{2}+6\kt)/2}(-4\kt^{6}-\ii\kt^{3}+4)\sqrt{(1-\ii)x}}{8\kt^{3}} \ .
\end{align}

Using $M$ and $N$ in Eq.~\eqref{eq:bogoRD} with the $x=\kt/(\at_\rh^{1/4}\mt^{1/2})$ substitution, we find the leading term in $|\beta_{k}|^{2}$ to be $1/(4\kt^6)$. The comoving number density is then
\begin{equation}
\ant=\frac{\kt^{3}}{2\pi^{2}}|\beta_k|^{2}=\frac{1}{8\pi^{2}\kt^{3}} \ .
\label{eq:minlateregionIresult}
\end{equation}
This result is given in Table \ref{tab:InterMin} and is the same result as that of Region I of late reheating in Eq.~\eqref{eq:minlateregionIresult}. 

\subsubsection{Early Reheat, Region II, $\at_{\rh}^{1/4}\mt^{1/2}<\kt<(2\at_{\rh})^{-1/2}$}

The modes of both Region II and III traverse the same $\etat$-intervals as the Region I modes discussed in Sec.~\ref{sssec:bdryEarlyRHMinI}, except that Interval 3 extends to the end of MD. This means that the boundary value of $\etat$ between Interval 3 and 4 is now $\etat_\rh=2\at_\rh^{1/2}-3$. Thus,
\begin{equation}
\omegat_k^{2}=\left\{ 
\begin{array}{llll}
\kt^{2}                                              & \quad\etat<\etat_1      & \quad\kt^2\mathrm{-dominant\ in\ dS}    & \quad \mathrm{Interval\ 1} \\
-2/\etat^2 & \quad \etat_1<\etat<\etat_e  & \quad \at^2\Rt\mathrm{-dominant\ in\ dS} & \quad \mathrm{Interval\ 2} \\
-\frac{1}{2}(\frac{3}{2}+\frac{1}{2}\eta)^{-2} & \quad \etat_e<\etat<\etat_\rh &
\quad\at^2\Rt\mathrm{-dominant\ in\ MD} & \quad \mathrm{Interval\ 3} \\
\kt^2 & \quad\etat_\rh<\etat<\etat_3 & \quad\kt^2\mathrm{-dominant\ in\ RD} & \quad \mathrm{Interval\ 4} \\
\at_\rh\etat^2 \mt^2 & \quad \etat_3<\etat & \quad \at^2\mt^2\mathrm{-dominant\ in\ RD} & \quad \mathrm{Interval\ 5} \ ,
\end{array}
\right.
\end{equation}
where $\etat_1=-\sqrt{2}/\kt$, $\etat_\rh=2\at_\rh^{1/2}-3$, and $\etat_3=\kt/(\at_\rh^{1/2} \mt)$. 

We begin our calculations from Interval 4, matching onto the result for Interval 3 of Sec.~\ref{sssec:bdryEarlyRHMinI}.

\begin{center}
\textbf{Interval 4 -- $\kt^2$-dominant in RD}
\end{center}

The mode equation in this region is $\partial_{\etat}^{2}\chit_{4}(\etat)=-\kt^{2}\chit_{4}(\etat)$, with solution $\chit_{4}(\etat)=Oe^{\ii\kt\etat}+Pe^{-\ii\kt\etat}$, where $\etat_\rh<\etat<\etat_3$ and $\etat_\rh=2\at_\rh^{1/2}-3$. The expressions for $\chit_3(\etat)$ is found in~\eqref{eq:bdryminlate2chi3} and~\eqref{eq:bdryminlate2coeff3}. By setting $\chit_{3}(\etat_\rh)=\chit_{4}(\etat_\rh)$, and $\partial_\etat\chit_3(\etat_\rh) \partial_\etat\chit_4(\etat_\rh)$,
we find $O$ and $P$ in the $\kt\ll1$ limit to be 
\begin{align}
    O&\approx\frac{(\sqrt{2}-\ii)\at_\rh^{1/2}e^{\ii\sqrt{2}}\left[-2\ii+\left(6+\at_\rh^{1/2}(-2-6\ii \kt)+9\ii \kt\right)\kt\right]}{6\kt^{2}},
    \nonumber
    \\
    P&\approx\frac{(\sqrt{2}-\ii)\at_\rh^{1/2}e^{\ii\sqrt{2}}\left[2\ii+\left(6+\at_\rh^{1/2}(-2+6\ii \kt)-9\ii \kt\right)\kt\right]}{6\kt^{2}}.
\end{align}

\begin{center}
    \textbf{Interval 5 -- $-\at^2\mt^2$-dominant in RD}
\end{center}

The mode equation in this time interval is $\partial_{\etat}^{2}\chit_5(\etat) = -\at_\rh\etat^{2}\mt^{2}\chit_5(\etat)$, with general solution $\chit_{5}(\etat) = Q\chit_{5Q}(\etat)+R\chit_{5R}(\eta)$, where $\chit_{5Q}(\etat)$ and $\chit_{5R}(\etat)$ are the parabolic cylinder functions defined in Eq.~\eqref{eq:parabolicCylinder}.

We find $Q$ and $R$ by matching the solution to that of Interval 4 at $\etat_3=\kt/\at_\rh^{1/2}\mt$.  The coefficients $Q$ and $R$ have dependence on parabolic cylinder functions whose arguments are in terms of $(\ii\pm1)\kt/(\at_\rh^{1/4}\mt^{1/2})$ The behavior of these functions depend on whether $\kt\gg \at_\rh^{1/4}\mt^{1/2}$ or $\kt\ll \at_\rh^{1/4}\mt^{1/2}$. The former corresponds to Region II, and the latter corresponds to Region III. To simplify the calculations, we again define $x=\kt/(\at_\rh^{1/4}\mt^{1/2})$. For Region II, we consider the limiting case $x\gg1$, and the coefficients become:
\begin{align}
Q  & \approx \dfrac{\ii^{5/4}(\sqrt{2}-\ii)e^{\ii\sqrt{2}-\ii x^{2}/2}(\at_\rh x)^{1/2}}{3\times2^{3/4}\kt^{2}} \bigg[2+\sqrt{2}e^{\ii x^{2}}\bigg( 2\ii+\Big(-6+\at_\rh^{1/2}\big(2+6\ii \kt\big)-9\ii \kt\Big)\kt \bigg)  \nonumber\\
&  + \kt\bigg(-6\ii-9\kt+\at_\rh^{1/2}\big(2\ii+6\kt\big)\bigg)\bigg] \ , 
\nonumber
\\
R  & \approx \frac{-\left(\frac{1}{6}+\frac{\ii}{6}\right)\ii^{1/4}(2\ii+\sqrt{2})e^{\ii(2\sqrt{2}+x^{2})/2}(\at_\rh x)^{1/2}}{2^{3/4}\kt^{2}}\bigg[2+\kt\bigg(6\ii-9\kt+\at_\rh^{1/2}\Big(-2\ii+6\kt\Big)\bigg)\bigg] \ .
\label{eq:bdryminearly2coeff5}
\end{align}

Substituting them into Eq.~\eqref{eq:bogoRD}, we find that the leading order term is $\at_\rh/3\kt^{4}$, and The comoving number density is then
\begin{equation}
\ant=\frac{\kt^{3}}{2\pi^{2}}|\beta_k|^{2}=\frac{\at_\rh}{6\pi^{2}\kt} .
\end{equation}
It is given in Table \ref{tab:InterMin}, Region II.

\subsubsection{Early Reheat, Region III, $\at_\rh\mt<\kt<\at_{\rh}^{1/4}\mt^{1/2}$}

The Region III modes differ from the Region II modes only in the relation between $\kt$ and $\at_\rh^{1/4}\mt^{1/2}$. Following the discussion in the paragraph preceding Eq.~\eqref{eq:bdryminearly2coeff5}, we now consider the case of $x\ll 1$, where $x=\kt/(\at_\rh^{1/4}\mt^{1/2})$. We simplify $Q$ and $R$ by series expanding them to first order in $x$ around $x=0$:
\begin{align}
    Q & \approx \frac{(\ii-\sqrt{2})\at_\rh^{1/2}e^{\ii\sqrt{2}}}{6\cdot2^{1/4}\sqrt{\pi}\kt^{2}}\cdot\left[x\left(2+\left(6\at_\rh^{1/2}-9\right)\kt^{2}\right)\Gamma[1/4]+\frac{(4-4i)(\at_\rh^{1/2}-3)\kt\pi}{\Gamma[1/4]}\right] \ , 
    \nonumber
    \\
    R & \approx \frac{(\sqrt{2}-\ii)\at_\rh^{1/2}e^{\ii\sqrt{2}}}{6\cdot2^{1/4}\sqrt{\pi}\kt^{2}}\cdot\left[x\left(2+\left(6\at_\rh^{1/2}-9\right)\kt^{2}\right)\Gamma[1/4]-\frac{(4+4i)(\at_\rh^{1/2}-3)\kt\pi}{\Gamma[1/4]}\right] \ .
\end{align}

Substituting them into Eq.~\eqref{eq:bogoRD}, and then replacing instances of $x$ with $\kt/\at_\rh^{1/4}\mt^{1/2}$, we find that the leading order term to be
\begin{equation}
    \ant=\frac{\kt^{3}}{2\pi^{2}}|\beta_k|^{2}=\frac{\at_\rh^{3/4}\Gamma^2[1/4]}{12\pi^{3}\sqrt{\mt}} \ .
\label{eq:RegionIIIEarly}
\end{equation}
It is given in Table \ref{tab:InterMin}, Region III.

\subsubsection{Early Reheat, Region IV, $0<\kt<\at_{\rh}^{1/4}\mt^{1/2}$}

These are the lowest $\kt$ modes of minimally-coupled scalars in early reheating, corresponding to $0<\kt<\at_\rh\mt$. The first three $\etat$-intervals of Region IV are again the same as those for all the Regions of minimal coupling. 
\begin{equation}
\omegat_k^{2}=\left\{ 
\begin{array}{llll}
\kt^{2}                                              & \quad\etat<\etat_1      & \quad\kt^2\mathrm{-dominant\ in\ dS}    & \quad \mathrm{Interval\ 1} \\
-2/\etat^2 & \quad \etat_1<\etat<\etat_e  & \quad \at^2\Rt\mathrm{-dominant\ in\ dS} & \quad \mathrm{Interval\ 2} \\
-\frac{1}{2}(\frac{3}{2}+\frac{1}{2}\eta)^{-2} & \quad \etat_e<\etat<\etat_\rh &
\quad\at^2\Rt\mathrm{-dominant\ in\ MD} & \quad \mathrm{Interval\ 3} \\
\at_\rh\eta^2 m^2 & \quad \etat_\rh<\etat & \quad \at^2\mt^2\mathrm{-dominant\ in\ RD} & \quad \mathrm{Interval\ 4} \ ,
\end{array}
\right.
\end{equation}
where $\etat_1=-\sqrt{2}/\kt$ and $\etat_\rh=2\at_\rh^{1/2}-3$. 

We begin with Interval 4, which matches to the end of Interval 3 of Late Reheat Region I in Eq.~\eqref{eq:bdryminlate2chi3} and Eq.~\eqref{eq:bdryminlate2coeff3}.

\begin{center}
    \textbf{Interval 4 -- $\at^2\mt^2$-dominant in RD}
\end{center}

The mode equation in this time interval is $\partial_{\etat}^{2}\chit_4(\etat)=-\at_\rh\etat^{2}\mt^{2}\chit_4(\etat)$, with general solution $\chit_{4}(\etat) = S\chit_{4S}(\etat)+T\chit_{4T}(\etat)$, where $\chit_{4S}(\etat)$ and $\chit_{4T}(\etat)$ are the parabolic cylinder functions defined in Eq.~\eqref{eq:parabolicCylinder}.

We find $S$ and $T$ by matching the solution to that of Interval 4 at $\etat_\rh=2\at_\rh^{1/2}-3$. Expanding the coefficients to zeroth order around $\kt=0$ and $\mt=0$, We find that
\begin{align}
    S&\approx \frac{2^{1/4}(\ii-\sqrt{2})\at_\rh^{1/4}e^{\ii\sqrt{2}}}{3\Gamma[3/4]\pi^{1/2}\mt^{1/2}\kt}\cdot\left[\pi+(1-\ii)\left(\at_\rh^{1/2}-3\right)\at_\rh^{1/4}\mt^{1/2}\Gamma^{2}[3/4]\right] \ ,
    \nonumber
    \\
    T & \approx \frac{2^{1/4}(\sqrt{2}-\ii)\at_\rh^{1/4}e^{\ii\sqrt{2}}}{3\Gamma[3/4]\pi^{1/2}\mt^{1/2}k}\cdot\left[\pi-(1+\ii)\left(\at_\rh^{1/2}-3\right)\at_\rh^{1/4}\mt^{1/2}\Gamma^{2}[3/4]\right] \ .
\end{align}

Substituting $S$ and $T$ into Eq.~\eqref{eq:bogoRD}, and keeping only the leading order term, we find the comoving number density to be
\begin{equation}
    \ant=\frac{\kt^{3}}{2\pi^{2}}|\beta_k|^{2}=\frac{\at_\rh^{3/4}\Gamma^2[1/4]}{12\pi^{3}\sqrt{\mt}} ,
\end{equation}
listed in Table \ref{tab:InterMin}, Region IV. It is identical to the result for Region III modes found in Eq.~\eqref{eq:RegionIIIEarly}. Therefore, for $\kt<\at_\rh^{1/4}\mt^{1/2}$, the $\ant$ versus $\kt$ spectrum is flat. 

\section{Stokes Phenomenon}
\label{app:Stokes}

In this Appendix, we provide further details of the results in Sec.~\ref{sec:methods-results} which employ the Stokes phenomenon. As discussed above, the mixing of mode functions leading to particle production can also be understood in terms of the Stokes phenomenon. We follow the procedure described in Ref.~\cite{Hashiba:2021npn} to apply the Stokes method to GPP. 

The complex $\etat$ plane can be divided into regions bound by Stokes lines, where the asymptotic behavior of the mode functions differ in each region. In our scenario, the Stokes line is a segment parallel to the imaginary-$\eta$ axis, connecting a pair of roots of the angular frequency $\omegat_k(\etat)$, which we denote as $\etat_c$ and $\overline{\etat_c}$, following the conventions of Ref.~\cite{Hashiba:2021npn}. We choose the pair of roots such that $\operatorname{Re}[\etat_c]>\etat_e$, where $\etat_e=-1$ denotes the end of inflation. The Stokes segment thus divides the complex $\etat$ plane into two regions. The region where $\operatorname{Re}[\etat]<\operatorname{Re}[\etat_c]$ corresponds to the ``early-time'' limit, where we have a single positive-frequency mode function with which we define the vacuum. The region where $\operatorname{Re}[\etat]>\operatorname{Re}[\etat_c]$ corresponds to the ``late-time'' limit, where the mode is now a mix of positive and negative frequency modes. We may consider the mode function as crossing the Stokes line at $\etat_c$. The Bogoliubov coefficient at the Stokes line crossing can be found as 
\begin{equation}
\beta_k=\ii\exp\left[\ii\int_{\overline{\etat_{c}}}^{\etat_{c}}d\etat\:\omegat_{k}\right].
\end{equation}

Finding $\beta_k$ as an integral along the Stokes segment is only valid for well-separated turning points. That is, the length of the Stokes segment cannot approach 0. Moreover, the resulting $|\beta_k|^{2}$ must be much less than unity. Both the length of the Stokes segment and the expression for $|\beta_k|^{2}$ depend on $\tilde{k}$ and $\tilde{m}$. Thus, the Stokes phenomenon is only applicable to regions of the parameter space where the Stokes segment does not collapse, and where the normalization condition is respected. 

\subsection{Conformal Coupling, Late Reheat, Region I, $\kt_\star<\kt$, $\mt<1$}

For conformally-coupled scalars in the late reheating scenario, the late-time epoch is MD, where 
\be
\omegat_k^2=\kt^2 + \mt^2 \left(\dfrac{3}{2} + \dfrac{1}{2}\etat\right)^4,
\ee
and where the turning points of interest are 
\be
\etat_c=-3+(1+\ii)\sqrt{\dfrac{2\kt}{\mt}} \ , \qquad \overline{\etat_c}=-3+(1-\ii)\sqrt{\dfrac{2\kt}{\mt}}\ .
\ee 
We find that
\begin{equation}
    \ant=\dfrac{\kt^3}{2\pi^2}|\beta_k|^{2}=\dfrac{\kt^3}{2\pi^2}\exp\left[-\sqrt{\dfrac{\kt^{3}}{\mt}}\,\dfrac{2\Gamma^2\left[1/4\right]}{3\sqrt{\pi}}\right] \ ,
    \label{eq:StokesLate}
\end{equation}
which agrees with the result in Ref.~\cite{Hashiba:2021npn}.

The length of the Stokes segment $\etat_c-\overline{\etat_c} = 2(2\kt/\mt)^{1/2}$, which is larger than unity for $\kt\gtrsim\mt$. The normalization of $|\beta_k|^2$ requires that $\kt\gtrsim\mt^{1/3}$. The Stokes segment crosses the real-$\etat$ axis at $\operatorname{Re}[\etat_c]>-1$ if $\kt>2\mt$. Therefore, Eq.~\eqref{eq:StokesLate} is applicable in Region I of conformal coupling, late reheating in Fig.~\ref{fig:ConfRegions} and Table~\ref{tab:LateConf}, where both $\kt>\mt^{1/3}$ and $\kt>\mt$. 

\subsection{Conformal Coupling, Early Reheat, Region I, $\kt_\star<\kt$, $\mt<1$}

For scalars undergoing early reheating, the late-time epoch is RD, where
\be
\omegat^2=\kt^2 + \at_\rh\mt^2\etat^2 \ ,
\ee
and where the turning points are 
\be
\etat_c=\dfrac{\ii \kt}{\at_\rh^{1/2}\mt}\ , \qquad \overline{\etat_c}=\dfrac{-\ii \kt}{\at_\rh^{1/2}\mt}\ .
\ee
We find that
\begin{equation}
    \ant=\dfrac{\kt^3}{2\pi^2}|\beta_k|^{2}= \dfrac{\kt^3}{2\pi^2}\exp\left[{-\dfrac{\kt^{2}\pi}{\at_\rh^{1/2}\mt}}\right].
\end{equation}
The length of the Stokes segment is $2\kt/(\at_\rh^{1/2}\mt)$, which is nonvanishing for $\kt>(\at_\rh^{1/2}\mt)$.  The normalization of $|\beta_k|^2$ requires that $\kt>\at_\rh^{1/4}\mt^{1/2}$. Therefore, the above result should be applicable in Region I of conformally-coupled scalars undergoing early reheating in Tab.~\ref{tab:InterConf} and Fig.~\ref{fig:ConfRegions}. However, we find from the numerical calculations that the spectrum instead scales as $\kt^{-6}$.

\section{Steepest Descent}
\label{app:steepest}
In this appendix we provide further details on the results in Sec.~\ref{sec:methods-results} arising from the steepest descent method. Calculating particle production by this method was introduced by Chung \cite{PhysRevD.67.083514}, and recently studied in detail by Racco, Verner, and Xue \cite{Racco:2024aac}.  Here we follow the analysis of Chung.

We start with the late-time mode function written in terms of the early-time mode functions: 
\begin{equation}
\chi_{k}^{\mathrm{IN}}=\alpha_{k}\chi_{k}^{\mathrm{OUT}}+\beta_{k}\chi_{k}^{\mathrm{OUT}*},
\end{equation}
where the coefficients $\alpha_{k}$ and $\beta_{k}$ are the Bogoliubov coefficients. From this expression and the mode equation, Eq.~\eqref{eq:modeeq}, we can write the differential equation for the Bogoliubov coefficient $\beta_k$ as (in this section, prime represents $\partial_\etat$) \cite{Kolb:2023ydq}: 
\begin{align}
    \beta_{k}' = \frac{\omegat_{k}'}{2\omegat_{k}}\exp\left(-2i\int^\etat\omegat_{k}(\etat_1)\dd\etat_1\right)\alpha_{k} \ .
\end{align}
For ease of notation we will express the dispersion relation as 
\begin{equation}
    \omegat_k^2(\etat)=\kt^2 + \mt^2 C(\etat) \ ,
\end{equation}
where $C(\etat)=\at^2(\etat)$ for conformal coupling and $C(\etat)=\at^2(\etat)(1+\Rt(\etat)/\mt^2)$ for minimal coupling. $C(\etat)$ is assumed to be $C^\infty$ on the real axis, and will be analytically continued to the complex plane.  We choose the initial conditions to be $\alpha_k(\etat)=1$ and $\beta_k(\etat)=0$ as $\etat \rightarrow -\infty$. As long as $|\beta(\etat)| \ll 1$, then $|\alpha(\etat)| \approx 1$, and we may make the following approximation (recall $|\alpha_k|^2-|\beta_k|^2=1$ \cite{Kolb:2023ydq}):
\begin{equation}
    \beta_{k}'(\etat)=\frac{\omegat_{k}'(\etat)}{2\omegat_{k}(\etat)}\exp\left(-2i\int^\etat\omegat_{k}(\etat_1)\dd\etat_1\right) \ ,
\end{equation}
so, at late time, $\beta_k$ can then be found as
\begin{align}
    \beta_k(\etat) & = \int^\etat \! \dd\etat_2 \, \frac{\omegat_{k}'(\etat_2)}{2\omegat_{k}(\etat_2)}\exp\left(-2\ii\int^{\etat_2}\omegat_{k}(\etat_1)\dd\etat_1\right)
    \ .
    \label{eq:sdbeta1}
\end{align}
For the rest of this section $\etat$ will be assumed to be in the complex plane and for the sake of simplicity in this section we drop the tilde on $\etat$.  The analytic structure of \eqref{eq:sdbeta1} in the complex-$\eta$ plane is clear: there is a pole at $\eta=\hat{\eta}$, determined by $C(\hat\eta)=-\kt^2/\mt^2$. Since there is a square-root in the exponent, the pole also corresponds to a branch point.  

\begin{figure}[htb!]
\begin{center}  
\includegraphics[width=0.8\linewidth]{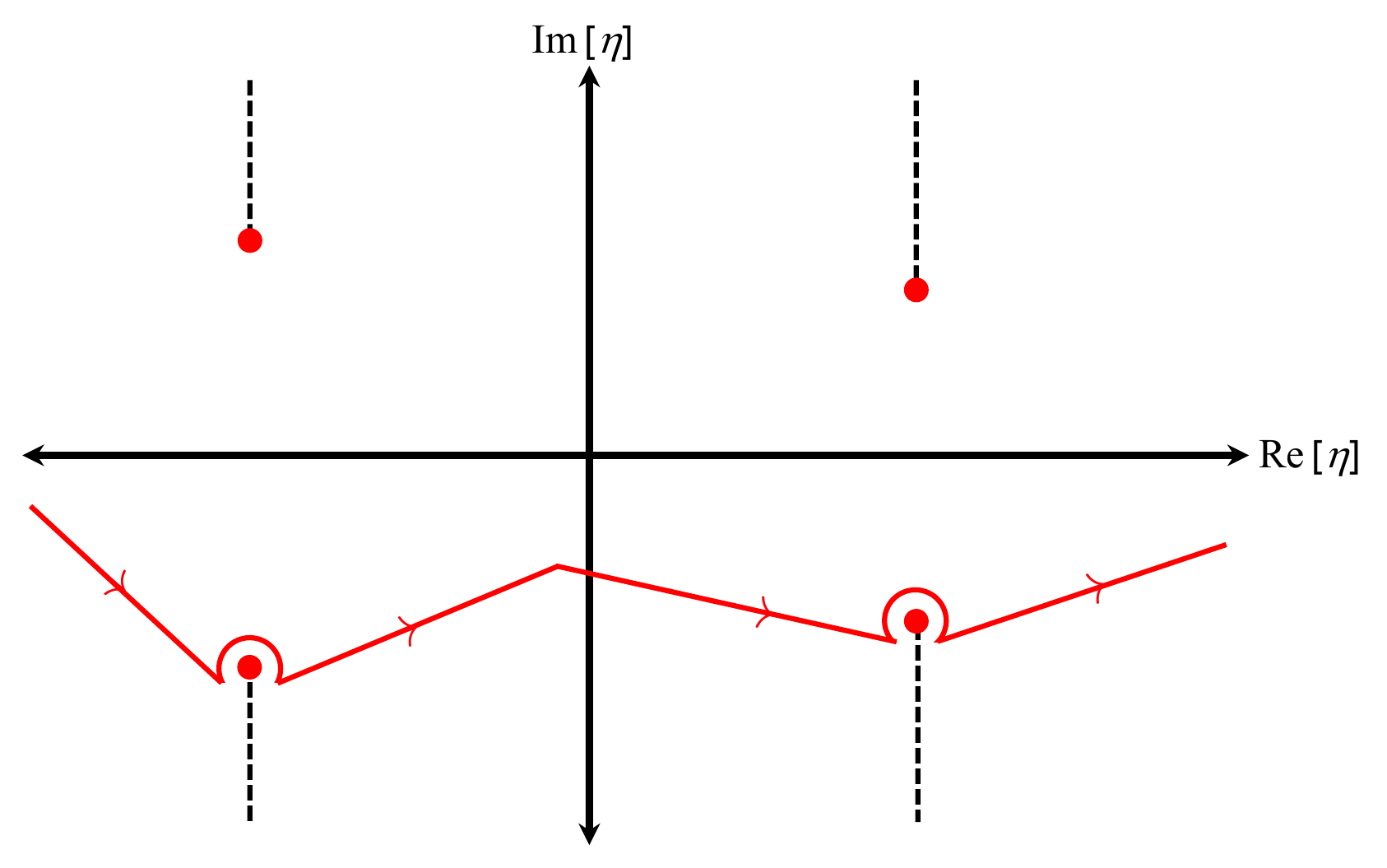}
\caption{A schematic of the contour for the integration in the complex $\eta$ plane is illustrated by the solid red line.  The red points are branch points, and the dashed lines indicates the branch cuts.  The contour path around the branch points covers an angle of $4\pi/3$ to avoid the branch cut \cite{PhysRevD.67.083514}.  }
\label{fig:Contour}
\end{center}
\end{figure} 

We use Cauchy's theorem and replace the integral in \eqref{eq:sdbeta1} by a contour in the lower half plane.  The steepest descent method uses the fact that the dominant contribution to the integral in \eqref{eq:sdbeta1} comes from near the branch points (the contribution from the end points will be suppressed as $\omegat_k^\prime/\omegat_k\to 0$ there).  For the moment, assume there is but one branch point $\hat\eta$. Near $\hat\eta$ we can write
\begin{align}
    \int_{-\infty}^\eta \dd \eta_1\, \omegat_k(\eta_1) = \int_{-\infty}^{\hat\eta} \dd \eta_1\, \omegat_k(\eta_1)  + \int_{\hat\eta}^{\eta} \dd \eta_1\, \omegat_k(\eta_1) \ . 
    \label{eq:expansion}
\end{align}
Expanding $\omegat_k$ about $\hat\eta$ yields $\omegat_k(\eta) = \mt\,\sqrt{C^\prime(\hat\eta)}\,(\eta-\hat\eta)^{1/2}$, and the second integral in \eqref{eq:expansion} becomes 
\begin{align}
\int_{\hat\eta}^{\eta} \dd \eta_1\, \omegat_k(\eta_1) = \frac{2\mt}{3} \sqrt{C^\prime(\hat\eta)}\, (\eta-\hat\eta)^{3/2}  = \frac{2\mt}{3} \sqrt{C^\prime(\hat\eta)}\, \delta^{3/2} \ ,
\end{align}
where $\delta = \eta-\hat\eta$.  Again using the expansion $C(\eta)= C(\hat\eta) + C'(\hat\eta)(\eta-\hat\eta)+\ldots = -k^2/m^2  + C'(\hat\eta)(\eta-\hat\eta)+\ldots$, the prefactor  in \eqref{eq:sdbeta1} becomes
\begin{align}
\frac{\omega'(\eta)}{\omega(\eta)} = \frac{1}{2} \frac{m^2C'(\tilde\eta)}{m^2C'(\tilde\eta)(\eta-\tilde\eta)} = \frac{1}{2}\frac{1}{\delta} \ .
\label{eq:twins}
\end{align}
Therefore $\beta_k$ becomes
\begin{align}
\beta_k(\eta) =   \left[\frac{1}{4}\int_\mathcal{C} \frac{\dd\delta}{\delta} \exp \left( -\ii\frac{4\mt}{3}\sqrt{C'(\hat\eta)} \ \delta^{3/2}\right) \right] \left[
 \exp \left( -2\ii \int_{-\infty}^{\hat\eta} \omegat_k(\eta_1)\,\dd\eta_1\right)\right] \ , 
 \label{eq:twosquares}
\end{align}
where $\mathcal{C}$ is an appropriately chosen contour in the complex $\eta$ plane. In Ref.~\cite{PhysRevD.67.083514} Chung gives details about the choice of the contour (also see Racco, et al.\ \cite{Racco:2024aac}) and finds 
\begin{align}
    \frac{1}{4}\int_\mathcal{C} \frac{\dd\delta}{\delta} \exp \left( -\ii\frac{4\mt}{3}\sqrt{C'(\hat\eta)} \ \delta^{3/2}\right) = \ii\frac{\pi}{3} \ .
    \label{eq:ipi}
\end{align}

It remains now to calculate the second factor in square brackets in Eq.\ \eqref{eq:twosquares}.  The integral will have a real and an imaginary part.  When exponentiated, the real part of the integral will contribute an irrelevant phase and will be ignored.  We will write the pole as $\hat\eta = \hat\eta_R-\ii\,\hat\eta_I$, where $\hat\eta_R$ and $\hat\eta_I\geq0$ are purely real.  Since $\omegat_k$ is analytic on the real axis, we can split the integral as
\begin{align}
    \int_{-\infty}^{\hat\eta} \omegat_k(\eta_1)\,\dd\eta_1 = \int_{-\infty}^{\hat\eta_R} \omegat_k(\eta_1)\,\dd\eta_1 + \int_{\hat\eta_R}^{\hat\eta} \omegat_k(\eta_1)\,\dd\eta_1 \ .
\end{align}
The first term is integration along the real axis, resulting in a purely real result, which can be ignored.  To calculate the second integral we expand $\omegat_k(\eta_1)$ about $\hat\eta_R$:
\begin{align}
    \omegat_k(\eta_1) = \omegat_k(\hat\eta_R) + \omegat_k'(\hat\eta_R)(\eta_1-\hat\eta_R) + \frac{1}{2}\omegat_k''(\hat\eta_R)(\eta_1-\hat\eta_R)^2 + \ldots \ ,
\end{align}
and find
\begin{align}
     \int_{\hat\eta_R}^{\hat\eta} \omegat_k(\eta_1)\,\dd\eta_1 = -\ii\,\hat\eta_I\omegat_k(\hat\eta_R) - \frac{1}{2}\hat\eta_I^2\, \omegat'_k(\hat\eta_R) + \frac{1}{6}\hat\eta_I^3\omegat''_k(\hat\eta_R) + \ldots \ .
\end{align}
The second term is purely real and can be ignored, so just keeping the first term, Eqs.\ \eqref{eq:twosquares} and \eqref{eq:ipi} yield
\be
\beta_k = \frac{\ii \pi}{3}\exp[-2\hat\eta_I \ \omegat_k(\hat\eta_R)] \ .
\label{eq:sdbeta2}
\ee

The final step is to express $\omegat_k(\hat\eta_R)$ and $\hat\eta_I$ in terms of $C(\hat\eta_R)$ and its derivatives.  This is accomplished by yet another expansion:
\begin{align}
    C(\hat\eta) = C(\hat\eta_R) + C'(\hat\eta_R)(\hat\eta-\hat\eta_R) + \frac{1}{2}C''(\hat\eta_R)(\hat\eta-\hat\eta_R)^2 + \frac{1}{6}C'''(\hat\eta_R)(\hat\eta-\hat\eta_R)^3 \ . 
\end{align}
Equating the real and imaginary parts of $C(\hat\eta) + \kt^2/\mt^2 = 0$ yields two equations:
\begin{align}
    \frac{\hat\eta_I^2}{6}C'''(\hat\eta_R) - C'(\hat\eta_R) &=0 \ , \nonumber \\
    \frac{\omegat_k^2(\hat\eta_R)}{\mt^2} -\frac{\hat\eta_I^2}{2}C''(\hat\eta_R)&=0 \ .
    \label{eq:etaI}
\end{align}
From Eq.~\eqref{eq:etaI}, the dispersion relation is found to be
\begin{equation}    \omegat_k(\hat\eta_{R})=\mt\sqrt{3}\sqrt{\frac{C''(\hat{\eta}_{R})C'(\hat{\eta}_{R})}{C'''(\hat{\eta}_{R})}} \ .
\label{eq:sdomega}
\end{equation}
Rearranging Eq.~\eqref{eq:sdomega}, we arrive at an useful expression for $\kt$:
\begin{equation}    
\tilde{k}^{2}=\tilde{m}^{2}\left(3\frac{C''(\hat\eta_R)C'(\hat\eta_R)}{C'''(\hat\eta_R)}-C(\hat\eta_R)\right) \ .
    \label{eq:sdk}
\end{equation}

Finally, from Eq.~\eqref{eq:sdbeta2} the squared Bogoliubov coefficient $|\beta_k|^2$ can be expressed as
\begin{equation}
    |\beta_k|^{2}=\frac{\pi^{2}}{9}\exp\left[-12\sqrt{2}\,\mt\frac{C'(\hat\eta_{R})\sqrt{C''(\hat\eta_{R})}}{C'''(\hat\eta_{R})}\right] \ .
    \label{eq:sdbeta3}
\end{equation}

The steepest descent method is only applicable to late reheating models, which do not evolve to the RD regime. The method is not applicable in RD, where $C(\hat\eta_R)\propto\hat\eta_R^2$, and so $C'''(\hat\eta_R)=0$. This method finds $|\beta_k|^2$ spectra that are constant in $\tilde{k}$ in the low-$\tilde{k}$ region, and exponentially decaying in $\tilde{k}$ in the high-$\tilde{k}$ region. 

We apply this method to the high-$\tilde{k}$ region ($\tilde{k}>\tilde{m}^{1/3}$) of spectra for light particles ($\tilde{m}<1$), where it agrees with the results of the Stokes phenomenon method, and to the low-$\tilde{k}$ region ($\tilde{k}<\tilde{m}$) of spectra for heavy particles ($\tilde{m}>1$). 

For calculations in the remainder of App.~\ref{app:steepest}, prime denotes derivative along the real-$\eta$ axis, and we define $\at(\hat\eta_R)\equiv \at_R$.

\subsection{Conformal Coupling, Late Reheat, Region I, $\kt_\star<\kt$, $\mt<1$}

For conformally-coupled scalars, $C(\hat\eta_R)= \at^2_R$.   First,  consider particle production when the background is matter-dominated.  In the MD regime, $\at_R' = \at_R^2\tilde{H}(\hat\eta_R) = \at_R^{1/2}$, since $\tilde{H}(\hat\eta_R)= \at_R^{-3/2}$.  Therefore, we find that $C(\hat \eta_R)=\at_R^2, \ C'(\hat \eta_R)= 2\at_R^{3/2}, \ C''(\hat \eta_R)= 3\at_R, \  C'''(\hat \eta_R)= 3\at_R^{1/2}$.  Substituting these expressions into Eq.~\eqref{eq:sdk}, we find that $\kt=\sqrt{5}\,\mt \, \at_R$. Using this relation and the expressions of $C(\hat \eta_R)$ and its derivatives with Eq.~\eqref{eq:sdbeta3}, we find the spectrum to be
\begin{equation}
    |\beta_k|^{2}=\dfrac{\pi^{2}}{9}\exp\left[-\sqrt{\dfrac{\tilde{k}^{3}}{\tilde{m}}}\dfrac{8\sqrt{6}}{5^{3/4}}\right] \ , \qquad \ant=\dfrac{\kt^3}{18}\exp\left[-\sqrt{\dfrac{\tilde{k}^{3}}{\tilde{m}}}\dfrac{8\sqrt{6}}{5^{3/4}}\right] \ .
    \label{eq:DescentLateRegionI}
\end{equation}

Under late reheating, the background remains in MD as long as $\at_R>1$, which yields $\tilde{k}>\sqrt{5}\tilde{m}$.  We use this result in Region I of conformal coupling, late reheating in Fig.~\ref{fig:ConfRegions} and Table~\ref{tab:LateConf}, where $\tilde{k}>\tilde{m}^{1/3}$ and $\mt<1$. Although there exists a gap in Region I where $\mt^{1/3}<\kt<\sqrt{5}\,\tilde{m}$ such that Eq.~\eqref{eq:DescentLateRegionI} is not technically applicable, we assume that Eq.~\eqref{eq:DescentLateRegionI} is valid for all of Region I. Equation~\eqref{eq:DescentLateRegionI} agrees up to a numerical factor with the result found by applying the Stokes Phenomenon in Eq.~\eqref{eq:StokesLate}. 

We do not apply this result to $\mt>1$, since the region where $\kt>\sqrt{5}\,\tilde{m}>\sqrt{5}$ is instead dominated by inflaton scattering.

\subsection{Conformal Coupling, Late Reheat, Region V, $0<\kt<\mt$, $\mt>1$}

Now, we consider particle production during the dS regime, where $\Ht(\hat\eta_R)=1$ and $\at_R'=\at_R^2$. Therefore, we have $C(\hat \eta_R)=\at_R^2$, $C'(\hat \eta_R) = 2\tilde{a}_R^{3}$, $C''(\hat \eta_R) = 6\tilde{a}_R^{4}$, and $C'''(\hat \eta_R) =24\tilde{a}_R^{5}$. Using these expressions and Eq.~\eqref{eq:sdbeta3}, we find:
\begin{align}
    |\beta_k|^{2}=\dfrac{\pi^{2}}{9}\exp\left[-2\sqrt{3}\,\tilde{m}\right] \ , \qquad \ant=\dfrac{\kt^3}{18}\exp\left[-2\sqrt{3}\,\tilde{m}\right] \ .
    \label{eq:descentConfdS}
\end{align}

Substituting the expressions for $C(\hat \eta_R)$ and its derivatives into Eq.~\eqref{eq:sdk} and imposing dS assumptions for the background ($\tilde{a}_R<1$), we find $\tilde{k}<\tilde{m}/\sqrt{2}$. We approximate this bound to be $\tilde{k}<\tilde{m}$. Note that for $\mt\ll1$, $|\beta_k|^2=\pi^2/9$, which violates the $|\beta_k|^2\rightarrow0$ normalization condition. We therefore apply Eq.~\eqref{eq:descentConfdS} to Region V of conformal coupling, late reheating in Fig.~\ref{fig:ConfRegions} and Table~\ref{tab:LateConf}, where $\kt<\mt$ and $\mt>1$.

\subsection{Minimal Coupling, Late Reheat, Region IV, $0<\kt<\mt$, $\mt>1$}

For minimally coupled scalars in the dS regime, $C(\hat\eta_R)=\at_R^{2}(1-2/\mt^{2})$ and again, $\at_R'=\at_R^2$. We find $C'(\hat \eta_R)= 2\tilde{a}_R^{3}(1-2/\mt^2)$, $C''(\hat \eta_R)= 6\tilde{a}_R^{4}(1-2/\mt^2)$, $C'''(\hat \eta_R)=24\tilde{a}_R^{5}(1-2/\mt^2)$. With these expressions and Eq.~\eqref{eq:sdbeta3}, we find the spectrum to be 
\be
|\beta_k|^2=\dfrac{\pi^{2}}{9}\exp\left[-2\sqrt{3(\mt^{2}-2)}\right] \ , \qquad \ant=\dfrac{\kt^3}{18}\exp\left[-2\sqrt{3(\mt^{2}-2)}\right] \ .
\label{eq:descentMindS}
\ee
Substituting the expressions for $C(\hat \eta_R)$ and its derivatives into Eq.~\eqref{eq:sdk}, we find that $\kt^2=\at_R^{2}(\mt^{2}/2-1)$. Imposing $\at_R<1$, we find the restriction that $\kt<\sqrt{\mt^2/2-1}$, where $\mt>\sqrt{2}$. We approximate this result's region of applicability to be $\tilde{k}<\tilde{m}$ and $\tilde{m}>1$, which corresponds to Region IV of minimal coupling, late reheating in Fig.~\ref{fig:MinRegions} and Table~\ref{tab:LateMin}. Notice that for $\mt\gg1$, Eq.~\eqref{eq:descentMindS} is the same as Eq.~\eqref{eq:descentConfdS} for Region V of conformal coupling, late reheating.

\section{Inflaton Scattering}
\label{app:scattering}

Finally, in this appendix, we provide further details on the $2\rightarrow 2$ and $ n\rightarrow 2$ inflaton scattering results shown in Sec.~\ref{sec:methods-results}. During reheating, the universe is assumed to be filled with a non-relativistic condensate of inflatons. The rapidly oscillating inflatons can produce $\chi$ particles through graviton exchange. The leading order effect is 2-to-2 scattering, $\varphi\varphi \rightarrow \chi \chi$, which is kinematically allowed if $m_\chi<m_\varphi$. For quadratic inflation, $m_\varphi=2H_e$, we therefore have 2-to-2 scattering for $\chi$ particles with mass $\tilde{m}_\chi<2$. For heavier $\chi$ particles, the leading order scattering mechanism will be $n\varphi \rightarrow \chi \chi$, where $n$ is the smallest number such that $m_\chi<nm_\varphi/2$. Inflaton scattering becomes the dominant effect in the region of the spectrum where $k>m_\varphi$. For the quadratic inflation model, the region of applicability is therefore $\tilde{k}>2$. Numerically, we find that for $\mt<1$, the effects of inflaton scattering dominates the spectrum after $\kt\approx1$. For $\mt>1$, scattering dominates after $\kt\approx2$, while the region $1<\kt<2$ transitions between the red-tilted and the decaying behavior. 

First consider $2\rightarrow 2$ scattering. The comoving number density of produced $\chi$ is 
\be 
\frac{d[a^{3}n]}{dt}=a^{3}n_{\varphi}^{2}|v| \left\langle\sigma_{\varphi\varphi\rightarrow\chi\chi}\right\rangle,
\label{eq:scatterdt}
\ee 
where $\left\langle|v|\sigma_{\phi\phi\rightarrow \chi\chi}\right\rangle$ is from Ref.~\cite{Tang_2017}, 
\begin{equation}
\left\langle |v|\sigma_{\varphi\varphi\rightarrow\chi\chi}\right\rangle = \frac{1}{512\pi M_{\mathrm{Pl}}^{4}}\sqrt{1-\frac{m^{2}}{m_{\varphi}^{2}}}\frac{[2(1-6\xi)m_{\varphi}^{2}+m^{2}]^{2}}{m_{\varphi}^{2}},
\label{eq:wutang}
\end{equation}
and the number density of inflatons $n_{\varphi}$ is 
\begin{equation}   
n_{\varphi}=\frac{3M_{\mathrm{Pl}}^{2}H_{e}^{2}a_{e}^{3}}{m_{\varphi}a^{3}}\Phi
\label{eq:scatternphi}\ ,
\end{equation}
where we work with the auxiliary variable $\Phi$ defined as 
\be 
\Phi = \frac{a^3\rho_\varphi}{a^3_e\rho_{\varphi,e}},
\ee 
where $\rho_{\varphi,e}=3H_e^2M_{\mathrm{Pl}}^2$ is the approximate value of the field energy density at the end of inflation. For late reheating, $\Phi = 1$, while for early reheating, $\Phi$ decays exponentially after reheating via 
\be 
\Phi=e^{-\Gamma_\varphi t}.
\label{eq:PhiDE}
\ee 
In the limit where $m \ll m_\varphi$, this resonant production mechanism produces $\chi$ particles with wavenumber $k$ when $a=k/m_{\varphi}$. Using $da/dt=Ha$ and $dk/da=m_\varphi$  in Eq.~\eqref{eq:scatterdt} and substituting in Eq.~\eqref{eq:scatternphi}, we find that 
\begin{equation}   
a^3n_k=k \frac{d[a^{3}n]}{dk}=\frac{9kH_{e}^{4}a_{e}^{6}M_{\mathrm{Pl}}^{4}}{m_{\varphi}^{3}a^{4}H}\Phi^2\left\langle|v|\sigma_{\varphi\varphi\rightarrow\chi\chi}\right\rangle.
\label{eq:scatterdk1}
\end{equation}

In the following subsections, we leave our results in terms of $m_\varphi$ and $\Gamma_\varphi$, although note that for the numerical spectra we present in Sec.s~\ref{sec:numerical} and~\ref{sec:methods-results}, we consider quadratic inflation where $m_\varphi=2H_e$, and we set $\Gamma/m_\varphi =0.1$ for early reheating. 

\subsection{Conformal Coupling, Late Reheat, Region O}

For conformally coupled scalars, using $\xi = 1/6$ in~\eqref{eq:scatterdk1} gives 
\begin{equation}
    \ant =\frac{9}{512\pi}\frac{\tilde{k}\tilde{m}^{4}}{\tilde{m}_{\varphi}^{5}\tilde{a}^{4}\tilde{H}}\Phi^2,
    \label{eq:scatterantConf}
\end{equation}
where we have converted to dimensionless quantities. 

For late reheating, $\Phi=1$ and inflaton scattering occurs during MD as a resonant production when $\at=\kt/\mt_\varphi$.  Using $\tilde{H}=\tilde{a}^{-3/2}$, Eq.~\eqref{eq:scatterantConf} gives
\begin{equation}
   \left( \at^3\nt_k\right)^{\rm conformal,\, late}_{2\rightarrow 2}=\frac{9}{512\pi}\frac{\tilde{m}^{4}}{\tilde{m}_{\varphi}^{5/2}\tilde{k}^{3/2}},
    \label{eq:scatterLateConf}
\end{equation}
which yields the expected $\tilde{a}^3\tilde{n}_k \propto \kt^{-3/2}$ dependence given in Region O of Tab.~\ref{tab:LateConf}.

For $m>m_\varphi$, we consider $n\varphi \rightarrow \chi \chi$ scattering. As described in Ref.~\cite{Basso_2022}, the higher-n scattering can be characterized as 
\be 
\left(\ant \right)^{\rm conformal, \, late}_{n\rightarrow2}\propto \kt^{-3[2(n-1)-1]/2}.
\label{eq:nto2}
\ee 

Additionally, there exists quantum interference between the leading order and sub-leading order $n\varphi \rightarrow \chi \chi$ processes, which introduce oscillations. These are discussed in detail in Ref.~\cite{Basso_2022}.

\subsection{Conformal Coupling, Early Reheat, Region O}

For early reheating, inflaton scattering again occurs as a resonant production when $\kt = \mt_\varphi \at$, which can occur during either matter or radiation domination depending on $\at$ compared to $\at_\rh$. If $\at < \at_\rh$, then the production becomes relevant during matter domination, while if $\at > \at_\rh$, it will be during radiation domination. 

Using $\tilde{H}=\tilde{a}^{-3/2}$ for MD and $\tilde{H}=\at_\rh^{1/2}/\tilde{a}^{2}$ for RD in Eq.~\eqref{eq:scatterantConf} gives 
\be
\left(\ant \right)^{\rm conformal,\, early}_{2\rightarrow 2} = \left\{
\begin{array}{ll}
    \dfrac{9}{512\pi}\dfrac{\tilde{m}^{4}}{\tilde{m}_{\varphi}^{5/2}\tilde{k}^{3/2}}\Phi_\md^2 & \qquad
    \kt<\mt_\varphi\at_\rh \quad (\mathrm{MD})
    \\ [3ex]
    \dfrac{9}{512\pi}\dfrac{\tilde{m}^{4}}{\tilde{m}_{\varphi}^{3}\tilde{k}\tilde{a}_\rh^{1/2}}\Phi_\rd^2 & \qquad \kt>\mt_\varphi\at_\rh \quad (\mathrm{RD})
    \ ,
\end{array} \right.
\ee
where we have used $\at=\kt/\mt_\varphi$. Using $t=2/(3H_e\Ht)$ for MD and $t=a^2\eta_\rh/2a_\rh \approx \tilde{a}^{2}/2\tilde{a}_\rh^{1/2}H_e$ for RD in Eq.~\eqref{eq:PhiDE} gives 
\begin{align}
    \Phi_\md &= \exp{\left(-\dfrac{2\Gamma_\varphi \tilde{k}^{3/2}}{3H_e\tilde{m}_\varphi^{3/2}}\right)} \ , \label{eq:PhiMD}\\
    \Phi_\rd &=\exp\left(-\dfrac{\Gamma_\varphi\tilde{k}^{2}}{2H_e\tilde{a}_\rh^{1/2}\tilde{m}_{\varphi}^{2}}\right) \ .
    \label{eq:PhiRD}
\end{align}
This result is presented in Region O of Tab.~\ref{tab:InterConf}. Since the early reheating model is only operative for $\mt<1$, the effects of higher-order scattering are subdominant. 

\subsection{Minimal Coupling, Late Reheat, Region O}

For minimally coupled scalars, using $\xi = 0$ in~\eqref{eq:scatterdk1} gives 
\be
\ant=\dfrac{9\tilde{k}}{128\pi\tilde{m}_{\varphi}\tilde{a}^{4}\tilde{H}}\Phi^{2},
\label{eq:scatterantMin}
\ee
where we have converted to dimensionless quantities. 

For late reheating, using $\Phi=1$ and $\tilde{H}=\tilde{a}^{-3/2}$ in Eq.~\eqref{eq:scatterantMin} gives
\begin{equation}
    \left(\ant\right)^{\rm minimal, \, late}_{2\rightarrow 2} = \dfrac{9\tilde{m}_{\varphi}^{3/2}}{128\pi\tilde{k}^{3/2}} \ , 
\end{equation}
which yields the expected $\tilde{a}^3\tilde{n}_k \propto \kt^{-3/2}$ dependence given in Region O of Tab.~\ref{tab:LateMin}.

The $\kt$ dependence of $n\rightarrow 2$ scattering for $n>2$ will again follow from Eq.~\eqref{eq:nto2}, with a slightly different numerical prefactor than in the conformal case. 

\subsection{Minimal Coupling, Early Reheat, Region O}

For early reheating, we again consider $2\rightarrow2$ scattering to take place in both MD and RD, and do not consider high-order scattering processes. Using $\tilde{H}=\tilde{a}^{-3/2}$ for MD and $\tilde{H}=\at_\rh^{1/2}/\tilde{a}^{2}$ for RD in Eq.~\eqref{eq:scatterantMin} gives
\be
\left(\ant \right)^{\rm minimal,\, early}_{2\rightarrow 2}= \left\{
\begin{array}{ll}
    \dfrac{9\tilde{m}_{\varphi}^{3/2}}{128\pi\tilde{k}^{3/2}} \Phi_\md^2 & \qquad 
    \kt<\mt_\varphi\at_\rh \quad (\mathrm{MD})
    \\ [3ex]
    \dfrac{9\tilde{m}_\varphi}{128\pi\tilde{k}\at_\rh^{1/2}} \Phi_\rd^2 & \qquad \kt>\mt_\varphi\at_\rh \quad (\mathrm{RD})
    \ , 
\end{array} \right.
\ee
where $\Phi_\md$ and $\Phi_\rd$ are again given by Eqs.~\eqref{eq:PhiMD} and ~\eqref{eq:PhiRD} respectively. This is the result given in Region O of Tab.~\ref{tab:InterMin}.

\bibliographystyle{JHEP} 
\bibliography{ref}

\end{document}